\documentclass[times,twocolumn,final,longtitle]{elsarticle}
\usepackage{aas_macros}
\usepackage{jasr}
\usepackage{framed,multirow}
\usepackage[utf8]{inputenc}
\usepackage[T1]{fontenc}
\usepackage{textcomp}
\usepackage{tabularx}
\usepackage{amssymb}
\usepackage{longtable}
\usepackage{latexsym}
\usepackage{amsmath}    %
\usepackage{bm}        
\usepackage{graphicx}  
\usepackage{appendix}
\usepackage{pdflscape}
\usepackage{booktabs}   
\usepackage{siunitx}  
\usepackage{nicematrix}
\usepackage{caption}
\usepackage{subfig}
\usepackage[switch]{lineno}
\usepackage{url}
\usepackage{xcolor}
\usepackage{orcidlink}
\usepackage{geometry}
\definecolor{newcolor}{rgb}{.8,.349,.1}
\usepackage{hyperref}
\usepackage{comment}

\journal{Advances in Space Research}
\begin{document}
    
\verso{Canbay \textit{et al.}}

\begin{frontmatter}

\title{Integrated Study of NGC 1662, NGC 1778, NGC 2422, and NGC 2437 Open Clusters: From Stellar Parameters to Orbital Trajectories utilizing Gaia DR3 archive}

\author[1]{R. \snm{Canbay}\orcidlink{0000-0003-2575-9892}}\ead{rmzycnby@gmail.com}
 
\author[2]{W. H. \snm{Elsanhoury}\orcidlink{0000-0002-2298-4026}\corref{cor1}}
\cortext[cor1]{Corresponding author}
\ead{elsanhoury@nriag.sci.eg}

\author[3]{D. C. \snm{{\c{C}}\i nar}\orcidlink{0000-0001-7940-3731}}\ead{denizcennetcinar@gmail.com}

\author[3]{S. \snm{Taşdemir}\orcidlink{0000-0003-1339-9148}}\ead{tasdemir.seval@ogr.iu.edu.tr}

\author[4,2]{A. \snm{Haroon}\orcidlink{0000-0002-8194-5836}}\ead{aaharoon@kau.edu.sa}

\author[5]{M. S. \snm{Alenazi}\orcidlink{0000-0003-4673-5187}}\ead{m.alenazi@nbu.edu.sa}

\author[6]{M. I. \snm{Elashiry}\orcidlink{0000-0003-1327-2876}}\ead{mustafa.elashiry@nbu.edu.sa}

\author[7]{A. \snm{Ahmed}\orcidlink{0000-0001-9070-4914}}\ead{ahamza@sci.cu.edu.eg}

\affiliation[1]{organization={Department of Space Sciences and Technologies, Faculty of Science, Akdeniz University},
                postcode={07058},
                city={Antalya},
                country={Türkiye}}

\affiliation[2]{organization={Department of Astronomy , National Research Institute of Astronomy and Geophysics (NRIAG), Helwan 11421},
                city={Cairo},
                country={Egypt}}

\affiliation[3]{organization={Programme of Astronomy and Space Sciences, Institute of Graduate Studies in Science, Istanbul University},
                postcode={34116},
                addressline={Beyazıt},
                city={Istanbul},
                country={Türkiye}}

\affiliation[4]{organization={Department of Astronomy and Space Science, Faculty of Science, King Abdulaziz University},
                city={Jeddah},
                country={Kingdom of Saudi Arabia}}

\affiliation[5]{organization={Department of Physics, College of Science, Northern Border University},
                city={Arar 91431},
                country={Saudi Arabia}}

\affiliation[6]{organization={Department of Mathematics, College of Science, Northern Border University},
                city={Arar 91431},
                country={Saudi Arabia}}

\affiliation[7]{organization={Department of Astronomy, Space Science and Meteorology, Faculty of Science, Cairo University},
                postcode={12613},
                city={Giza},
                country={Egypt}}

\received{\today}
\finalform{}
\accepted{}
\availableonline{}
\communicated{}

\begin{abstract}

We present a comprehensive photometric, astrometric, and dynamical study of four Galactic open clusters, NGC~1662, NGC~1778, NGC~2422, and NGC~2437 using the high-precision data from the \textit{Gaia} DR3 catalog. Stellar samples were extracted from circular regions centered on the literature cluster coordinates, and probable cluster members were identified through astrometric and photometric criteria. Radial density profiles were constructed and modelled with a King-type surface-density distribution to derive the structural parameters of the clusters. Mean parallaxes and proper motions were determined from the cleaned member samples, yielding astrometric distances of 410 $\pm$ 12 pc (NGC~1662), 1761 $\pm$ 335 pc (NGC~1778), 477 $\pm$ 18 pc (NGC~2422), and 1681 $\pm$ 172 pc (NGC~2437). Fundamental astrophysical parameters were estimated through Bayesian isochrone fitting using \textsc{PARSEC} evolutionary models within a Markov Chain Monte Carlo framework. The resulting ages ($\log(t/\text{yr})$) of the target clusters are $8.91_{-0.33}^{+0.24}$ (NGC~1662), $8.20_{-0.11}^{+0.18}$ (NGC~1778), $7.80_{-0.11}^{+0.06}$ (NGC~2422), and $8.56_{-0.34}^{+0.27}$ (NGC~2437), respectively. Luminosity functions constructed from the member samples were transformed into present-day mass functions using an isochrone-based mass–luminosity relation. The derived mass-function slopes are broadly consistent, within uncertainties, with the classical Salpeter form. Mean systemic radial velocities were obtained from \textit{Gaia} DR3 measurements and complementary spectroscopic surveys, enabling the determination of Galactic space-velocity components $(U,~V,~W)$ and the corresponding total space velocities in the Local Standard of Rest frame. Backward orbital integrations over a 3~Gyr baseline were performed using the \textsc{galpy} package within the \textsc{MWPotential2014} Galactic potential. The resulting orbital parameters indicate nearly circular trajectories ($e \le 0.106$) with limited vertical excursions ($Z_{\rm max}<0.3$~kpc), consistent with thin-disc kinematics. The inferred birth radii suggest moderate radial displacements along the Galactic disc, with NGC~2422, NGC~2437, and NGC~1778 showing ossible outward radial displacement and NGC~1662 indicating a possible inward displacement. Overall, the orbital properties and kinematic parameters suggest that all four clusters have evolved within relatively stable Galactic disc orbits.

\end{abstract}


\end{frontmatter} 


\section{Introduction}
A group of stars, usually a few dozen to several thousand in number, that originated almost simultaneously from the same Molecular Cloud (MC) is known as an open cluster (OC). The usual distances of OCs, primarily because of observational bias, are one to a few kpc. Since many of these objects are young, they are only found in the Galactic plane. \cite{2004A&A...414..163S} go into much detail about the age distribution of OCs. It exhibits a metallicity gradient as a function of Galactocentric distance \citep[e.g.][] {2004A&A...414..163S}, but no discernible age-metallicity relation \citep[e.g.][]{1979ApJS...39..135J, Friel95}. Several groups' careful examinations of OCs show that the Galactocentric metallicity gradient was approximately 0.1 dex kpc$^{-1}$ within $\sim$ 8 Gyr ago, flattened with time, and is currently of order 0.04 dex kpc$^{-1}$ \citep{2004ApJ...617.1115D}. This finding illustrates the Galactic disk's enrichment and mixing timescale.

OCs have similar ages and initial chemical compositions because they have a shared genesis. OCs are useful tools for improving models of the Milky Way's (MW's) genesis and evolutionary history because of their similar characteristics. Astronomers can estimate the chemical properties and approximate location of the parent gas cloud at the time of the cluster's formation by figuring out important parameters like age, spectral features, and correct movements. The structure, history, and movement of gas clouds throughout our Galaxy \citep{2005astro.ph..9152V} can be better understood by examining clusters that span a broad range of ages, masses, and developmental stages.

NGC~1662, NGC~1778, NGC~2422, and NGC~2437 are included in recent open-cluster compilations, including the Unified Cluster Catalog (UCC; \citet{Perren_2023}). The UCC provides an important catalog-level reference by collecting representative parameters for Galactic open clusters (OCs) from recent literature and homogeneous catalog analyses. The aim of the present work is not to duplicate these catalog entries, but to provide an independent and internally homogeneous Gaia DR3-based analysis of the four clusters. In this framework, membership determination, structural modelling through radial density profiles, Bayesian PARSEC isochrone fitting, luminosity and mass-function analysis, member-filtered radial-velocity estimation, Galactic space velocities, internal dynamical diagnostics, and orbit integrations are treated using the same methodology for all clusters.

The UCC values are therefore used here as an external benchmark for the basic astrometric and astrophysical parameters. The comparison shows that our parallaxes, proper-motion components, photometric distances, metallicities, and systemic radial velocities are generally consistent with the representative UCC values. The added value of the present study is that these catalog-level parameters are connected to a single, self-consistent Gaia DR3 analysis that also provides member-based kinematical quantities, relaxation and mass-segregation diagnostics, and Galactic orbital parameters for the four clusters.

\subsection{NGC 1662}
The constellation Orion involves the loosely bound open cluster NGC 1662, additionally referred to Collinder 55. NGC 1662 locates with \cite{2023A&A...674A..37G} at ($\alpha_{\rm J2000}=04^{\rm h}\,48^{\rm m}\,48^{\rm s}.45$, $\delta_{\rm J2000}=10^{\circ}\,52^{\prime}\,46^{\prime\prime}.30$), situated in the directionof the anticenter ($l = 187^{\circ}.7992$) and relatively distant from the Galactic plane ($b=-21^{\circ}.0750$). Corresponding to $Z=140$ pc for a distance of 378 pc \citep{1972MmSAI..43..117H}. This distance to the plane may seem surprising for such a young cluster $\log({\rm t}/yr)$ = 8.11 \citep{1972MmSAI..43..117H}.In accordance with \citep[e.g.][]{1930LicOB..14..154T,1931AnLun...2....1C,1931AnLun...2...64C}, the cluster has angular diameters of 14, 20, 16.5, and 19 (arcmin) and the corresponding distances of about 840, 725, 725, and 760 pc. 

In addition, Trumpler identified the cluster as \textsc{I2p}. The distance, color excess $E(B-V)$, and angular diameter of NGC 1662 were determined to be 415 pc, 0.34 mag, and 14 arcmin based on photometric observations carried out at the U.S. Naval Observatory later in 1961 \citep{1961PUSNO..17..343H}. These are inconsistent with the earlier findings. According to \cite{1997AJ....114.2556T}, NGC 1662 OC has distance modulus $(m-M)=9.14$ mag and their color excess $E(B-V)=0.30$. Moreover, its proper motion components are $(\mu_{\alpha}\cos{\delta},~\mu_{\delta};~\text{mas yr}^{-1})=(-3.20 \pm 0.60, -1.60 \pm 0.60)$ \citep{2000A&A...357..149D}. \cite{2021A&A...647A..19T} investigated that NGC 1662 radial velocity as $\langle R_V \rangle=-13.21 \pm 0.16$ km s$^{-1}$.

\subsection{NGC 1778}
In the direction of the Perseus arm with a diameter of roughly 10 arcmin, an open cluster NGC 1778 is fixed with equatorial coordinates by \cite{2023A&A...674A..37G} with $\alpha_{\rm J2000}=05^{\rm h}\,08^{\rm m}\,07^{\rm s}.86$, $\delta_{\rm J2000}=37^{\circ}\,01^{\prime}\,00^{\prime\prime}.70$ corresponds to Galactic coordinates ($l,~b$) = (168$^{\circ}$.9160,~-2$^{\circ}$.0090).

Using Johnson photometry \cite{1965ApJS...12..215H} reported a distance $d = 1380$ pc, reddening of $E(B-V) = 0.34$ mag, and distance modulus of $(m-M) = 10.7$. A distance $\sim$ 1670 pc, $E(B-V) = 0.33$ mag, and $(m-M) = 11.11$ were estimated by \cite{1973A&AS...10....1B}. Moreover, $d = 1100$ pc, $(m-M) = 10.2$, and age as $\log({\rm t}/yr)$ of 11.3 were bold attention by \cite{1998BASI...26..539P}. On the other hand, \cite{2007ApJ...660..319X} and his co-authors investigated for NGC 1778 that $(m-M) = 11.53$, $E(B-V) = 0.23$ mag, $d = 2023$ pc, and $\log({\rm t}/yr)$ = 11.18. Also, NGC 1778 OC exhibits $\langle R_V \rangle=7.50 \pm 1.40$ km s$^{-1}$ by \cite{2017A&A...600A.106C}.

\subsection{NGC 2422}
The age of the comprehensive Pleiades cluster, i.e., $\sim110-160$ Myr \cite{2018ApJ...863...67G} is similar to that of the southern Galactic open cluster NGC 2422. NGC 2422's Galactic and equatorial coordinates are ($l,~b$) = (230$^{\circ}$.9780,~03$^{\circ}$.1230), and $\alpha_{\rm J2000}=07^{\rm h}\,36^{\rm m}\,33^{\rm s}.82$, $\delta_{\rm J2000}=-14^{\circ}\,30^{\prime}\,15^{\prime\prime}.10
$, respectively via \cite{2023A&A...674A..37G}. Its $\langle R_V \rangle=36.14 \pm 0.46$ km s$^{-1}$ by \cite{2021A&A...647A..19T}.

Several authors have already provided estimates of the basic parameters of this cluster, including age $\sim$ 10$^8$ years \citep{1997A&AS..123...25R}, $d=498^{+135}_{-88}$ pc, corresponding to a distance modulus $(m-M)=8.48^{+0.52}_{-0.42}$ \citep{1999A&A...345..471R}, and $E(B-V)=0.088$ mag was reported by \cite{1999AstL...25....7D}.

\subsection{NGC 2437}
NGC 2437 or Messier 46, is a rich and fairly compact lie near the Galactic equator in Monoceros \citep{1941ApJ....94...55C,1962AJ.....67..224F}. NGC 2437 is well suited with angular size $\sim$ 1 deg$^{2}$ \citep{2013PASP..125..115D}. Its recent location is reported by \cite{2023A&A...674A..37G} as ($\alpha_{\rm J2000}=07^{\rm h}\,41^{\rm m}\,45^{\rm s}.38$, $\delta_{\rm J2000}=-14^{\circ}\,50^{\prime}\,42^{\prime\prime}.40$) and ($l,~b$) = (231$^{\circ}$.8869,~04$^{\circ}$.0451), with relatively high $\langle R_V \rangle$ as $54.54 \pm 0.3446$ km s$^{-1}$ \cite{2021MNRAS.504..356D}. In comparison to Solar metallicity isochrones, a color excess $E(B-V)=0.115 \pm 0.035$ mag and $(m-M)=11.05 \pm 0.05$. Moreover, $\log({\rm t}/yr)$ = 8.35 $\pm$ 0.15 \citep{2013PASP..125..115D}.

\begin{table*}[!ht]
\centering
\caption{Comparison of literature parameters for the investigated clusters. Listed are equatorial coordinates ($\alpha$, $\delta$; sexagesimal), cluster radius ($r$; arcmin), Age, distance ($d$; pc), color excess ($E(B-V)$; mag), trigonometric parallax ($\varpi$; mas), proper motion components ($\mu_\alpha \cos\delta$,~$\mu_\delta$; mas\,yr$^{-1}$), number of members ($N$), metallicity ([Fe/H]; dex), and references.}
\label{tab:literature}
\setlength\tabcolsep{2pt}
\scalebox{0.85}{
\begin{tabular}{cccccccccccc}
\hline\hline
$\alpha$ & $\delta$ & $r$ & Age & $d$ & $E(B-V)$ & $\varpi$ & $\mu_\alpha\cos\delta$ & $\mu_\delta$ & $N$ & [Fe/H] & Ref. \\
(hh:mm:ss) & (dd:mm:ss) & (arcmin) & $\log({\rm t}/yr)$ & (pc) & (mag) & (mas) & (mas\,yr$^{-1}$) & (mas\,yr$^{-1}$) &  & (dex) &  \\
\hline
\multicolumn{12}{c}{NGC 1662} \\
\hline
04:48:27.36  & 10:54:13.54 & 131.80$^+$ & 8.77 & 405 & -- & $2.439 \pm 0.085$ & $-1.154 \pm 0.245$ & $-1.966 \pm 0.210$ & 384 & -- & 1 \\
04:48:47.52 & 10:52:55.20 & 36.12$^*$ & 8.89 & 428 & -- & $2.435 \pm 0.060$ & $-1.123 \pm 0.181$ & $-1.961 \pm 0.198$ & 228 & -- & 2 \\
04:48:59.42 & 10:51:25.92 & 36.18$^*$ & 8.77 & 410 & -- & $2.405 \pm 0.107$ & $-1.121 \pm 0.278$ & $-1.971 \pm 0.240$ & 236  & $-0.10 \pm 0.01$ & 3 \\
04:48:47.52 & 10:52:55.20 & 36.12$^*$ & -- & 412 & -- & $2.400 \pm 0.096$ & $-1.128 \pm 0.225$ & $-1.962 \pm 0.229$ & 238 & -- & 4 \\
04:48:47.52 & 10:52:55.20 & 36.12$^*$ & 8.89 & 428 & -- & $2.400 \pm 0.096$ & $-1.128 \pm 0.225$ & $-1.962 \pm 0.229$ & 231 & -- & 5 \\
04:48:27.00 & 10:56:12.01  & 8.32 & 8.62 & 437 & 0.30 & -- & $-0.090 \pm 3.130$ & $-2.440 \pm 2.550$ & 55 & -- & 6 \\
04:48:27.00 & 10:56:12.01 & 11.00 & -- & -- & -- & -- & $0.790 \pm 2.860$ & $-2.160 \pm 1.800$ & 151 & -- & 7 \\
04:48:26.88 & 10:55:30.00& 29.10  & 8.70 & 437 & 0.30 & -- & $-2.500 $ & $-2.670$ & -- & $-0.095$ & 8 \\
04:48:46.59 & 10:53:21.37 & $38.50^{\ast}$ & 8.68 & 430 & $0.31^{\dagger}$ & 2.434 & $-1.149$ & $-1.976$ & 455 & {$-0.100$} & 9 \\
\hline
\multicolumn{12}{c}{NGC 1778} \\
\hline
05:08:06.22 &  37:01:06.31 & 34.06$^+$ & 8.41 & 1594 & -- & $0.583 \pm 0.060$ & $0.472 \pm 0.116$ & $-3.208 \pm 0.106$ & 243 & -- & 1 \\
05:08:07.92 &  37:01:12.00 & 6.72$^*$ & 8.25 & 1663 & -- & $0.586 \pm 0.035$ & $0.460 \pm 0.071$ & $-3.215 \pm 0.069$ & 151 & -- & 2 \\
05:08:09.10 &  37:00:55.44& 6.30$^*$ & 7.99 & 1480  & -- & $0.581 \pm 0.076$ & $0.497 \pm 0.184$ & $-3.139 \pm 0.132$ & 171 & $-0.07 \pm 0.11$ & 3 \\
05:08:07.92 &  37:01:12.00 & 6.72$^*$ & -- & 1657 & -- & $0.581 \pm 0.066$ & $0.506 \pm 0.139$ & $-3.145 \pm 0.115$ & 175 & -- & 4 \\
05:08:07.92 &  37:01:12.00 & 6.72$^*$ & 8.25 & 1663 & -- & $0.581 \pm 0.066$ & $0.506 \pm 0.139$ & $-3.145 \pm 0.115$ & 115 & -- & 5 \\
05:08:04.00 &  37:01:23.99 & 5.00 & 8.15 & 1469 & 0.34 & -- & $-0.700 \pm 4.530$ & $-3.650 \pm 4.480$ & 113  & -- & 6 \\
05:08:04.00 &  37:01:23.99& 5.00 & -- & -- & -- & -- & $-0.510 \pm 0.100$ & $-3.930 \pm 1.150$ & 116 & -- & 7 \\
05:07:58.80  & 37:01:40.80& 10.20 & 8.45 & 1500 & 0.30 & -- & $-0.760$ & $-7.590$ & -- & -- & 8 \\
05:08:07.77 & 37:01:10.81 & $7.50^{\ast}$ & 8.18 & 1500 & $0.34^{\dagger}$ & 0.577 & 0.465 & $-3.190$ & 332 & $-0.101$ & 9 \\
\hline
\multicolumn{12}{c}{NGC 2422} \\
\hline
07:36:40.58 & $-$14:29:33.56  & 56.58$^+$ & 8.29 & 469 & -- & $2.100 \pm 0.048$ & $-7.069 \pm 0.241$ & $1.042 \pm 0.222$ & 1129 & -- & 1 \\
07:36:35.28 & $-$14:29:20.4 & 15.54$^*$ & 8.52 & 504 & -- & $2.101 \pm 0.041$ & $-7.037 \pm 0.234$ & $1.028 \pm 0.238$ & 411 & -- & 2 \\
07:36:34.06 & $-$14:29:00.24& 15.36$^*$ & 8.19 & 477  & -- & $2.069 \pm 0.073$ & $-7.053 \pm 0.272$ & $0.968 \pm 0.274$ & 433 & $0.13 \pm 0.05$ & 3 \\
07:36:35.28  & $-$14:29:20.40 & 15.54$^*$ & -- & 477 & -- & $2.070 \pm 0.066$ & $-7.057 \pm 0.259$ & $0.993 \pm 0.247$ & 442 & -- & 4 \\
07:36:35.28  & $-$14:29:20.40 & 15.54$^*$ &  8.52 & 504 & -- & $2.070 \pm 0.066$ & $-7.057 \pm 0.259$ & $0.993 \pm 0.247$  & 416 & -- & 5 \\
07:36:35.00 & $-$14:28:59.99 & 13.00 & 7.86 & 490 & 0.07 & -- & $-2.960 \pm 4.580$ & $0.570 \pm 4.730$ & 984 & -- & 6 \\
07:36:35.00 & $-$14:28:59.99 & 13.50 & -- & -- & -- & -- & $-3.350 \pm 3.140$ & $0.770 \pm 2.080$ & 1293 & -- & 7 \\
07:36:39.60 & $-$14:29:24.00 & 12.60 & 8.12 & 471 & 0.10 & -- & $-6.630$ & $2.110$ & -- & $-0.081$ & 8 \\
07:36:33.48 & $-14$:29:09.35 & $23.50^{\ast}$ & 8.04 & 480 & $0.07^{\dagger}$ & 2.093 & $-7.046$ & 1.024 & 1348 & 0.110 & 9 \\
\hline
\multicolumn{12}{c}{NGC 2437} \\
\hline
07:41:50.35 & $-$14:50:00.73 & 39.41$^+$ & 8.84 & 1556 & -- & $0.599 \pm 0.057$ & $-3.871 \pm 0.120$ & $0.407 \pm 0.118$ & 2577 & -- & 1 \\
07:41:46.80 & $-$14:50:38.40 & 12.66$^*$ & 8.48 & 1511 & -- & -- &-- &-- & 1717 & -- & 2 \\
07:41:44.71& $-$14:54:55.44 & 11.82$^*$ & 8.71& 1521 & -- & $0.609 \pm 0.065$ & $-3.837 \pm 0.171$ & $0.371 \pm 0.151$ & 952 & $0.01 \pm 0.07$ & 3 \\
07:41:46.80 & $-$14:50:38.40 & 12.66$^*$ & -- & 1586 & -- & $0.603 \pm 0.060$ & $-3.838 \pm 0.149$ & $0.365 \pm 0.144$   & 1797 & -- & 4 \\
07:41:46.80 & $-$14:50:38.40& 12.66$^*$ & 8.48 & 1511 & -- & $0.603 \pm 0.060$ & $-3.838 \pm 0.149$ & $0.365 \pm 0.144$  & 1733 & -- & 5 \\
07:41:46.00  & $-$14:48:36.00 & 11.00 & 8.40 & 1510  & 0.10 & -- & $-2.850 \pm 4.150$ & $0.330 \pm 4.160$ & 1952 & -- & 6 \\
07:41:46.00 & $-$14:48:36.00 & 11.00 & -- & -- & -- & -- & $-2.600 \pm 1.400$ & $0.090 \pm 1.070$ & 2712 & -- & 7 \\
07:41:49.20  & $-$14:48:18.00  & 13.20 & 8.35 & 1375   & 0.15 & -- & $-3.050$ & $0.380$ & -- & $-0.47$ & 8 \\
07:41:46.93 & $-14$:50:36.67 & $11.30^{\ast}$ & 8.39 & 1510 & $0.15^{\dagger}$ &0.596 & $-3.867$ & 0.399 & 2014 & 0.055 & 9 \\
\hline
\end{tabular}}
\\
\noindent{\footnotesize
Ref.: (1) \cite{Hunt2024}; (2) \cite{Pog21}; (3) \cite{Dias2021}; (4) \cite{Cant20a}; (5) \cite{CantatGaudin2020}; (6) \cite{sampedro2017multimembership}; (7) \cite{dias2014proper}; (8) \cite{kharchenko2013global}; (9) UCC \citet{Perren_2023}. \\
$^*$ Radii enclosing 50\% of the cluster members; $^+$ Tidal radii of the clusters.$^{\dagger}$ The UCC catalog lists $A_V$; the corresponding $E(B-V)$ values shown here were computed as $E(B-V)=A_V/3.1$ for consistency with the table columns.}
\end{table*}
\section{Data}

This study is based on the astrometric and photometric data provided by the third data release of the \textit{Gaia} mission \citep{GaiaDR3}. The \textit{Gaia} DR3 catalog delivers five-parameter astrometric solutions for nearly 1.8 billion sources, including celestial coordinates ($\alpha$,~$\delta$), trigonometric parallaxes ($\varpi$), and proper motion components ($\mu_{\alpha}\cos\delta$, $\mu_{\delta}$), together with homogeneous photometry in the $G$, $G_{\rm BP}$, and $G_{\rm RP}$ bands.

The proper motion measurements are available down to a limiting magnitude of $G = 21$ mag. Compared to \textit{Gaia} DR2, the DR3 release represents a significant improvement in astrometric precision, providing more accurate parallax and proper motion measurements \citep{evans2018}.

\begin{figure*}[!ht]
    \centering
    \includegraphics[width=0.31\linewidth]{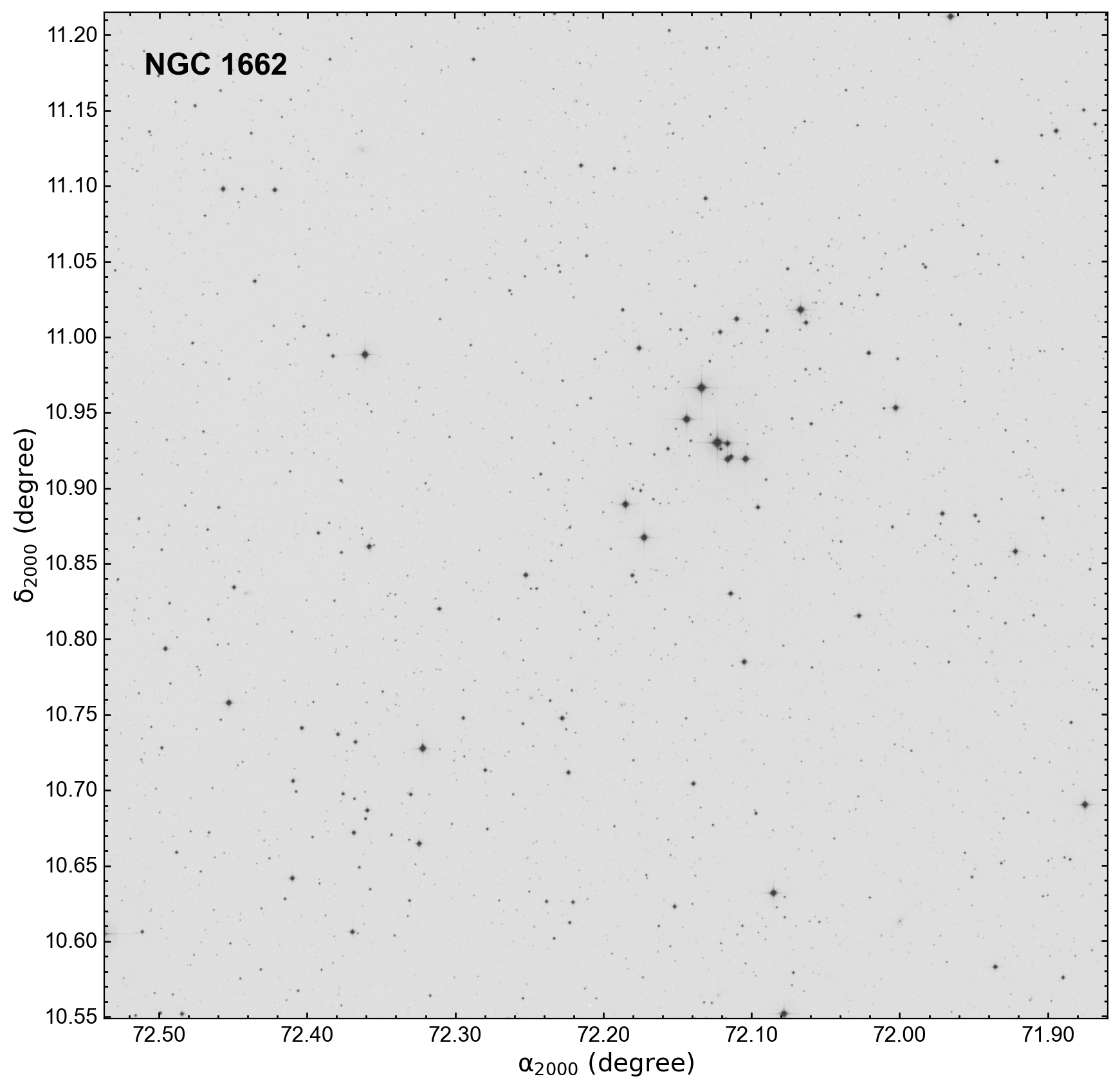}
    \includegraphics[width=0.31\linewidth]{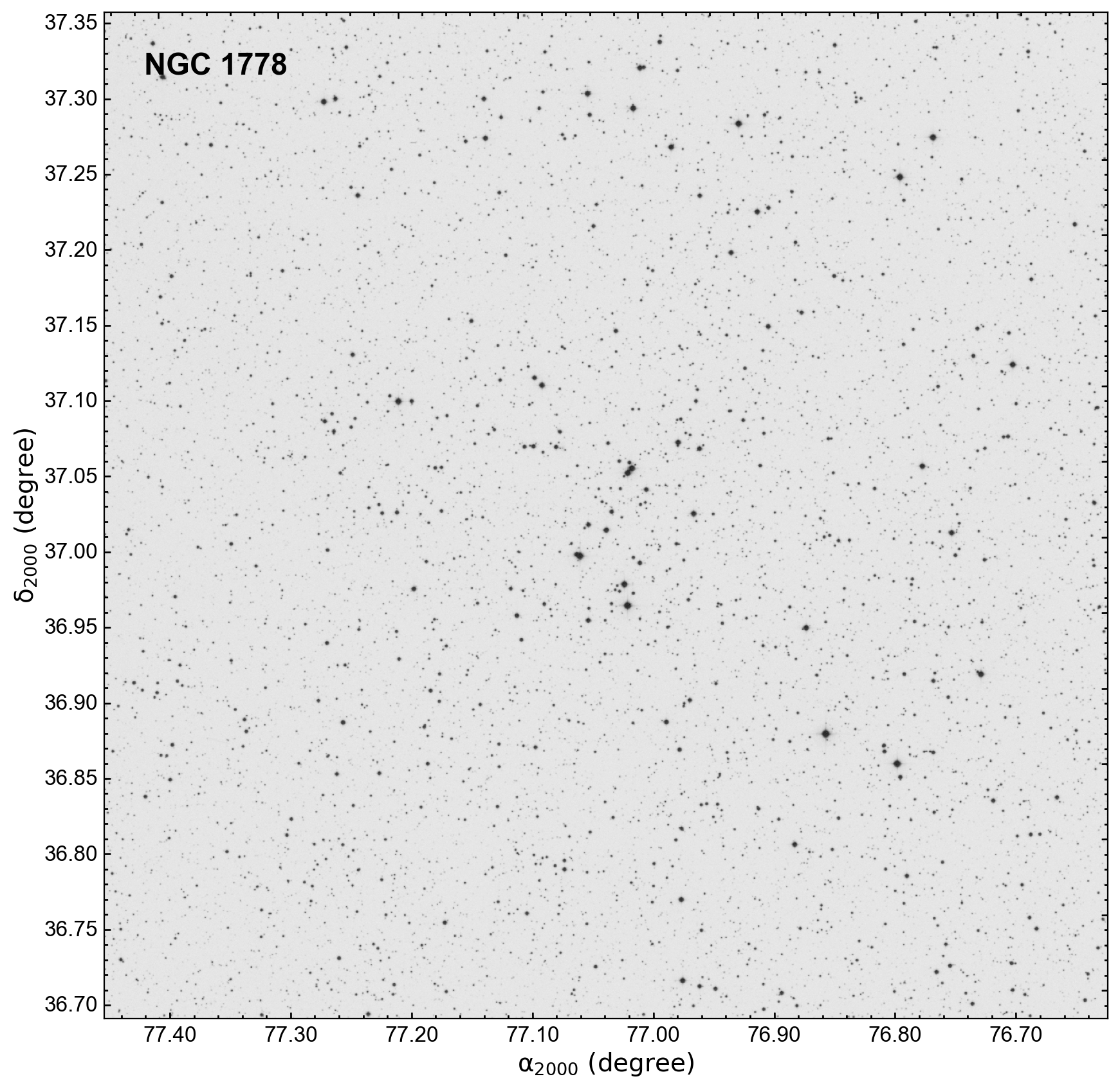}\\
    \includegraphics[width=0.31\linewidth]{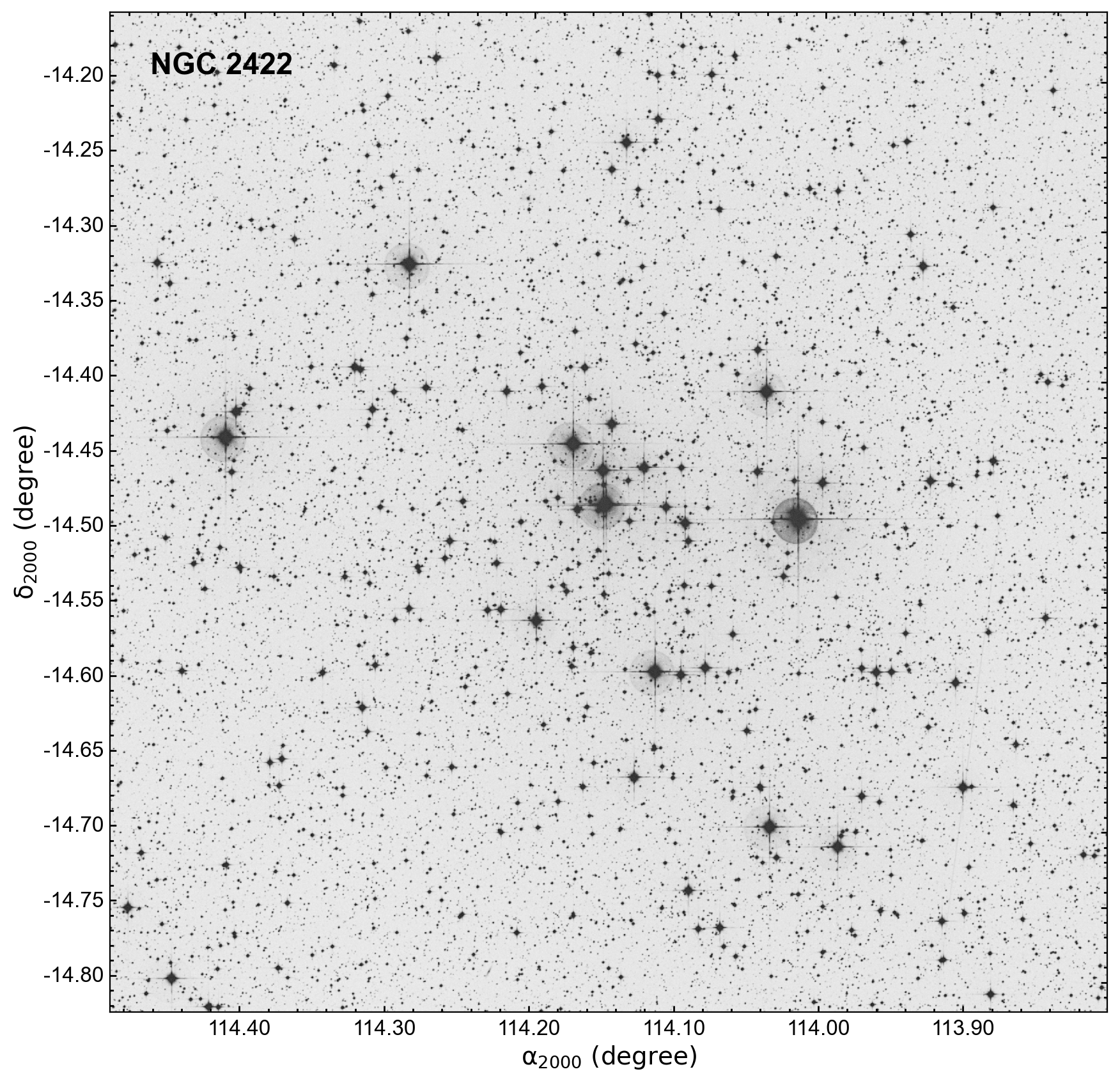}
    \includegraphics[width=0.31\linewidth]{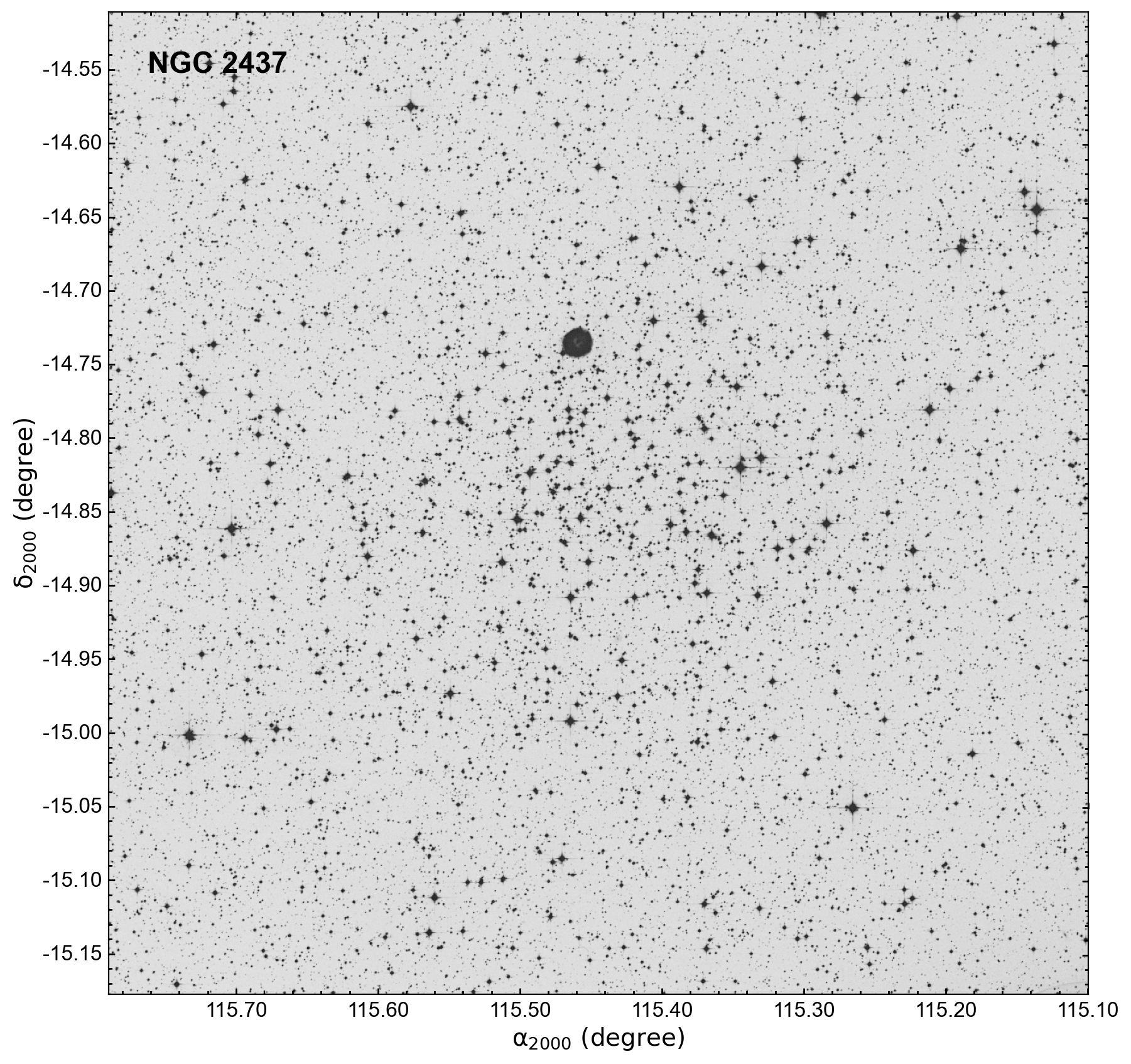}
    \caption{DSS images illustrating the locations of the target clusters.}
    \label{fig:finderchart}
\end{figure*}

For each of the clusters NGC 1662, NGC 1778, NGC 2422, and NGC 2437, the data were extracted from circular regions centeredd on the literature cluster coordinates. The adopted extraction radii were 75 arcmin for NGC 1662, 40 arcmin for NGC 1778, 80 arcmin for NGC 2422 and 50 arcmin for NGC 2437. These radii were chosen to be larger than the reported cluster sizes in order to fully sample four cluster members and the surrounding field population.

Figure~\ref{fig:finderchart} presents the Digitized Sky Survey (DSS) finder charts of the target clusters, displayed using the same angular sizes as the adopted extraction radii. The DSS images provide a visual confirmation that the selected fields fully encompass the clusters and their immediate surroundings. These extraction regions were used consistently throughout the subsequent analyses.

The photometric uncertainties in \textit{Gaia} DR3 show a strong dependence on the $G$ magnitude. For bright stars ($G<15$ mag), the typical uncertainties in the $G$ band are of the order of $\sim~0.003$ mag, while the uncertainties in the color index $(G_{\rm BP}-G_{\rm RP})$ remain below $\sim0.006$ mag for all clusters. In this magnitude range, the photometric quality is nearly uniform among the clusters considered.

At intermediate magnitudes ($15 \leq G \leq 17$ mag), the $G$-band uncertainties remain almost constant at $\sim0.003$ mag, whereas the color uncertainties gradually increase and reach values of about $0.009$ mag. For fainter stars in the range $17 \leq G \leq 19$ mag, the uncertainties in $G$ increase slightly to $\sim~0.003$--$0.004$ mag, while the color uncertainties rise more rapidly, reaching $\sim0.04$ mag at $G \sim 19$ mag.

Beyond $G \approx 19$ mag, the photometric precision decreases significantly. In the interval $19 \leq G \leq 20$ mag, the typical $G$-band uncertainties are $\sim~0.004$--$0.005$ mag, and the color uncertainties increase to about $0.08$ mag. Close to the faint limit of the catalog ($20 \leq G \leq 21$ mag), the $G$-band errors rise sharply to $\sim0.009$--$0.010$ mag, while the color uncertainties reach values of $\sim0.18$--$0.21$ mag, depending on the cluster.

At the faintest magnitudes ($G>21$ mag), the uncertainties increase substantially, with typical values of $\sim0.025$ mag in $G$ and $\sim0.40$ mag in $(G_{\rm BP}-G_{\rm RP})$. These stars are therefore strongly affected by photometric noise and were not considered in the subsequent analysis.

These magnitude-dependent photometric uncertainties were fully taken into account during the membership determination and the subsequent isochrone-fitting procedures. As a result, the final cleaned samples consist of stars with reliable photometric measurements up to the adopted completeness limit. The mean photometric uncertainties as a function of magnitude for each cluster are summarized in Table~\ref{tab:photometric_errors}.

\begin{table*}[h]
\centering
\small
\caption{Mean photometric uncertainties in Gaia apparent magnitudes ($G$) and colors ($G_{ \rm BP}-G_{ \rm RP}$) for stars in the direction of the target OCs. The values are presented as a function of $G$ magnitude intervals.}
\label{tab:photometric_errors}
\resizebox{\linewidth}{!}{%
\begin{tabular}{c|ccc|ccc|ccc|ccc}
\toprule
\hline
 & \multicolumn{3}{c|}{\textbf{NGC 1662}} 
 & \multicolumn{3}{c|}{\textbf{NGC 1778}} 
 & \multicolumn{3}{c|}{\textbf{NGC 2422}} 
 & \multicolumn{3}{c}{\textbf{NGC 2437}} \\
\cmidrule(lr){2-4} \cmidrule(lr){5-7} \cmidrule(lr){8-10} \cmidrule(lr){11-13}
$G$ (mag) 
& $N$ & $\sigma_{{\rm G}}$ & $\sigma_{{G_{{\rm BP}}-G_{{\rm RP}}}}$ 
& $N$ & $\sigma_{{\rm G}}$ & $\sigma_{{G_{{\rm BP}}-G_{{\rm RP}}}}$ 
& $N$ & $\sigma_{{\rm G}}$ & $\sigma_{{G_{{\rm BP}}-G_{{\rm RP}}}}$ 
& $N$ & $\sigma_{{\rm G}}$ & $\sigma_{{G_{{\rm BP}}-G_{{\rm RP}}}}$ \\
\hline
6--14 & 1057 & 0.0028 & 0.0052 & 927 & 0.0029 & 0.0056 & 6046 & 0.0029 & 0.0056 & 2800 & 0.0028 & 0.00570.09 \\ 
14--15 & 1040 & 0.0028 & 0.0056 & 1065 & 0.0028 & 0.0053 & 6409 & 0.0029 & 0.0055 & 3142 & 0.0029 & 0.0054 \\ 
15--16 & 1862 & 0.0029 & 0.0062 & 2097 & 0.0029 & 0.0065 & 12458 & 0.0029 & 0.0064 & 5673 & 0.0029 & 0.0062 \\ 
16--17 & 3163 & 0.0029 & 0.0089 & 3830 & 0.0029 & 0.0095 & 22070 & 0.0030 & 0.0091 & 10140 & 0.0030 & 0.0091 \\ 
17--18 & 5119 & 0.0030 & 0.0174 & 6672 & 0.0030 & 0.0174 & 35674 & 0.0031 & 0.0169 & 16543 & 0.0031 & 0.0169 \\ 
18--19 & 7670 & 0.0033 & 0.0405 & 10371 & 0.0033 & 0.0380 & 51499 & 0.0035 & 0.0382 & 23130 & 0.0035 & 0.0380 \\ 
19--20 & 10723 & 0.0042 & 0.0815 & 14418 & 0.0043 & 0.0820 & 68564 & 0.0047 & 0.0824 & 30343 & 0.0046 & 0.0820 \\ 
20--21 & 14505 & 0.0084 & 0.1775 & 19031 & 0.0091 & 0.1944 & 85590 & 0.0102 & 0.2062 & 37307 & 0.0104 & 0.2129 \\ 
21--23 & 4371 & 0.0265 & 0.3970 & 2677 & 0.0244 & 0.4255 & 7426 & 0.0255 & 0.4169 & 2747 & 0.0260 & 0.4167 \\ 
\hline
Total/Error & 49512 & 0.0069 & 0.1138 & 61089 & 0.0062 & 0.1079 & 295739 & 0.0061 & 0.0991 & 131826 & 0.0060 & 0.0978 \\ 
\bottomrule
\end{tabular}}
\end{table*}
\subsection{Photometric Completeness Analysis}

Determining the photometric completeness limit is a crucial step for deriving reliable cluster parameters. To estimate the completeness level for each cluster, we constructed $G$-band magnitude histograms using all stars located within the selected cluster regions. In all cases, the number of detected stars increases steadily toward fainter magnitudes until a clear turnover point is reached, beyond which the counts decrease due to observational incompleteness.

\begin{figure}[!ht]
    \centering
    \includegraphics[width=0.9\linewidth]{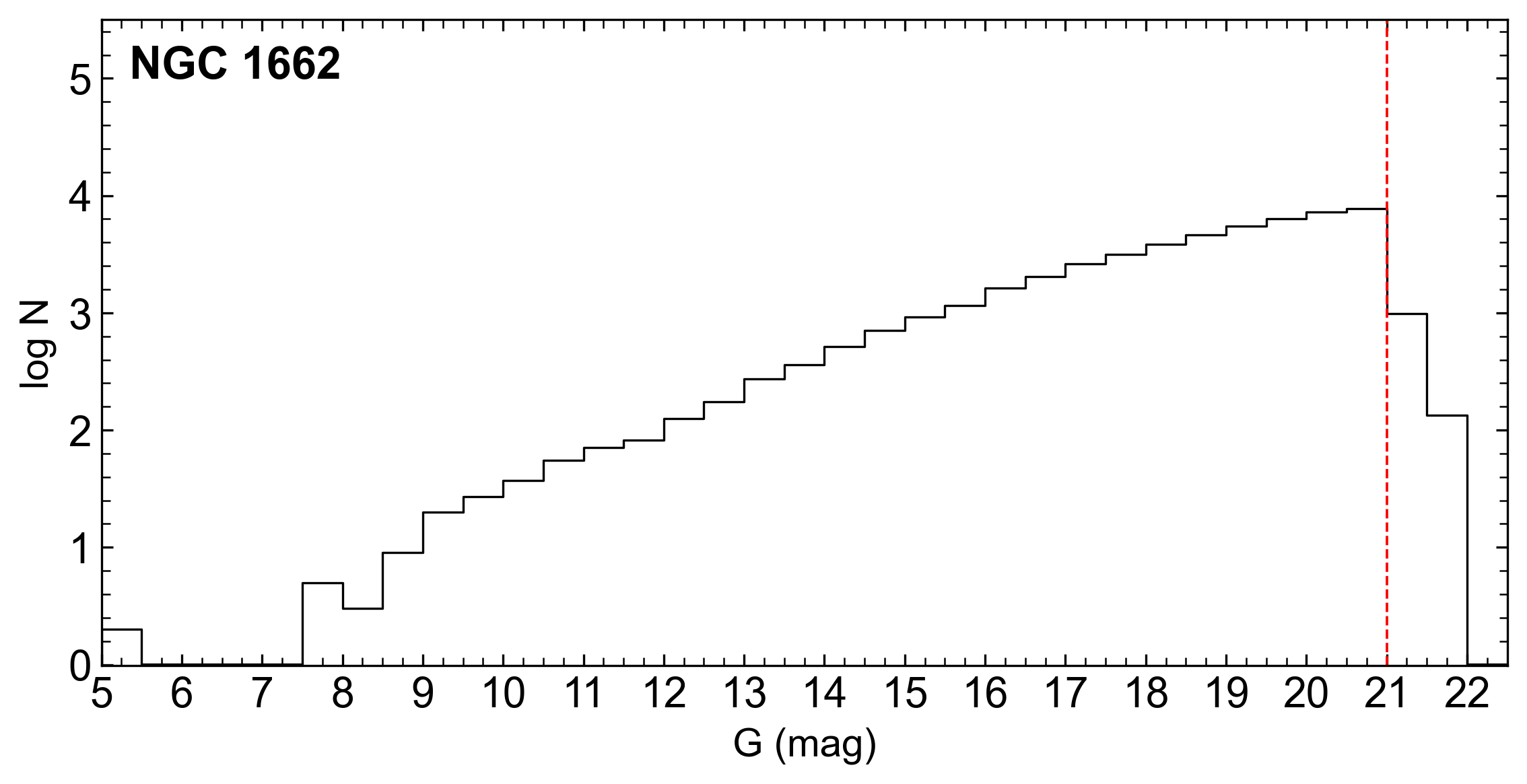}
    \includegraphics[width=0.9\linewidth]{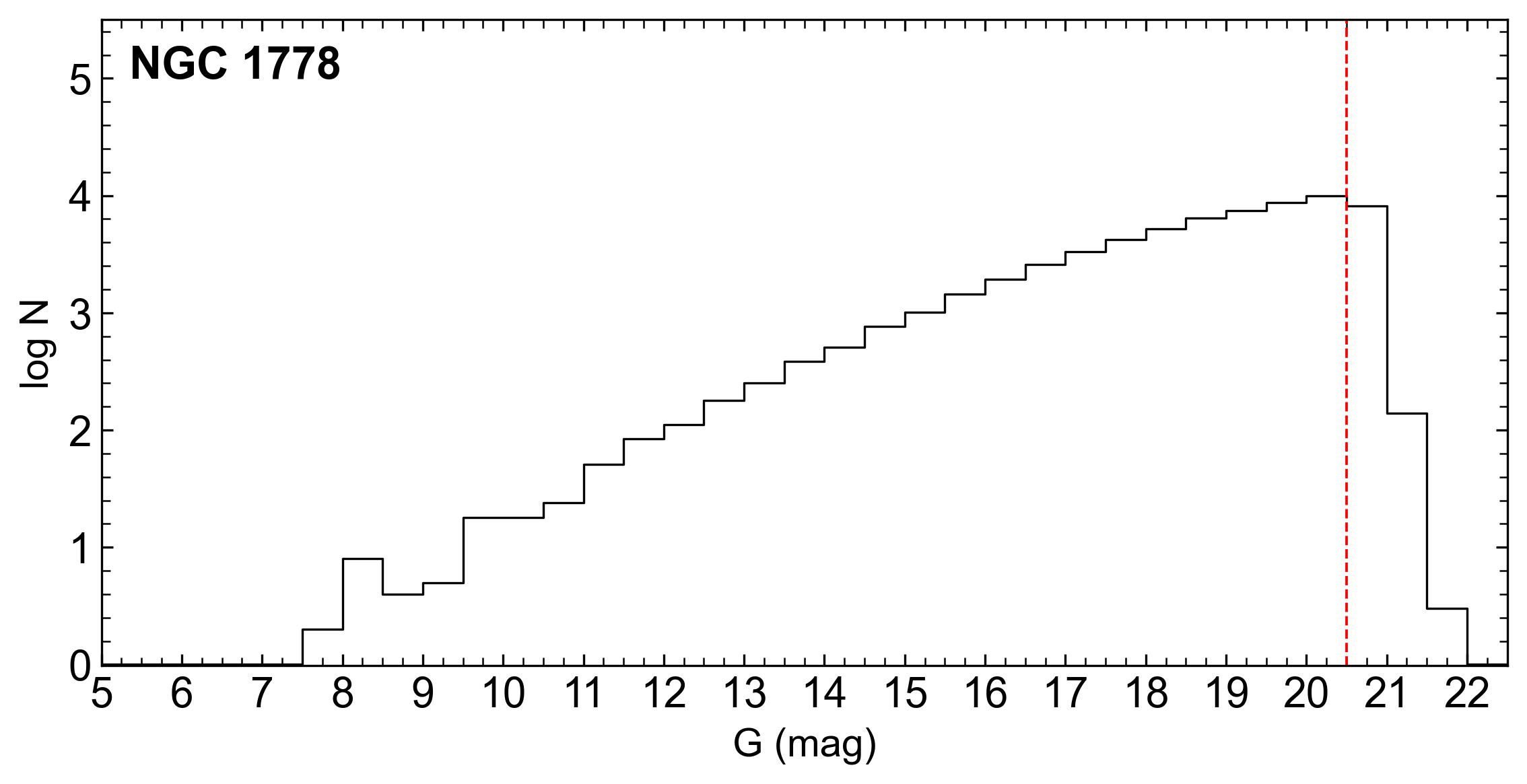}
    \includegraphics[width=0.9\linewidth]{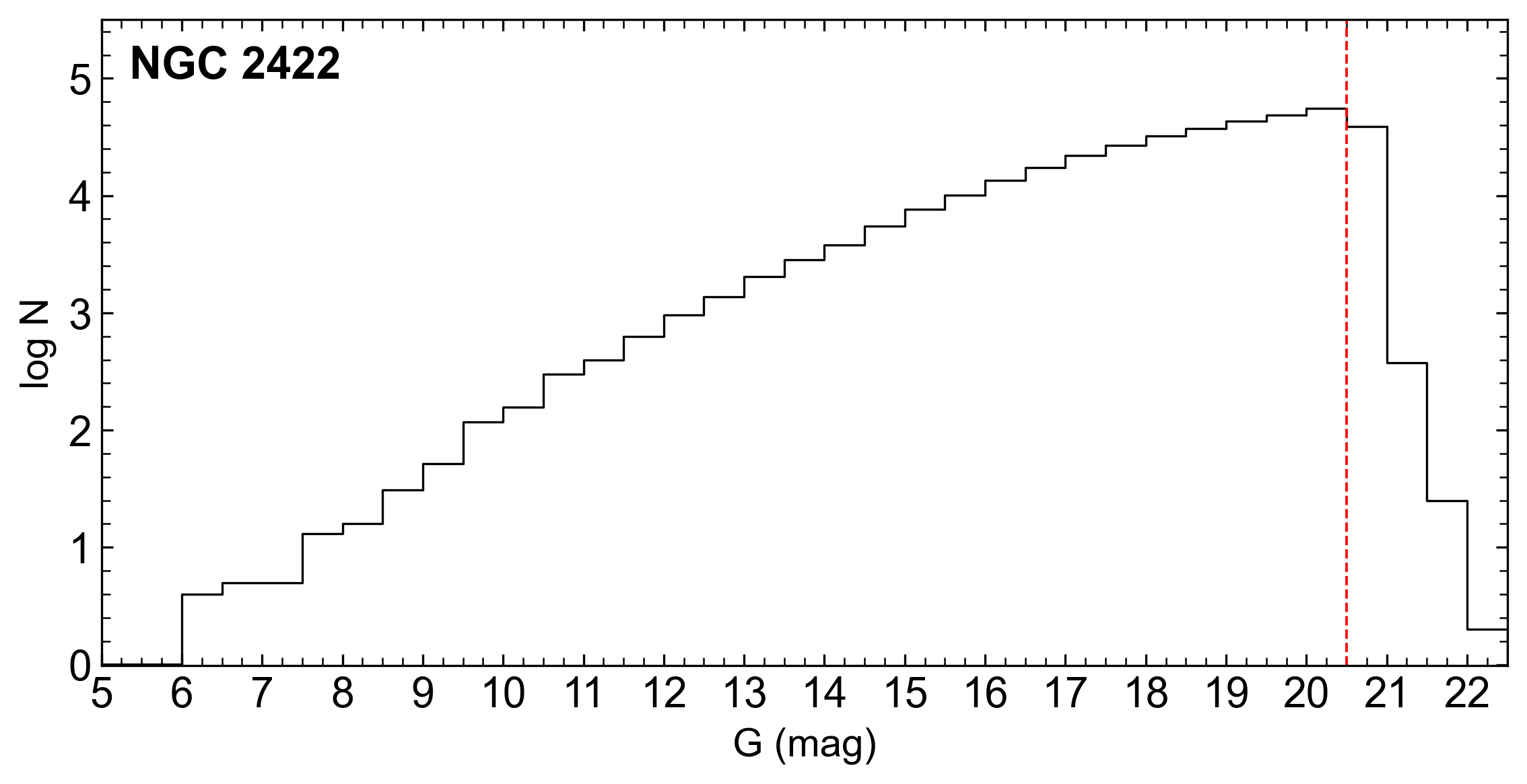}
    \includegraphics[width=0.9\linewidth]{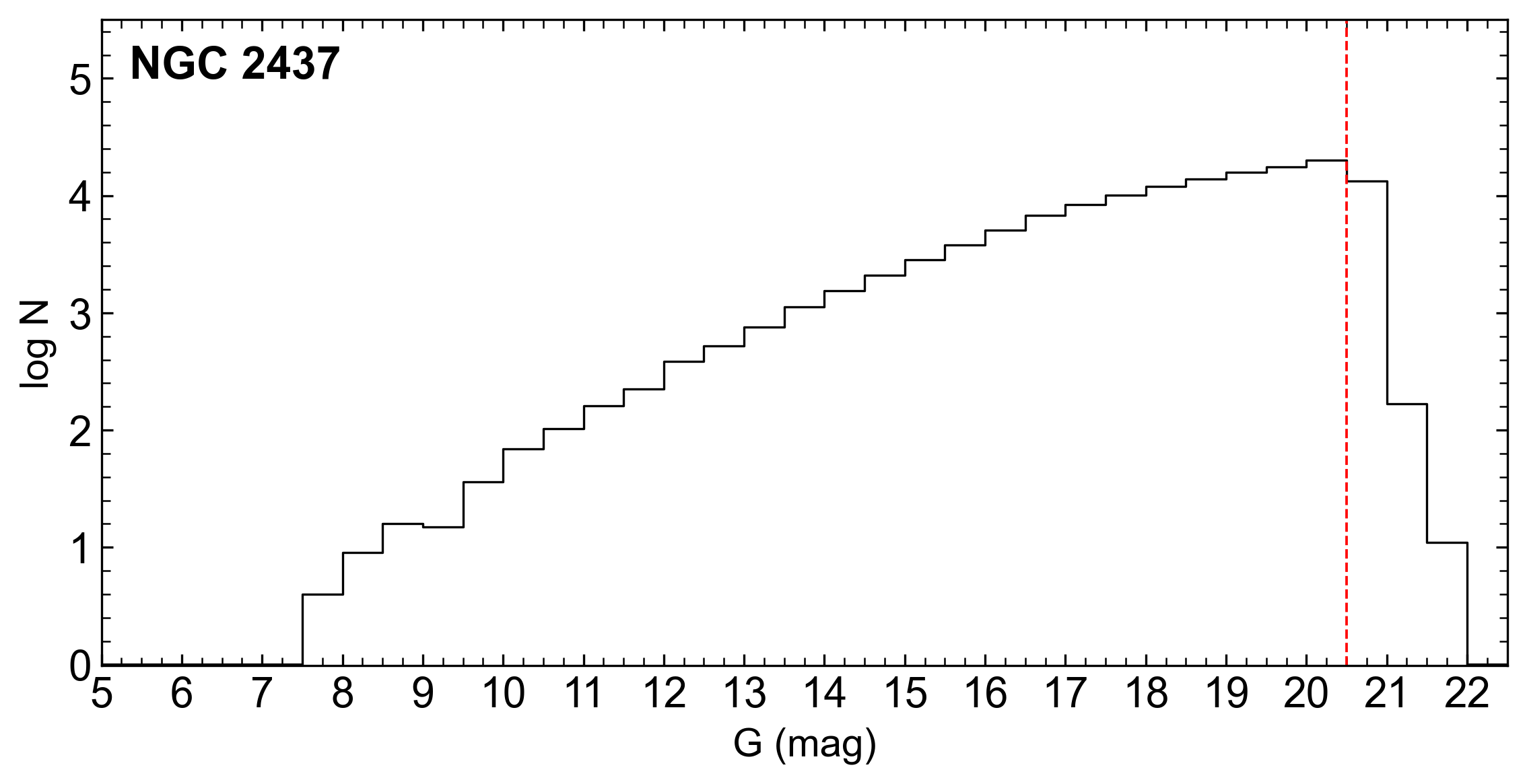}
    \caption{Completeness histograms for the target clusters. The red dashed line indicates the \textit{Gaia} $G$-band completeness limit.}
    \label{fig:completness}
\end{figure}

The photometric completeness analysis was carried out individually for each cluster using the $G$-band magnitude distributions. The adopted completeness limits were found to be $G = 20.5$ mag for NGC 1778, NGC 2422, and NGC 2437, while a slightly fainter limit of $G = 21.0$ mag was obtained for NGC 1662. After applying these magnitude cuts, the final stellar samples consist of 68,039 stars for NGC 1662, 49,474 stars for NGC 1778, 303,076 stars for NGC 2422, and 112,854 stars for NGC 2437. The corresponding completeness histograms are shown 
in Figure~\ref{fig:completness}.

\section{Structure Analysis}
\subsection{Clusters' Centers}
The center of a stellar cluster is commonly defined as the location of maximum projected stellar density on the sky. For NGC 1662, NGC 1778, NGC 2422, and NGC 2437, we determined this position using the equatorial coordinates ($\alpha,~\delta$) of stars extracted from the {\it Gaia} archive. 

For each cluster, one-dimensional stellar count histograms were constructed separately along the right ascension and declination directions. The stars were grouped into consecutive coordinate bins, and Gaussian functions were fitted to the resulting distributions to model the density peaks. The centroid of each Gaussian profile was adopted as the refined central coordinate in the corresponding direction.

\begin{figure*}[!ht]
    \centering
    \includegraphics[width=0.4\linewidth]{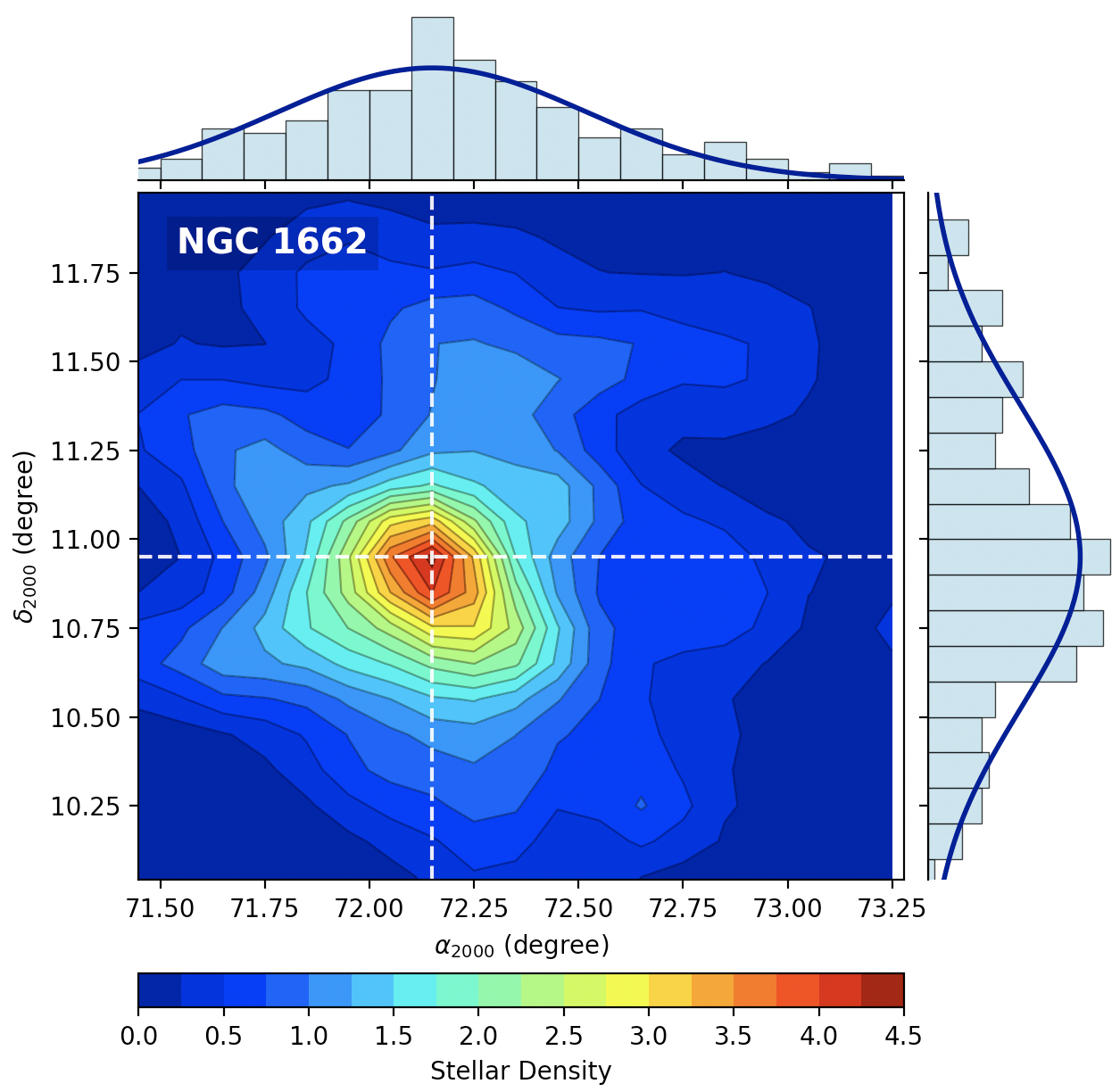}
    \includegraphics[width=0.4\linewidth]{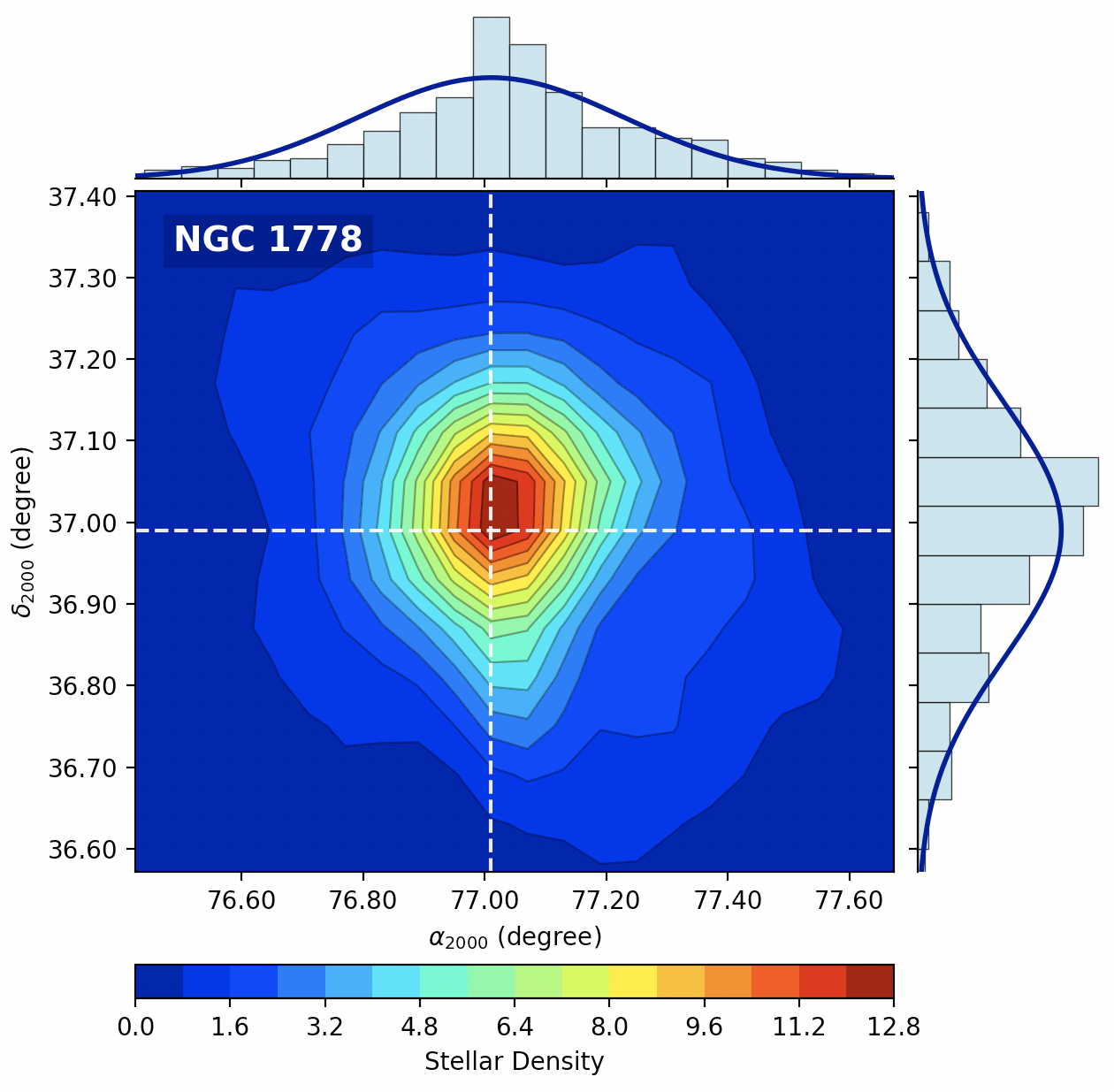}\\
    \includegraphics[width=0.4\linewidth]{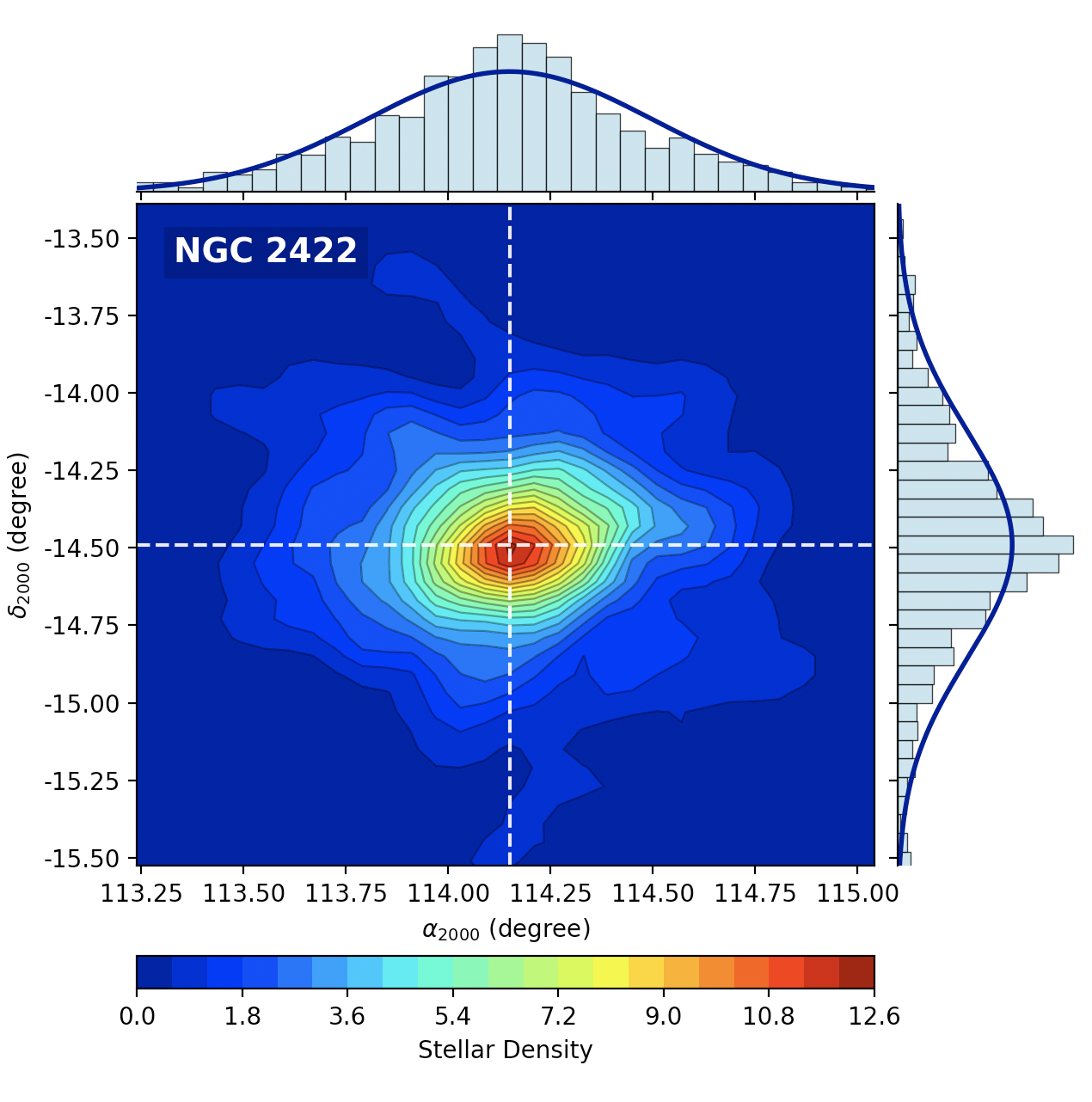}
    \includegraphics[width=0.4\linewidth]{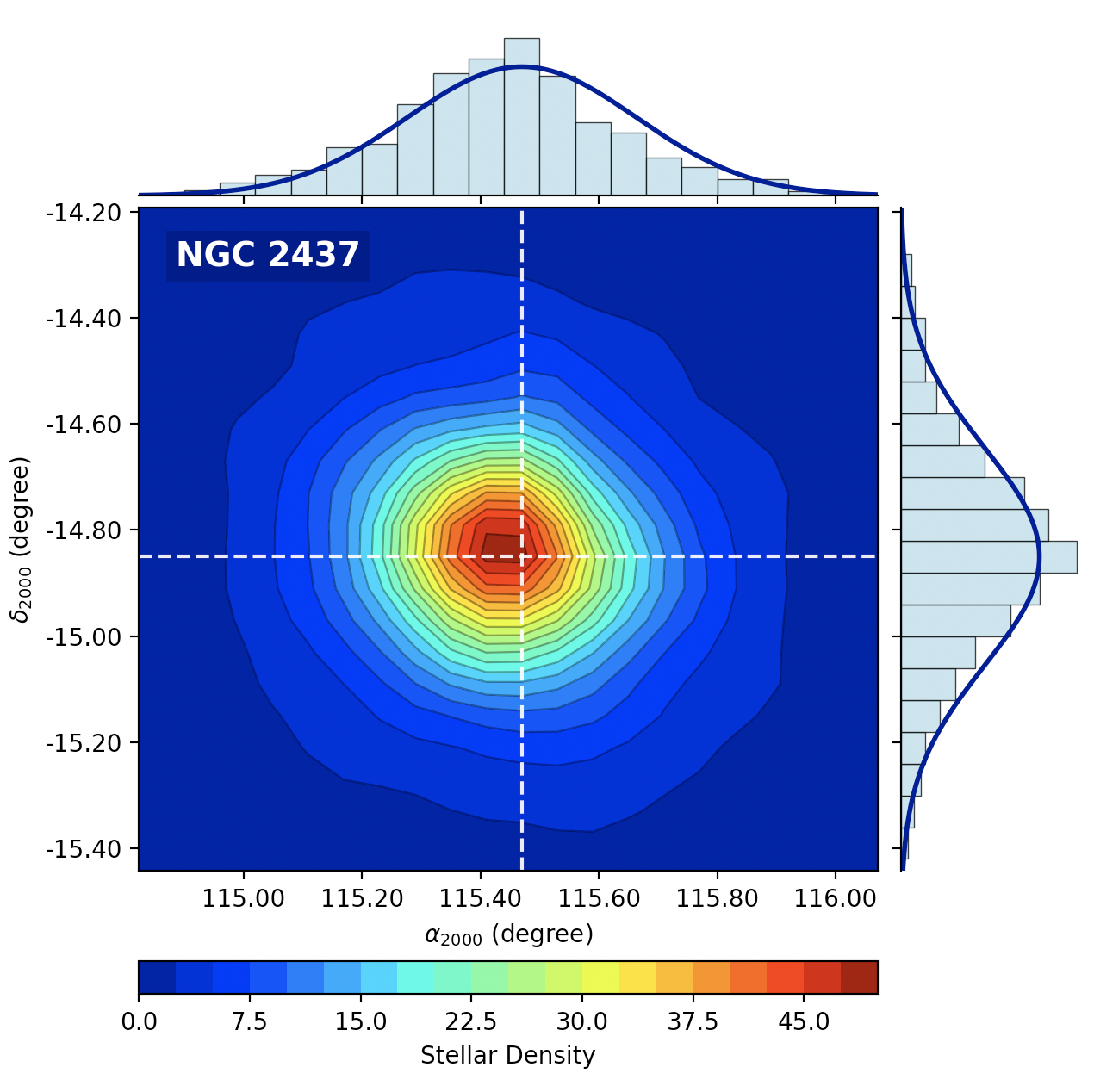}
    \caption{Stellar surface density maps of the target clusters. The color scale represents the projected stellar density, with red tones indicating regions of higher stellar concentration. The cluster centers were determined by fitting Gaussian profiles to the stellar count histograms along the right ascension and declination directions. The dashed lines mark the derived central coordinates obtained from the Gaussian peaks.}
    \label{fig:cluster_centers}
\end{figure*}

The intersection of the Gaussian peaks in $\alpha$ and $\delta$ defines the cluster center, corresponding to the highest stellar concentration. The derived central positions for all four clusters are illustrated in Figure~\ref{fig:cluster_centers}. The final equatorial coordinates, together with the corresponding Galactic coordinates, are listed in Table~\ref{all_results}.

\subsection{Radial Density Profiles}\label{sec:RDP}

\begin{figure*}[!ht]
    \centering
    \includegraphics[width=0.5\linewidth]{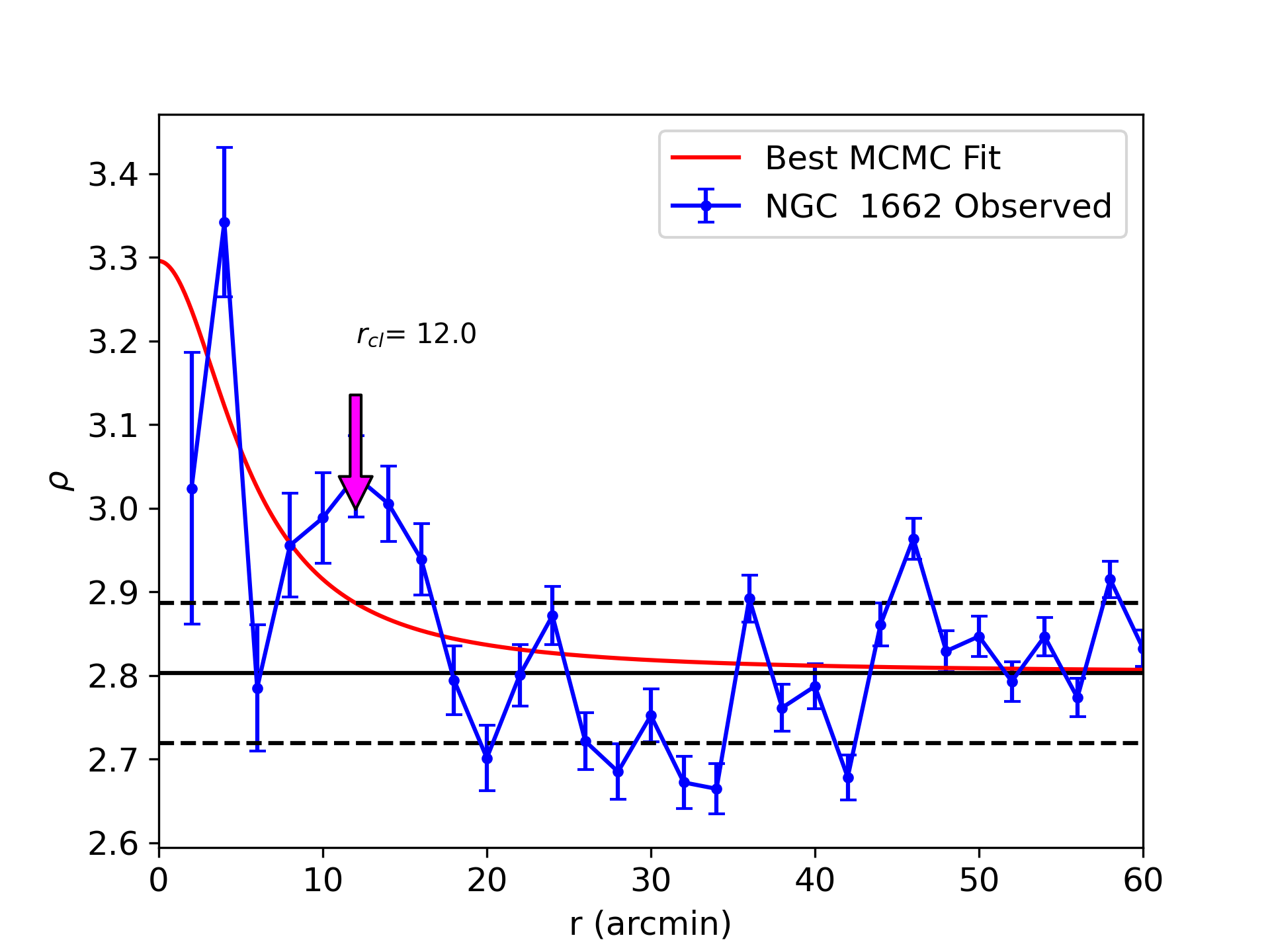}
    \includegraphics[width=0.45\linewidth]{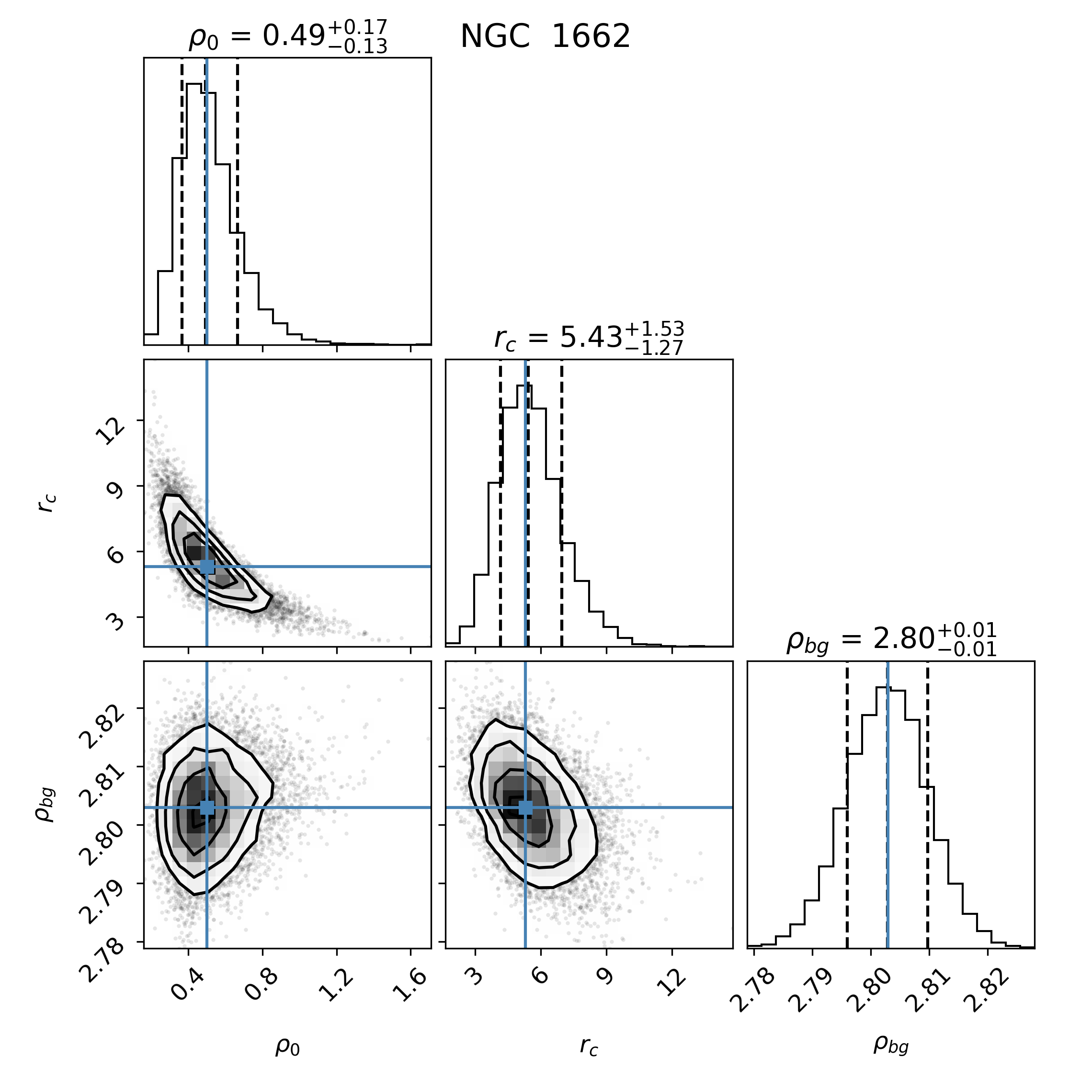}
        \caption{The RDPs of the NGC 1662 OC represented by the blue solid line. The red solid line represents the \citet{King1962} fitting while the black solid line and black dashed line represent the background density $\rho_{\mathrm{bg}}$ and the uncertainties of the background density, respectively.}
    \label{fig:rdps}
\end{figure*}

The structural properties of star clusters can be investigated using their Radial density profiles (RDPs), such as their degree of central concentration and overall richness. For each cluster, the region around the center was divided into concentric annuli, and the stellar surface density in each ring was calculated as
\begin{equation}
\rho(r_i) = \frac{N_i}{A_i},
\label{radial_density}
\end{equation}
where $N_i$ is the number of stars within the $i$th annulus and $A_i$ is its area. The associated uncertainties were estimated assuming Poisson statistics, $1/\sqrt{N_i}$. The resulting RDP of NGC~1662, together with its best-fitting King model, is shown in Figure~\ref{fig:rdps}, while the rest are presented in \ref{rdps-append}. The figure illustrates the observed stellar surface densities as a function of radius, along with the corresponding background levels and model fits for four clusters.

To get the structural parameters, we fitted the observed RDPs with the empirical surface-density model introduced by \citet{King1962}. We adopted the following form of the model
\begin{equation}
\rho(r) = \rho_{\mathrm{bg}}+\frac{\rho_0}{1+(r/r_c)^2},
\label{eq:king_model_short}
\end{equation}
where $\rho_0$ denotes the central stellar density above the background level, $r_c$ is the core radius, and $\rho_{\mathrm{bg}}$ represents the background density.

The model parameters were obtained following the maximum-likelihood approach, with the log-likelihood defined as
\begin{equation}
\ln \mathcal{L} = -\sum_i
\left( \frac{\rho_i - \rho_{i,\mathrm{model}}}{\sigma_{\rho_i}} \right)^2 .
\end{equation}

Parameter sampling was performed with the \texttt{emcee} Markov Chain Monte Carlo sampler \citep{Foreman-Mackey2013}, assuming uniform priors. Convergence of the chains was verified using the Gelman–Rubin statistic \citep{gelman1992}. The resulting structural parameters for four clusters are listed in Table~\ref{King_para}.

Following \citet{King1966}, the concentration parameter was computed as $C = r_{\rm cl} / r_{\rm c}$, which provides a quantitative measure of cluster compactness. Our results are generally in agreement with previous studies, with small differences likely arising from variations in the adopted datasets, profile construction methods, and fitting techniques, as well as from the improved quality and homogeneity of the {\it Gaia} DR3 data used in this work.

\begin{table*}
\centering
\caption{Derived structural parameters for NGC~1662, NGC~1778, NGC~2422 and NGC~2437 obtained from fitting the \citet{King1962} model to the radial density profiles. The columns list the cluster radius ($r_{\rm cl}$), core radius ($r_{\rm c}$), central density ($\rho_0$), background density ($\rho_{\rm bg}$), and concentration parameter ($C$).}
\label{King_para}
\renewcommand{\arraystretch}{1.3}
\begin{tabular}{lccccc}
\hline\hline

Cluster & $r_{\rm cl}$ & $r_{\rm c}$ & $\rho_0$ & $\rho_{\rm bg}$ & $C$  \\
\cline{2-3} \cline{4-5}
 & \multicolumn{2}{c}{(arcmin $\&$ pc)} & \multicolumn{2}{c}{(stars arcmin$^{-2}$)} &    \\
\hline
NGC~1662 & 12.00 $\pm$ 3.46 $\&$ 1.43 $\pm$ 0.01 & 5.43$^{+1.53}_{-1.27}$ $\&$ 0.65 $\pm$ 0.01 &  $0.49^{+0.17}_{-0.13}$  & $2.80^{+0.01}_{-0.01}$   &  2.21 \\
NGC~1778 & 43.81 $\pm$ 6.62 $\&$ 22.46 $\pm$ 4.74 & 9.06$^{+0.08}_{-0.08}$ $\&$ 4.65 $\pm$ 0.46 & $4.46^{+0.04}_{-0.04}$ & $11.96^{+0.00}_{-0.00}$   & 4.84 \\
NGC~2422 & 18.01 $\pm$ 4.24 $\&$ 2.50 $\pm$ 0.06 & 6.25$^{+0.10}_{-0.10}$ $\&$ 0.87 $\pm$ 0.01 &  $2.69^{+0.00}_{-0.00}$  & $18.42^{+0.00}_{-0.00}$   &  2.88 \\
NGC~2437 &39.61 $\pm$ 6.30 $\&$ 19.36 $\pm$ 4.40 & 6.85$^{+0.03}_{-0.03}$ $\&$ 3.35 $\pm$ 1.83 &  $8.57^{+0.04}_{-0.04}$  & $16.96^{+0.00}_{-0.00}$ &  5.78 \\
\hline
\end{tabular}
\end{table*}

\section{The Photometric Analysis}

\subsection{Membership Determination}

The accurate determination of cluster members from the dense stellar background of the Galactic plane is a fundamental prerequisite for any robust astrophysical characterization. OCs are physical groupings of stars that exhibit common kinematic signatures, specifically in their proper motion components and trigonometric parallaxes due to their shared origin. To exploit these features, we performed a membership analysis of NGC 1662, NGC 1778, NGC 2422, and NGC 2437 using the Unsupervised Photometric Membership Assignment in Stellar Clusters (\texttt{UPMASK}) algorithm \citep{Krone-Martins_2014}. 

Unlike traditional methods that rely on restrictive assumptions about cluster distributions, \texttt{UPMASK} employs an automated, data-driven strategy. The algorithm identifies spatially and kinematically coherent groups within a multidimensional parameter space through an iterative process involving $k$-means clustering and statistical resampling. This methodology is particularly well-suited for high-precision \textit{Gaia} datasets, as it effectively disentangles cluster populations from field star contamination even in crowded regions \citep{Cantat-Gaudin2018}.

\begin{figure*}
    \centering
    \includegraphics[width=0.7\linewidth]{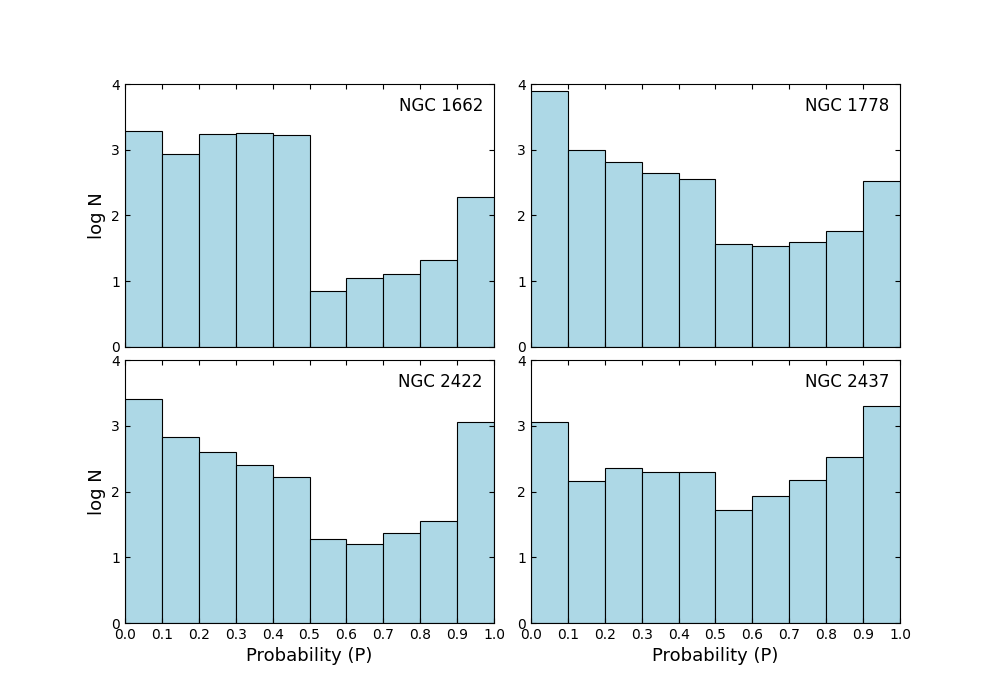}
    \caption{Membership probability ($P$) distributions obtained with \texttt{UPMASK} for the investigated clusters.}
    \label{fig:prob}
\end{figure*}

In this study, the membership probabilities were computed using five-dimensional astrometric data from the \textit{Gaia} DR3 catalog: equatorial coordinates ($\alpha,~\delta$), proper motion components ($\mu_{\alpha} \cos \delta,~\mu_{\delta}$), and trigonometric parallaxes ($\varpi$), along with their respective measurement uncertainties. To robustly separate true cluster members from the dense field star contamination, we utilized the \texttt{UPMASK} open-cluster membership assignment method. The algorithm operates by iteratively partitioning the multidimensional astrometric space into $k$ distinct clusters using a $k$-means clustering approach. In each iteration, the spatial distribution of these $k$-groups is evaluated against a uniform random distribution. Since real open clusters form tightly bound structures in proper motion ($\mu_{\alpha} \cos \delta,~\mu_{\delta}$) and parallax ($\varpi$) spaces compared to the randomly distributed field stars, \texttt{UPMASK} effectively identifies these overdensities without relying on any prior assumptions about the cluster's geometry or stellar evolution models. This iterative process yields a continuous membership probability ($P$) for each star, ranging from $0$ to $1$. To determine an objective and reliable membership threshold, we constructed a membership probability histogram for each cluster (see Figure~\ref{fig:prob}). This histogram reveals a clear bimodal or segregated distribution, separating the heavily populated field star regime (low probabilities) from the distinct cluster member population (high probabilities). Guided by this empirical probability distribution, we adopted a strict threshold of $P \ge 0.50$ (or $50\%$). Only stars satisfying this rigorous mathematical criterion were considered true cluster members and utilized in the subsequent astrophysical analyses, including CMD isochrone fitting, mass function estimations, and galactic orbital trajectories. This thresholding approach ensures that foreground and background field contamination is minimized, even for poorly populated galactic clusters like NGC~1662 and NGC~1778.

According to the membership analysis via \texttt{UPMASK}, the final catalogs of probable cluster members consist of 242 stars for NGC 1662, 502 stars for NGC 1778, 1226 stars for NGC 2422, and 2604 stars for NGC 2437. These high-confidence samples provide the necessary empirical basis for the precise derivation of the clusters' fundamental parameters.

The effectiveness of this unsupervised approach in providing reliable member lists has led to its extensive application in recent Galactic studies \citep[e.g.,][]{CantatGaudin2020, Yontan2022, Karagoz2025}. By applying a probability threshold (e.g., $P \geq 0.5$), we secured a clean sample of member stars for each cluster, which serves as the foundation for the subsequent isochrone fitting and parameter estimation.

\subsection{Astrometric Properties of the OCs}

\begin{figure*}[htb!]
\centering
\includegraphics[width=0.9\linewidth]{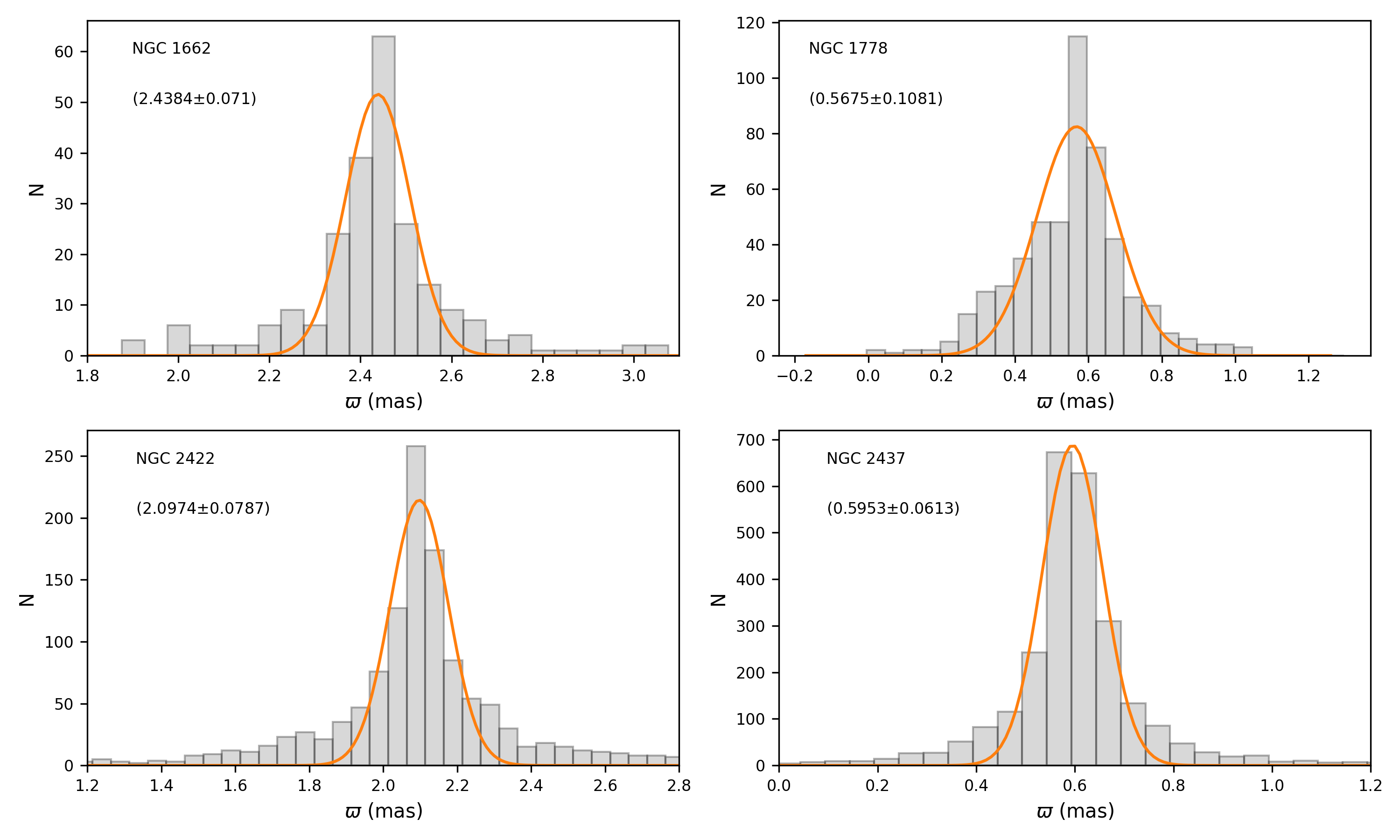}
\caption{Histograms of the trigonometric parallaxes for all member stars of the clusters investigated in this work. The mean values and standard deviations are shown in each panel, while the orange curves represent the Gaussian fits to the observed trigonometric parallax distributions.
\label{fig: Plx}}
\end{figure*}

\begin{figure*}[htb!]
\captionsetup[subfigure]{labelformat=empty}
\centering
\subfloat[]{\scalebox{.4}{\includegraphics{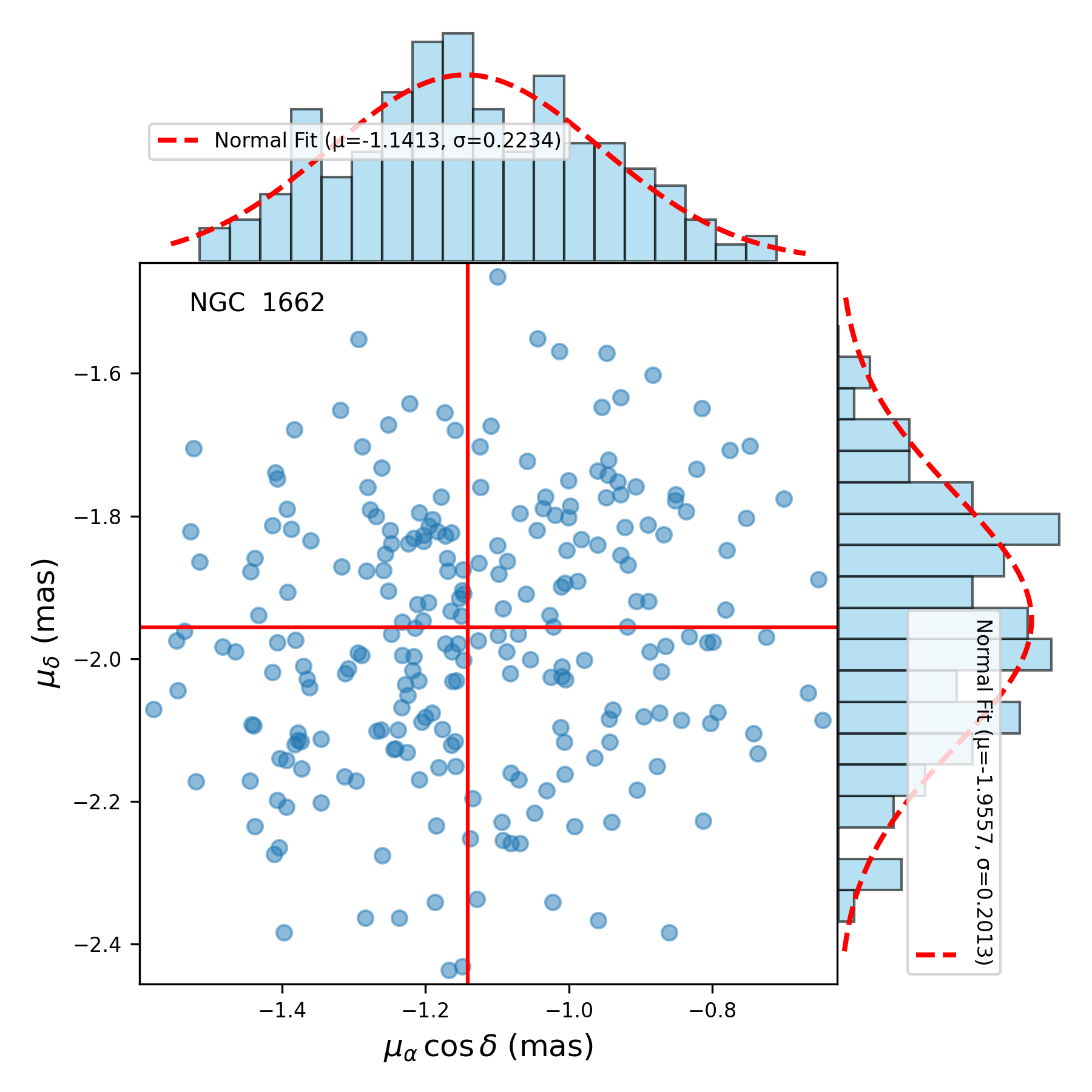}}}
\subfloat[]{\scalebox{.4}{\includegraphics{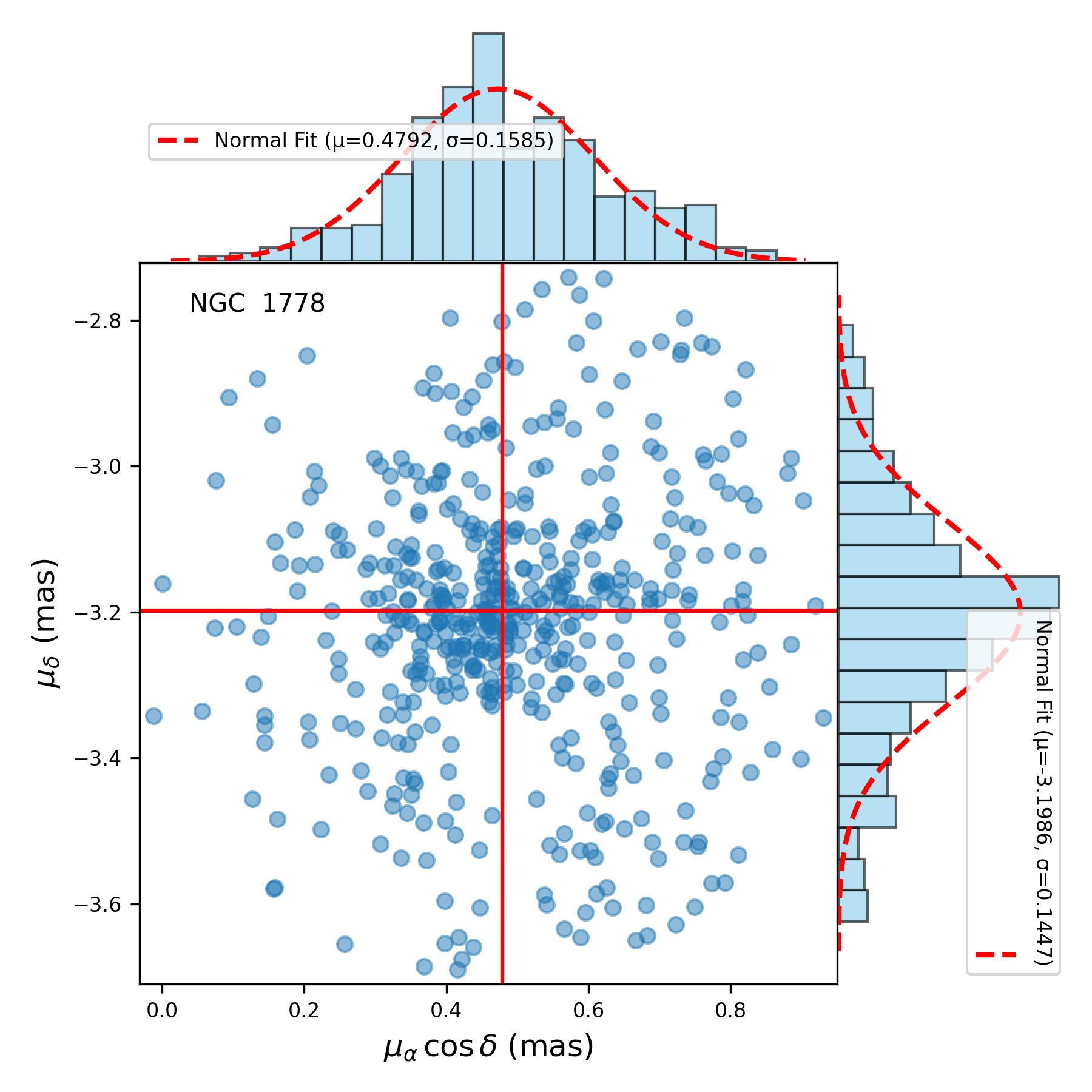}}}\\
\subfloat[]{\scalebox{.4}{\includegraphics{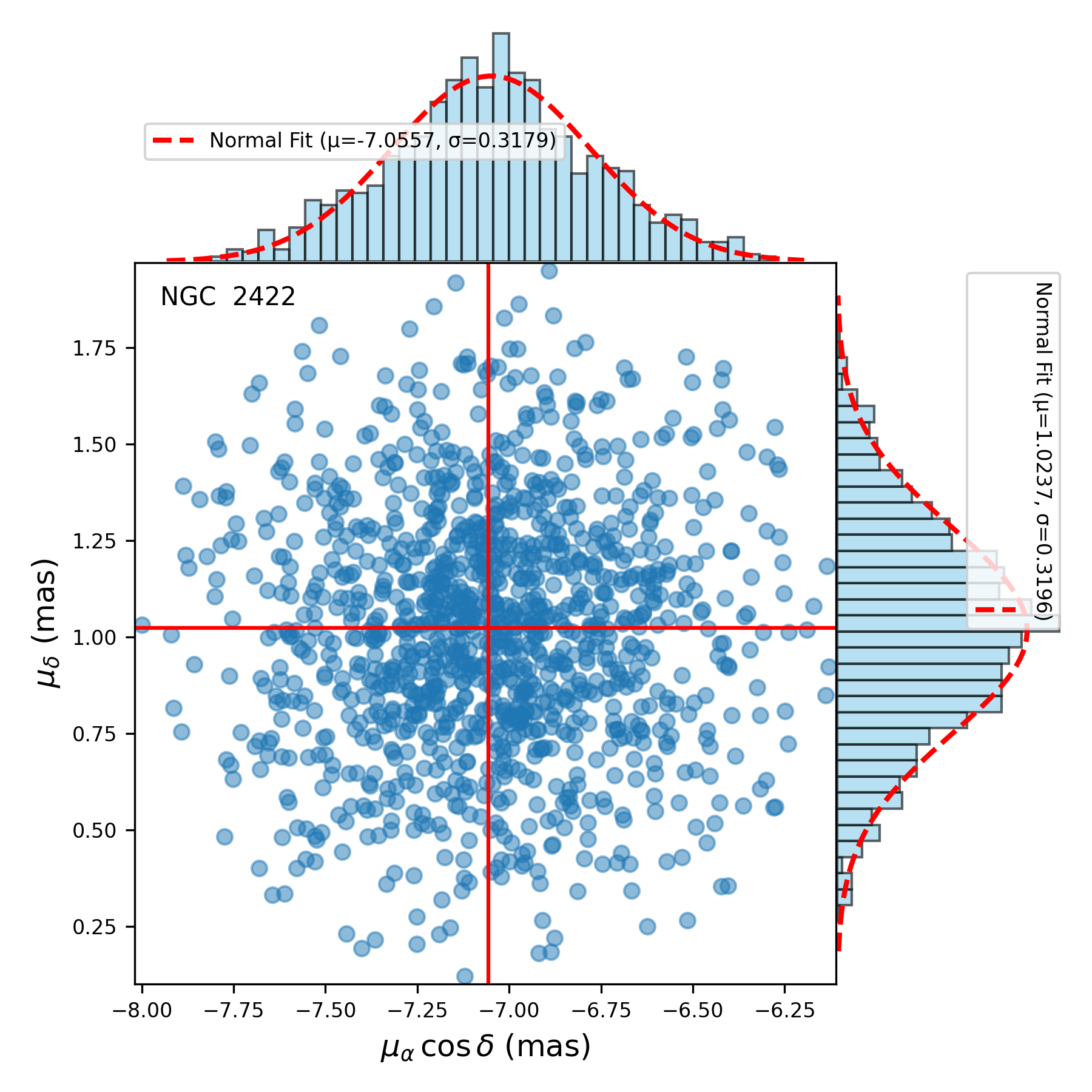}}}
\subfloat[]{\scalebox{.4}{\includegraphics{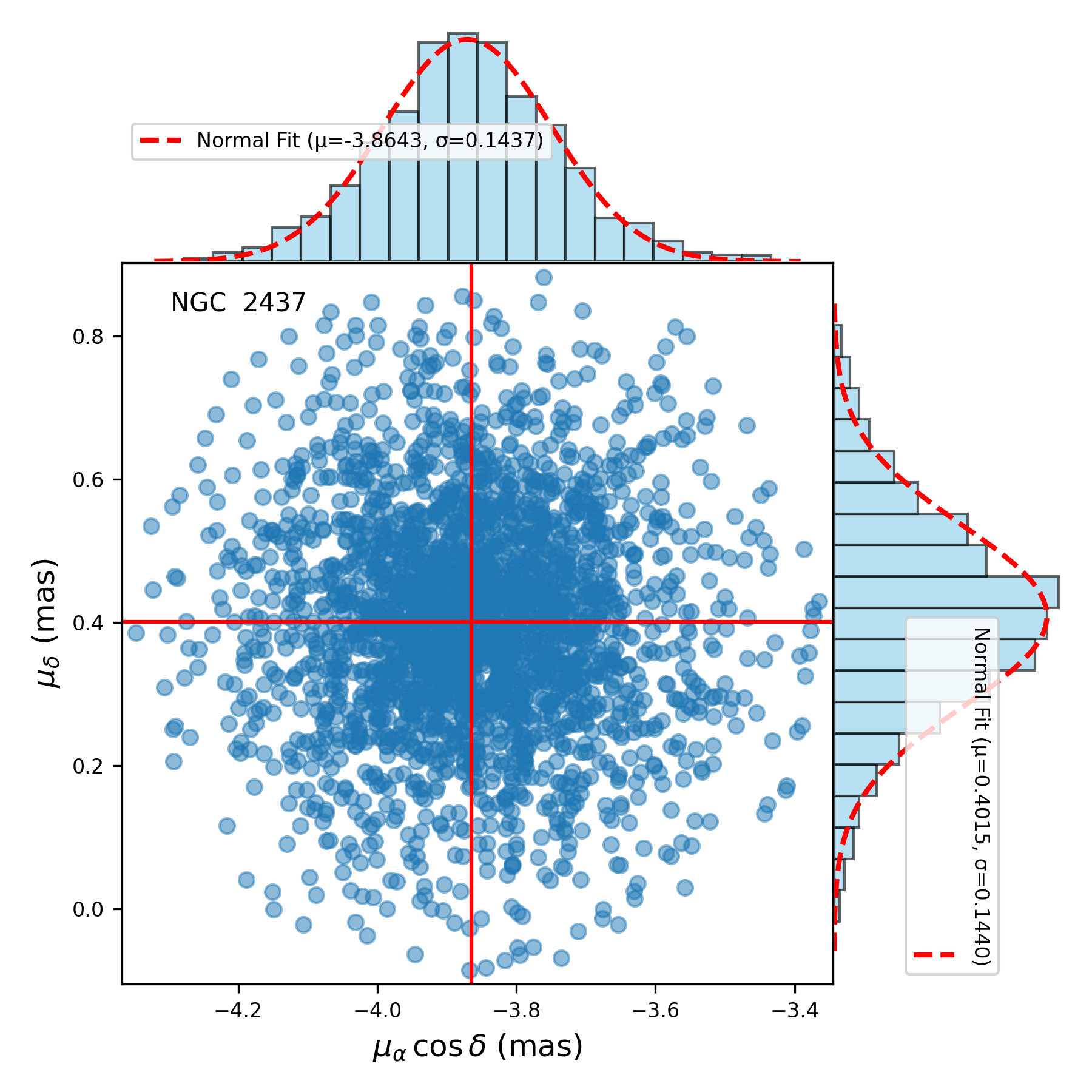}}}
\caption{Histograms of the proper motion components ($\mu_{\alpha}\cos\delta$,~$\mu_{\delta}$) for the member stars of each target cluster. The mean value and standard deviation for each distribution are noted in the panels.
\label{fig: PM}}
\end{figure*}

Estimates of the mean proper-motion components in right ascension and declination
($\mu_{\alpha}\cos\delta$ and $\mu_{\delta}$), as well as the mean trigonometric parallaxes ($\varpi$), were obtained using the astrometric data of the selected member stars of each cluster. In order to achieve this, we constructed histograms of the proper-motion components and parallaxes, which were fitted with Gaussian profiles. The mean values and standard deviations of the best-fitting Gaussians were adopted as the representative astrometric parameters of the clusters (see Figures~\ref{fig: Plx} and \ref{fig: PM}). 
Our measured values are in good agreement with the results of    \cite{Hunt2024}, \cite{Pog21}, \cite{Dias2021}, \cite{Cant20a} and \cite{CantatGaudin2020} which used either {\it Gaia} DR3 or DR2 database as shown in Table \ref{tab:literature}, while they are significantly different from those of \cite{sampedro2017multimembership}, \cite{dias2014proper} and \cite{kharchenko2013global} which relied on less precise astrometric catalogs. 

Next, the mean parallaxes were used to compute cluster distances, yielding 410 $\pm$ 12 pc (NGC~1662), 1761 $\pm$ 335 pc (NGC~1778), 477 $\pm$ 18 pc (NGC~2422), and 1681 $\pm$ 172 pc (NGC~2437). These distances are consistent, within uncertainties, with values reported in the literature based on the most recent {\it Gaia} releases \citep[e.g.][]{Pog21, Dias2021}. All estimated astrometric parameters are summarized in Table~\ref{all_results}.

\subsection{Bayessian Isochrone-Fitting Method}

Accurate determination of fundamental parameters, such as age, chemical abundance, distance, and interstellar extinction, is critical for defining the evolutionary status and Galactic positioning of OCs. In this study, the astrophysical properties of NGC 1662, NGC 1778, NGC 2422, and NGC 2437 were derived using a Bayesian framework based on the Markov Chain Monte Carlo (\texttt{MCMC}) sampling technique. Our analysis utilizes the \texttt{PARSEC} stellar evolutionary models \citep{Bressan12}, following the computational methodology described in \citet{Bilir2026}.

\subsection{Statistical Methodology and Model Fitting}

The estimation of cluster parameters was conducted by sampling the joint posterior probability distribution, assuming that photometric errors follow a Gaussian distribution. The likelihood function, which compares observed magnitudes with model-predicted values in the \textit{Gaia} passbands ($G,~G_{\rm BP},~G_{\rm RP}$), is defined as follows:

\begin{equation}
\ln \mathcal{L}(\boldsymbol{\theta}) = - \frac{1}{2} \sum_{i=1}^{N_{\mathrm{stars}}} \sum_{X} \left[ \frac{(m_{X,i} - m^{\mathrm{mod}}_{X,i})^2}{\sigma_{X,i}^2} + \ln(2\pi \sigma_{X,i}^2) \right]
\end{equation}

where $m_{X,i}$ and $\sigma_{X,i}$ represent the observed magnitude and its associated uncertainty for the $i$-th star in filter $X$, while $m^{\mathrm{mod}}_{X,i}$ is the synthetic magnitude interpolated from the theoretical grid. The parameter vector $\boldsymbol{\theta}$ includes the $\log$ (t), metallicity ($Z$), distance ($d$), and extinction ($A_{\rm G}$). To mitigate the impact of field star contamination, the fitting procedure was restricted to stars with high membership probabilities.

\begin{figure*}[!ht]
\centering
\includegraphics[width=0.85\linewidth]{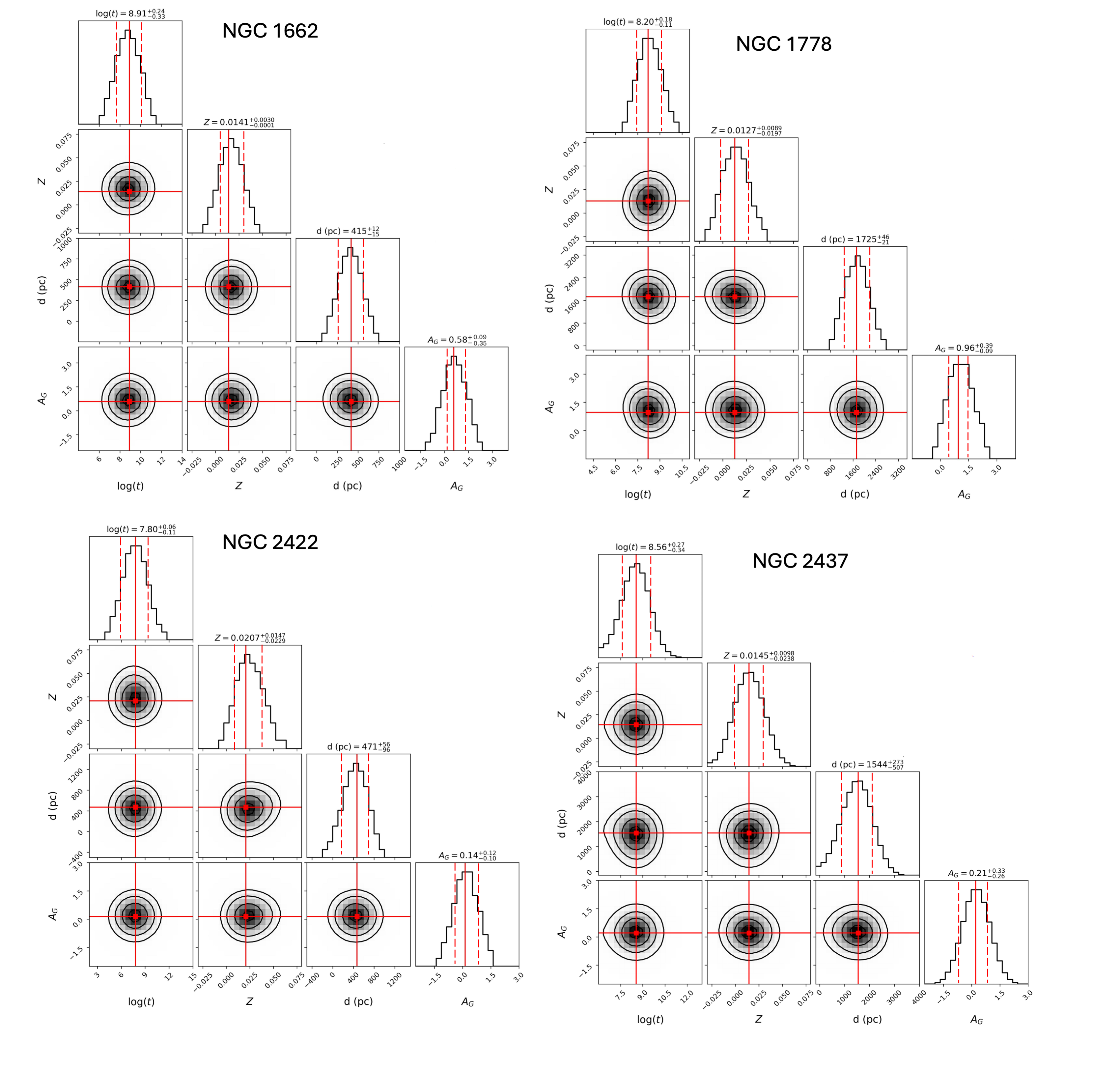}
\caption{Visual representation of the MCMC sampling results for the targeted OCs. The corner plots illustrate the multidimensional joint posterior probability distributions for the derived astrophysical parameters ($\log$~(t),$~Z,~d,~A_{\rm G}$). In each panel, the central solid red line identifies the median value of the sampled distribution, representing the most probable parameter estimate. The associated $1\sigma$ uncertainties are indicated by the dashed vertical red lines, which correspond to the $16^{\rm th}$ and $84^{\rm th}$ percentiles of the posterior samples. These distributions demonstrate the degree of parameter correlation and the convergence of the Markov chains for each cluster.}
\label{fig:mcmc_corner}
\end{figure*}

The posterior space was explored using the affine-invariant ensemble sampler provided by the \texttt{emcee} package \citep{Foreman-Mackey2013}. The number of walkers was scaled according to the number of identified member stars in each cluster, and each chain was evolved for 5,000 steps to ensure convergence. Following the removal of the burn-in phase, the final parameters were extracted as the median of the distributions, with uncertainties defined by the $16^{\rm th}$ and $84^{\rm th}$ percentiles ($1\sigma$ confidence intervals).

\begin{figure*}[htb!]
\captionsetup[subfigure]{labelformat=empty}
\centering
\subfloat[]{\scalebox{.4}{\includegraphics{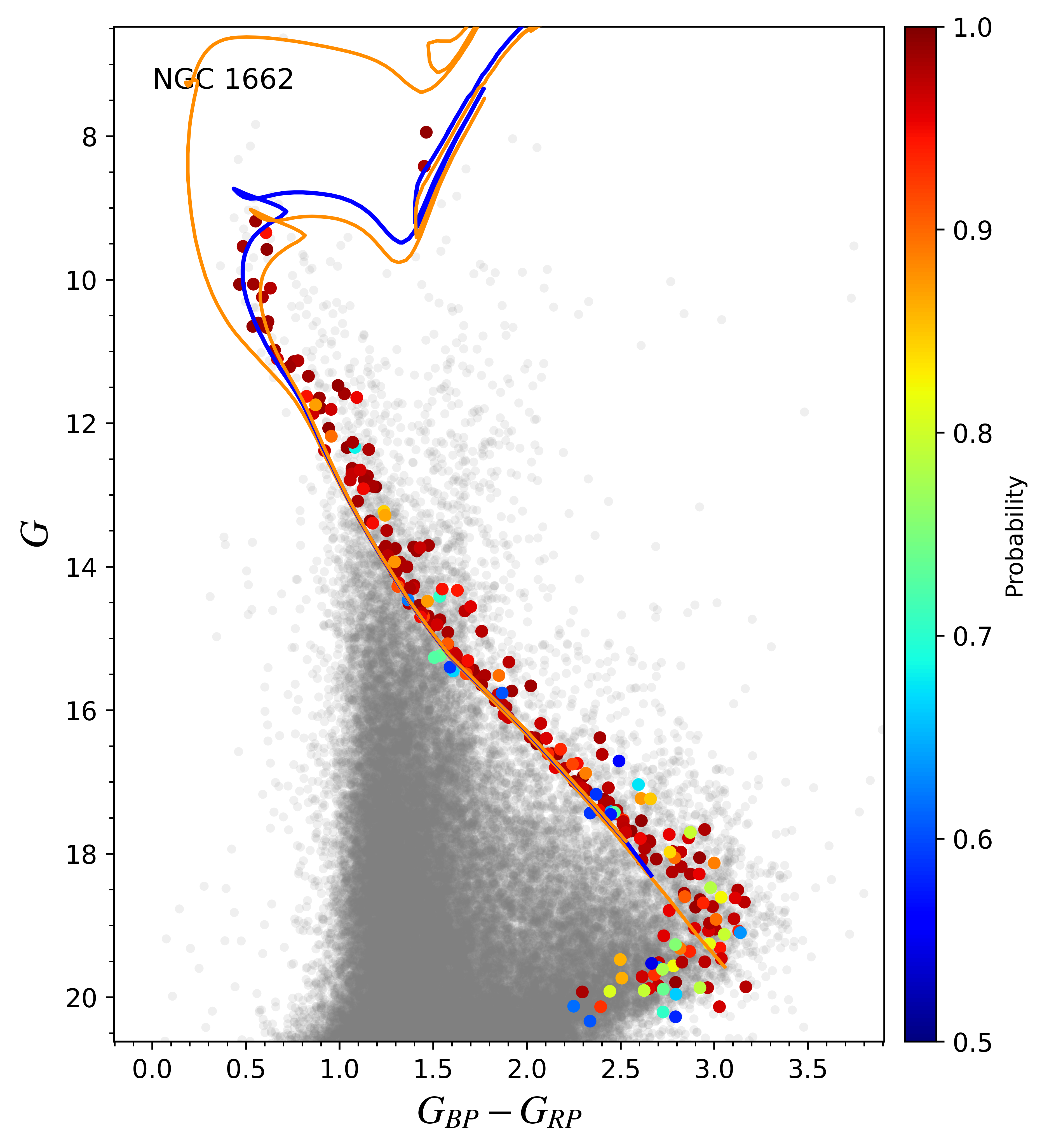}}}
\subfloat[]{\scalebox{.4}{\includegraphics{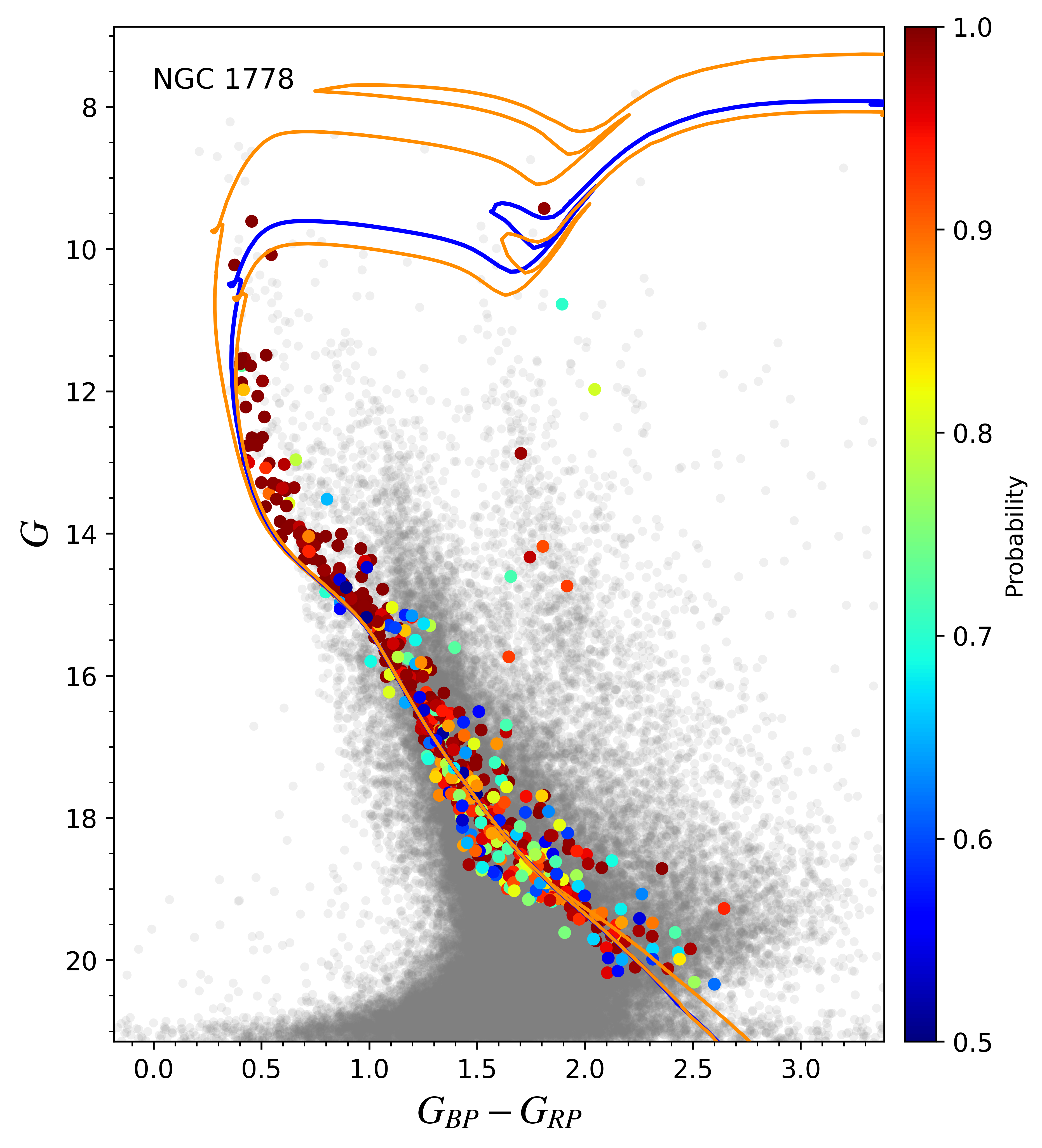}}}\\
\subfloat[]{\scalebox{.4}{\includegraphics{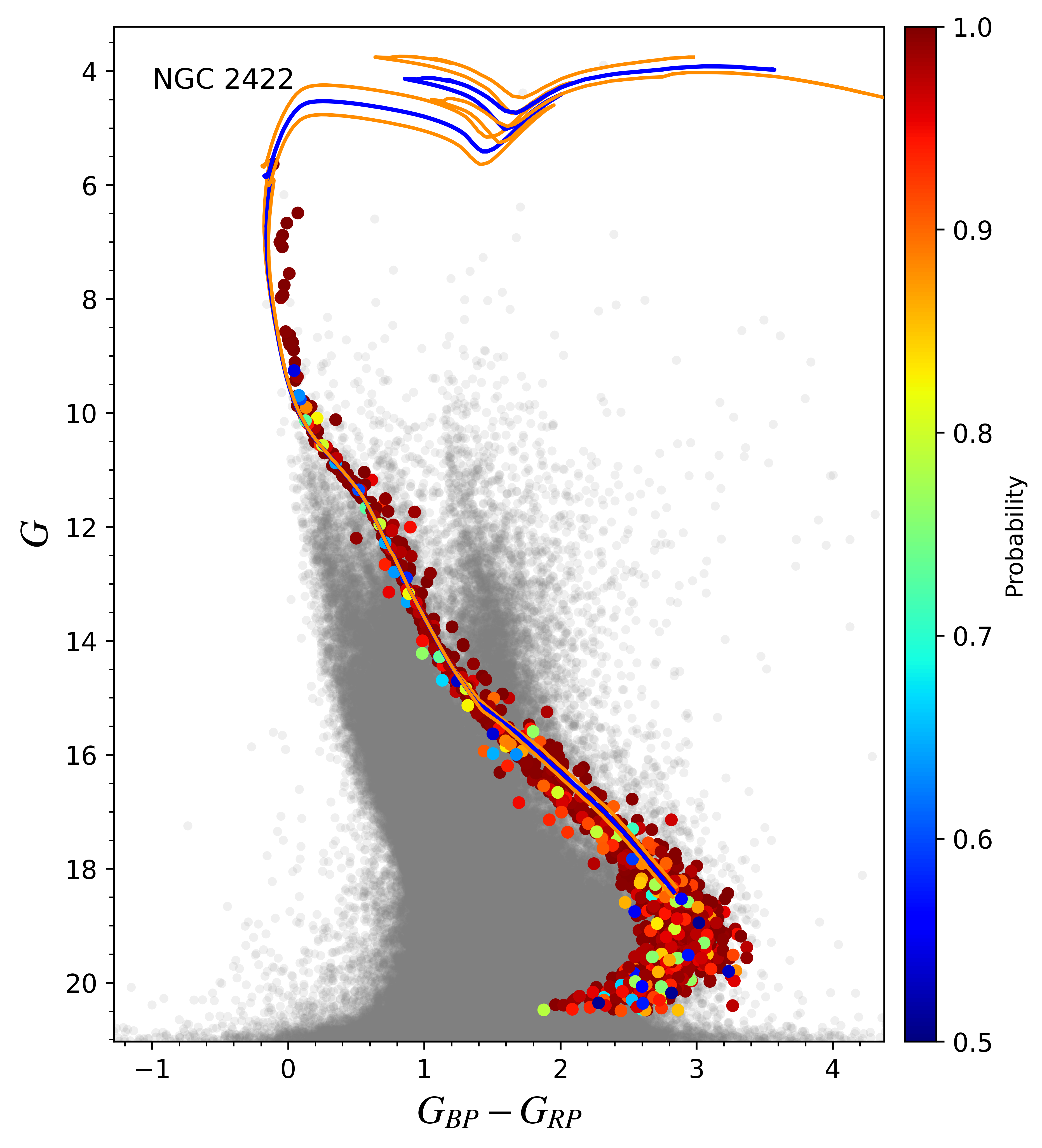}}}
\subfloat[]{\scalebox{.4}{\includegraphics{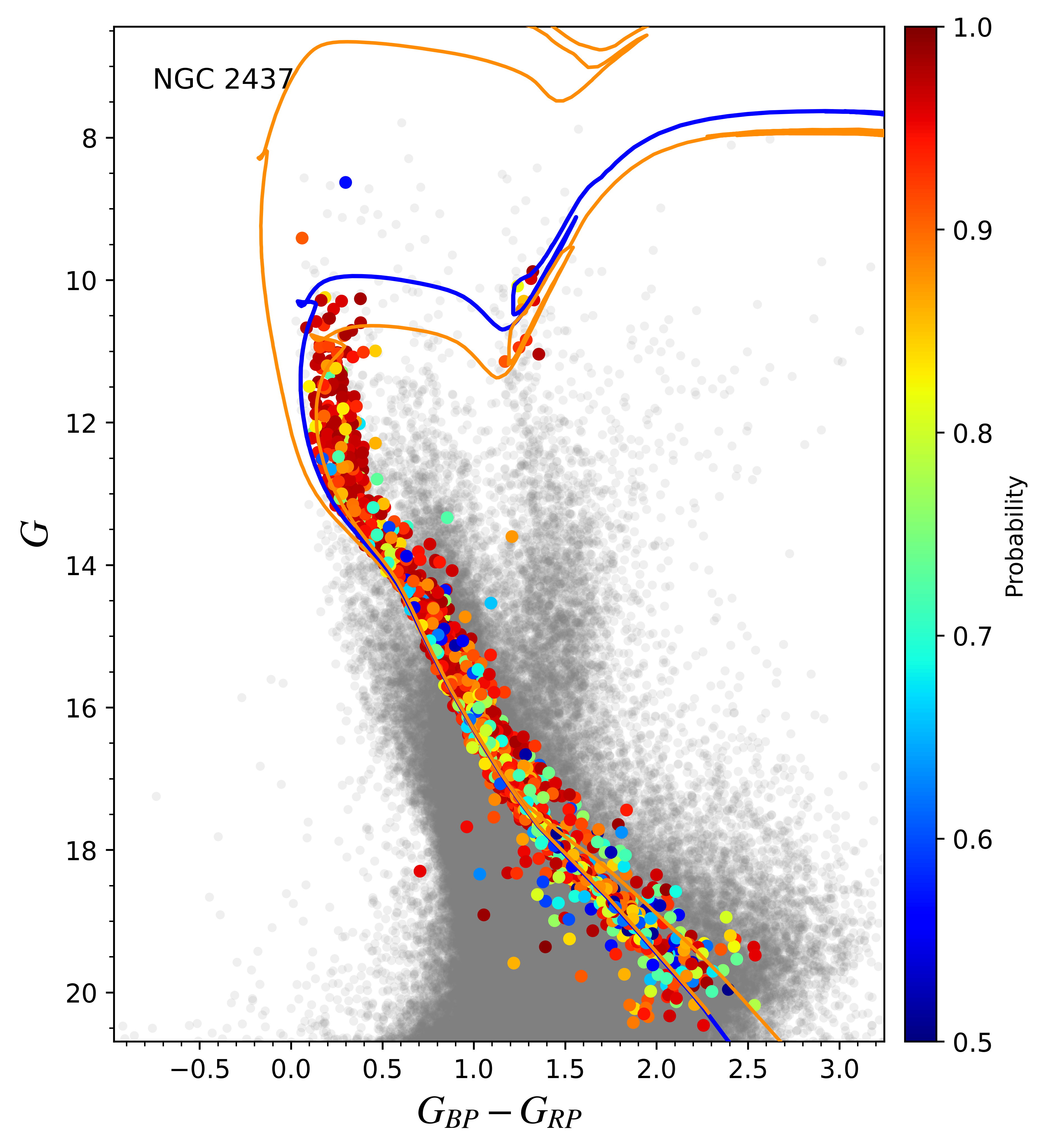}}}
\caption{CMDs of the OCs NGC 1662, NGC 1778, NGC 2422  and NGC 2437. Probable cluster members with $0.5 \leq P \leq 1$ are shown in distinct color codes, while field stars are represented by grey circles. The best-fitting {\sc PARSEC} isochrones are represented by blue lines, with their corresponding uncertainty ranges represented by orange lines.}
    \label{CMDs}
\end{figure*}

To account for the influence of heavy-element content on age estimates, the metallicity ($Z$) was converted into the iron abundance ([Fe/H]) using the relations provided by \citet{Bovy2012}. This approach is consistent with the \texttt{PARSEC} isochrones and has been validated in several recent studies \citep[e.g.,][]{2024AN....34540054C}. We first calculate the intermediate variable $Z_{\rm x}$
\begin{equation}
Z_{\rm x} = \frac{Z}{0.7515 - 2.78 \times Z}
\end{equation}
The cluster iron abundance is then derived as
\begin{equation}
\text{[Fe/H]} = \log (Z_{\rm x}) - \log \left( \frac{Z_{\odot}}{1 - 0.248 - 2.78 \times Z_{\odot}} \right)
\end{equation}
where $Z_\odot = 0.0152$ denotes the solar metallicity. Based on this procedure, the estimated iron abundances for NGC 1662, NGC 1778, NGC 2422, and NGC 2437 are $-0.03$, $-0.08$, $+0.14$, and $-0.02$ dex, respectively. The corner plots illustrating the marginal posterior distributions, along with the best-fit isochrones overlaid on the color-magnitude diagrams (CMDs), are shown in Figure~\ref{CMDs}. A summary of all derived values is provided in Table~\ref{all_results}.

The fundamental astrophysical parameters of the investigated clusters can be derived from their observed CMDs, constructed using {\it Gaia} photometry for the selected cluster members. The observed CMDs were fitted with theoretical stellar evolutionary models to determine the best-matching cluster properties. The {\sc PARSEC} isochrones of \citep{Bressan2012} were adopted in this work, which were computed for the metallicity values, $Z$, of 0.0141, 0.0127, 0.0207,  0.0145 for NGC 1662, NGC 1778, NGC 2422 and NGC 2437, respectively. For clusters with available iron abundance measurements, the reported [Fe/H] values were converted into the corresponding heavy-element mass fraction $Z$ to ensure consistency with the metallicity scale of the {\sc PARSEC} models. This conversion follows the analytical prescriptions calibrated for the PARSEC framework \footnote{\url{https://github.com/jobovy/isodist/blob/master/isodist/Isochrone.py}}:
\begin{equation}
z_{\rm x} = 10^{{\rm [Fe/H]} + \log\left(\frac{Z_{\odot}}{1 - 0.248 - 2.78~Z_{\odot}}\right)},
\end{equation}
\citep{2023AJ....165...79Y, Haroon2025, Elsanhoury2025}
where $Z_{\odot}$ is the solar metallicity, which is equal to 0.0152, the heavy-element mass fraction $Z$ is obtained from
\begin{equation}
Z = \frac{z_{\rm x} (1 - 0.2485)}{1 + 2.78~z_{\rm x}}.
\end{equation}
These relations provide a consistent conversion between iron abundance and mass fraction, allowing a uniform isochrone fitting procedure across the sample \citep{YontanCanbay2023, Tasdemir2023, Yontan2022, Tasdemir2025a}.

Figure~\ref{CMDs} presents the $G$ versus $(G_{\rm BP}-G_{\rm RP})$ CMDs of the analysed clusters, together with their best-fitting isochrones. The isochrone data provide absolute magnitudes ($M_G$) and intrinsic colors as a function of stellar mass at a fixed age. These quantities were used to estimate the observed distance modulus and color excess of each cluster. The observed distance modulus was obtained as $(m-M)$ in $G$ band, while the color excess was derived as $E(G_{\rm BP}-G_{\rm RP})$, defined as the difference between the observed stellar colors and the intrinsic isochrone colors.

It is worth noting that the simultaneous determination of age, distance, extinction, and metallicity via photometric isochrone fitting carries an obvious risk of multi-parameter degeneracy. To overcome this concern and evaluate the reliability of leaving the metallicity ($Z$) as a free parameter in our MCMC simulations, we cross-checked our results with independent estimators. Since spectroscopic measurements are currently lacking in the literature for these clusters, we compared our final $\text{[Fe/H]}$ estimations with the independent photometric catalog of \cite{Dias2021}, finding an excellent agreement. More importantly, our MCMC-derived photometric distances ($d_{(m-M)}$) shows a  consistency with the purely geometric distances ($d_\varpi$) extracted from the \textit{Gaia} DR3 parallaxes (e.g., $415^{+12}_{-15}$ pc vs. $410 \pm 12$ pc for NGC~1662, and $1725^{+46}_{-21}$ pc vs. $1761 \pm 335$ pc for NGC~1778). This cross-consistency suggests that the inclusion of geometric parallax constraints helps mitigate the multi-parameter degeneracy, guiding the MCMC algorithm toward physically reasonable parameters. Consequently, the broad confidence intervals reflected by the upper boundaries in the CMD fits (see Figure ~\ref{CMDs}) can be interpreted as a reflection of realistic error propagation under localized observational uncertainties, rather than a systematic underestimation of the cluster ages.

The adopted ages of the clusters range between $63$ and $813$~Myr. The corresponding distance moduli range between 8.09 and 11.18~mag. The clusters have color excesses ranging between 0.05 and 0.37~mag. The $E(B-V)$ color excesses of the clusters can be computed from the {\it Gaia} colors using the relation $E(B-V) = 0.775 \times E(G_{\rm BP}-G_{\rm RP})$ \citep{cardelli1989relationship}. Also, the line-of-sight extinction in the {\it Gaia} $G$ band was then computed using $A_G = 2.74 \times E(B-V)$ \citep{casagrande2018use, zhong2019substructure, 2023AJ....165..163C}. Results show that the investigated clusters have $E(G_{\rm BP}-G_{\rm RP})$ color excess values range between 0.08 and 0.52 mag, while the corresponding $A_G$ values range between 0.14 and 0.96 mag. The individual values of each cluster are listed in Table~\ref{all_results}.

We also got estimates of the clusters' distances using their observed distance moduli using
\begin{equation}
d = 10^{\left[(m-M)_{\rm obs}-A_G+5\right]/5}.
\end{equation}
The resulting photometric distances are $415_{-15}^{+12}$, $1725_{-21}^{+46}$, $471_{-96}^{+56}$ and $1544_{-507}^{+273}$ pc for NGC~1662, NGC~1778, NGC~2422 and NGC~2437, respectively, that are in agreement within uncertainties with the corresponding astrometric distances.

The distances of the clusters from the Galactic plane ($Z_{\odot}$), their projected distances from the Sun ($X_{\odot}$ and $Y_{\odot}$), and their Galactocentric distances ($R_{\rm GC}$) were computed using
\begin{align}
X_{\odot} = d \cos b \cos l,~~~ Y_{\odot} = d \cos b \sin l,~~~
Z_{\odot} = d \sin b,
\end{align}
and
\begin{equation}
R_{\rm GC} = \sqrt{R_{\odot}^2 + d^2 - 2 R_{\odot} d \cos b \cos l},
\end{equation}
where $d$ is the heliocentric distance in parsecs and $R_{\odot}$ is the distance of the Sun from the Galactic center. We adopted $R_{\odot} = 8.20 \pm 0.10$ kpc \citep{bland2019galah}. The resulting Galactocentric distances are $\sim$8.60 kpc for NGC 1662, $\sim$9.90 kpc for NGC 1778, $\sim$8.50 kpc for NGC~2422 and $\sim$9.30 kpc for NGC~2437. All values of $X_{\odot}$, $Y_{\odot}$, $Z_{\odot}$, and $R_{\rm GC}$ are summarized in Table~\ref{all_results}.

\begin{table*}[ht]
\centering
\caption{Summary of the results obtained in the present work.}
\label{all_results}
\scalebox{0.85}{
\begin{tabular}{lcccc}
\hline\hline
Parameter & NGC 1662 & NGC 1778 & NGC 2422 & NGC 2437 \\
\hline

$\alpha$ (hh:mm:ss)  
& 04:48:53.22  
& 05:08:10.85  
& 07:36:33.05  
& 07:41:45.25 \\

$\delta$ (dd:mm:ss)  
& +10:57:24.59  
& +36:59:58.02  
& -14:28:55.31  
& -14:50:33.97 \\

$l$ (deg)  
& 187.7430  
& 168.9359  
& 230.9532  
& 231.8847 \\

$b$ (deg)  
& -21.0148  
& -2.0114  
& +3.1240  
& +4.0457 \\
\hline
\multicolumn{5}{c}{Astrometric Parameters} \\
\hline

$\varpi$ (mas) & $2.438 \pm 0.071$ & $0.568 \pm 0.108$ & $2.097 \pm 0.079 $ & $0.595 \pm 0.061$\\
$d_{\varpi}$ (pc) & $410 \pm 12$ & $1761 \pm 335 $ & $477 \pm 18$ & $1681 \pm 172$  \\
$\mu_{\alpha}\cos\delta$ (mas\,yr$^{-1}$) & $-1.141 \pm 0.223$ & $0.479 \pm 0.159$ & $-7.056 \pm 0.318$ & $-3.864 \pm 0.144$ \\
$\mu_{\delta}$ (mas\,yr$^{-1}$) & $-1.956 \pm 0.201$ & $-3.199 \pm 0.145$ & $1.024 \pm 0.320$ & $0.402 \pm 0.144$\\
$N_{\rm stars}$ & 242& 502 & 1226& 2604  \\
\hline
\multicolumn{5}{c}{Astrophysical Parameters} \\
\hline 

$(m-M)_{\rm obs}$ (mag) & $8.67_{-0.59}^{+0.71}$ & $12.14_{-0.44}^{+0.75}$ & $8.51_{-0.14}^{+0.17}$ & $11.15_{-0.40}^{+0.48}$ \\[3pt]
$E(G_{\rm BP}-G_{\rm RP})$ (mag) & $0.31_{-0.18}^{+0.05}$ & $0.52_{-0.05}^{+0.21}$ & $0.08_{-0.05}^{+0.06}$ & $0.11_{-0.14}^{+0.18}$ \\ [3pt]
$E(B-V)$ (mag) & $0.22_{-0.13}^{+0.03}$ & $0.37_{-0.03}^{+0.15}$ & $0.05_{-0.04}^{+0.05}$ & $0.08_{-0.10}^{+0.13}$ \\ [3pt]
$A_{G}$ (mag) & $0.58_{-0.35}^{+0.09}$ & $0.96_{-0.09}^{+0.39}$ & $0.14_{-0.10}^{+0.12}$ & $0.21_{-0.26}^{+0.33}$ \\ [3pt]
$d_{(m-M)}$ (pc) & $415_{-15}^{+12}$ & $1725_{-21}^{+46}$ & $471_{-96}^{+56}$ & $1544_{-507}^{+273}$ \\ [3pt]
$Z$ & $0.0141_{-0.0001}^{+0.0030}$ & $0.0127_{-0.0197}^{+0.0089}$ & $0.0207_{-0.0229}^{+0.0147}$ & $0.0145_{-0.0238}^{+0.0098}$ \\ [3pt]
$[{\rm Fe/H}]$ (dex) & $-0.03$ & -$0.08$ & $+0.14$ & $-0.02$ \\ [3pt]
$\log({\rm t}/yr)$ & $8.91_{-0.33}^{+0.24}$ & $8.20_{-0.11}^{+0.18}$ & $7.80_{-0.11}^{+0.06}$ & $8.56_{-0.34}^{+0.27}$ \\ [3pt]

\hline
\multicolumn{5}{c}{Distances from the Galactic Plane and center} \\
\hline
$X_{\odot}$ (kpc) & $-0.379$ & $-1.728$ & $-0.300$ & $-1.034$ \\
$Y_{\odot}$ (kpc) & $-0.052$ & $0.338$ & $-0.370$ & $-1.318$\\
$Z_{\odot}$ (kpc) & $-0.147$ & $-0.062$ & 0.026 &0.119\\
$R_{\rm GC}$ (kpc) & 8.581 & 9.934 & 8.508 & 9.329 \\
\hline
\end{tabular}}
\end{table*}

\section{Luminosity and Mass Functions}

\begin{figure*}
\centering
\includegraphics[width=0.33\linewidth]{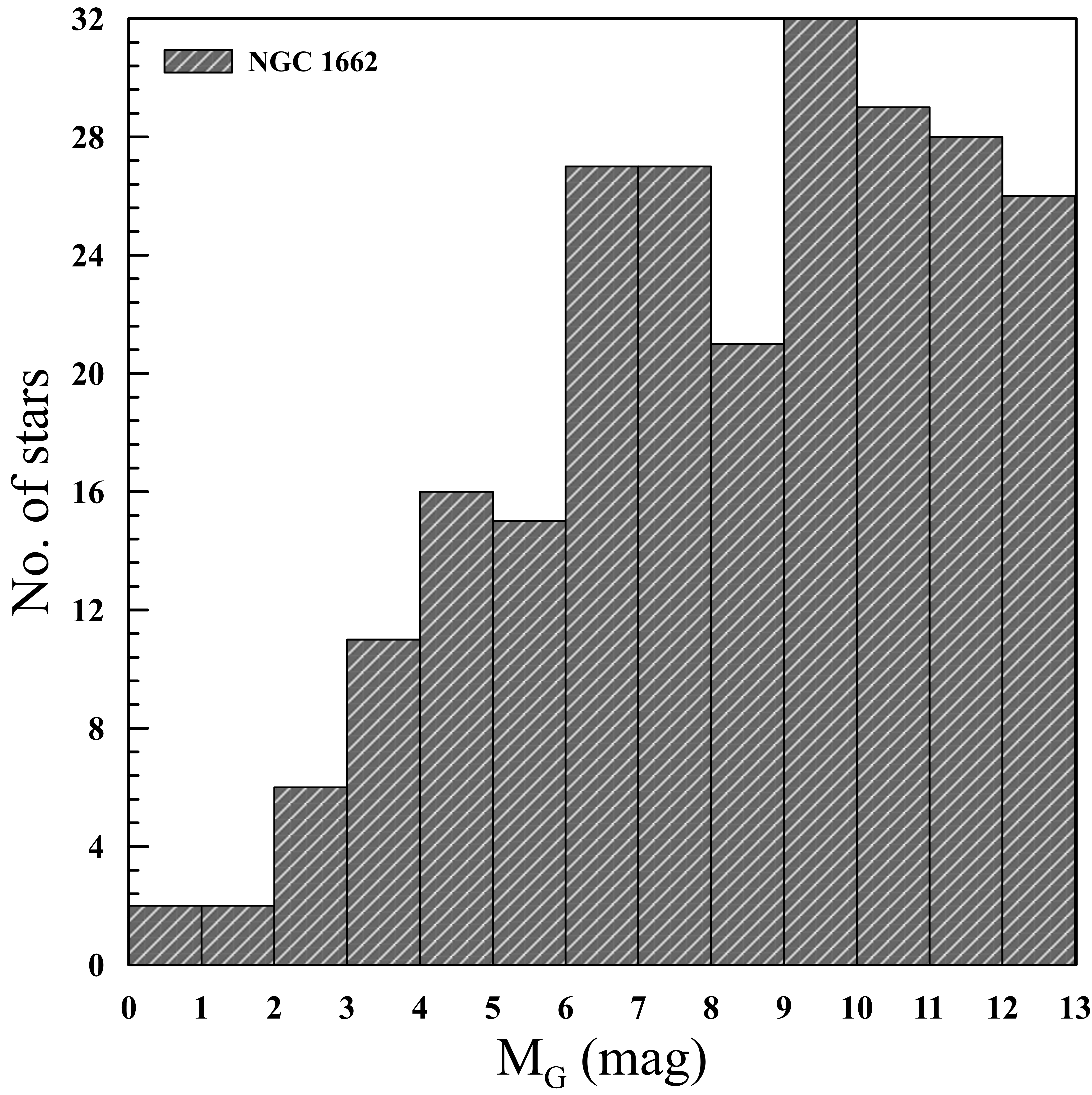}
\includegraphics[width=0.33\linewidth]{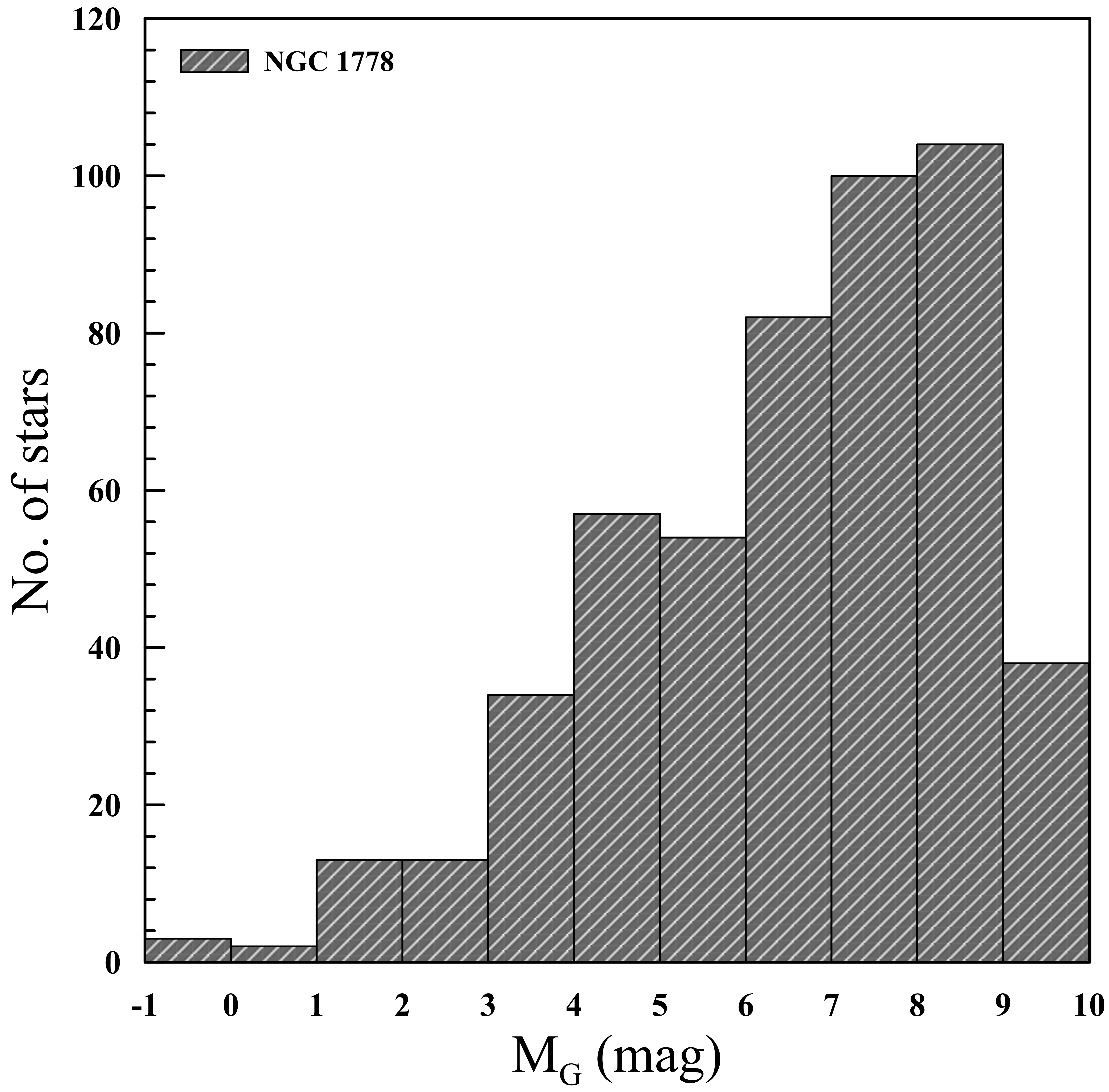}\\
\includegraphics[width=0.33\linewidth]{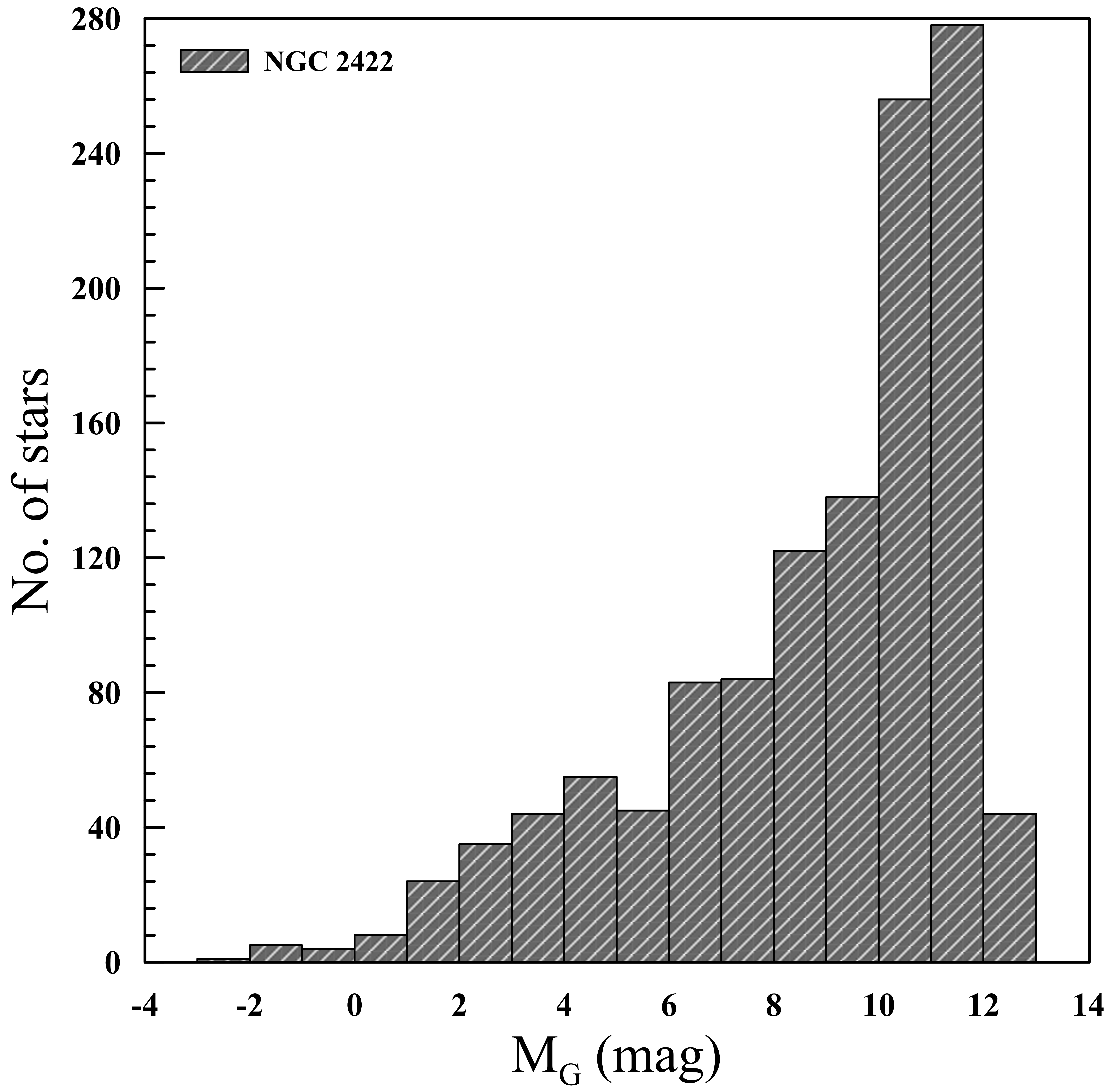}
\includegraphics[width=0.33\linewidth]{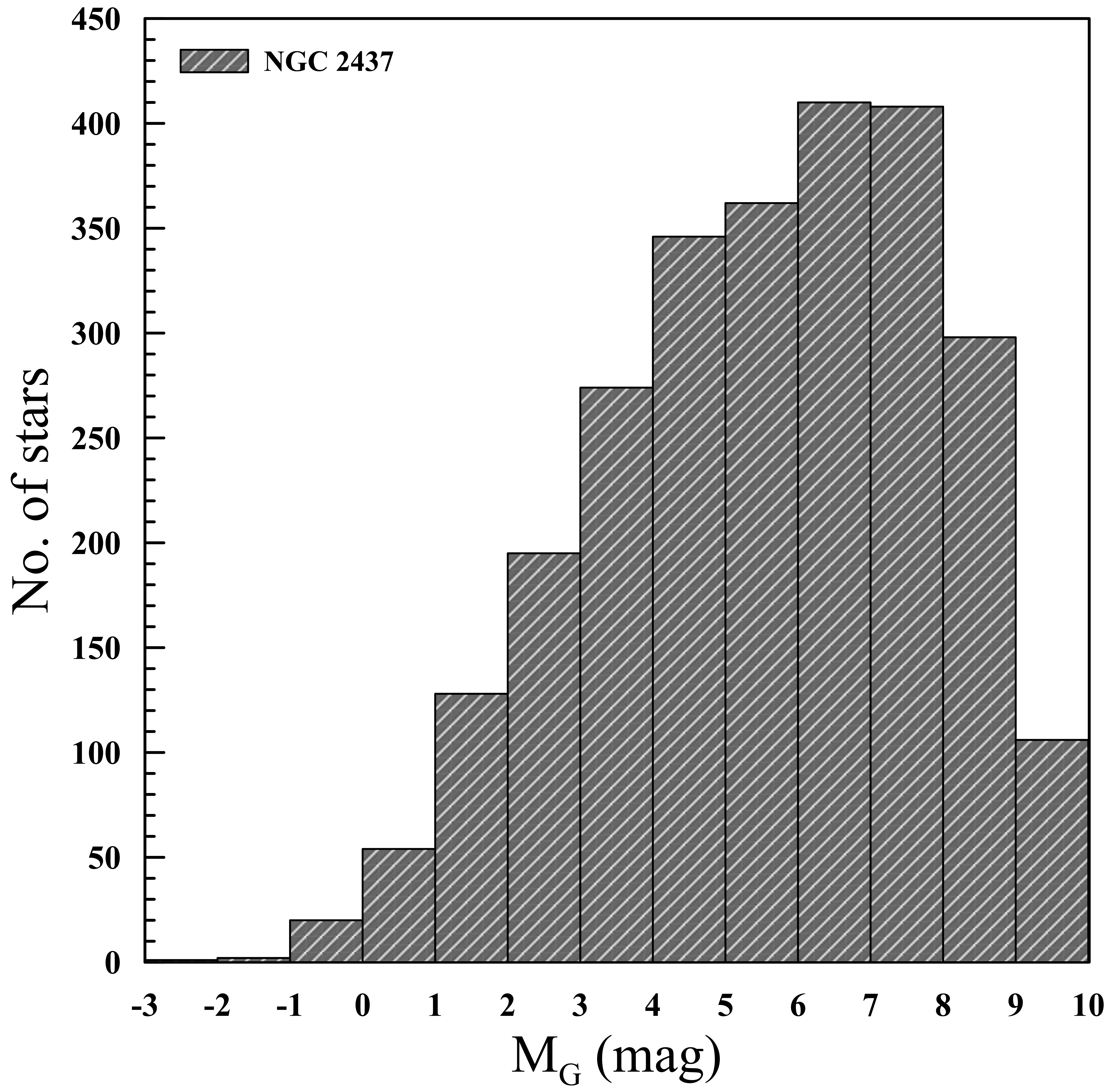}
\caption{Luminosity functions of NGC 1662, NGC 1778, NGC 2422, and NGC 2437 OCs in the $G$ absolute band.}
\label{LFs}
\end{figure*}

The luminosity function (LF) describes how the brightness of cluster members is distributed in terms of absolute magnitude and provides direct insight into the observable stellar content of a system \citep{2015Ap&SS.355..267Y, 2015MNRAS.453.1095B, Bostanci2018}.  For NGC 1662, NGC 1778, NGC 2422, and NGC 2437, the $G$-band absolute magnitude range was divided into equal intervals, and the number of probable members in each bin was counted. Absolute magnitudes were calculated from the observed apparent magnitudes using the distance moduli derived from the isochrone analysis \citep{Alzahrani2025b,Alzahrani2025a, Bisht2021, Bisht2025}. The resulting luminosity functions are presented in Figure~\ref{LFs}. 

The LF reflects the observed brightness distribution of the cluster population. When combined with theoretical stellar evolution models, it can be transformed into the mass function (MF), which provides a more physically meaningful description of the stellar content. The MF is particularly valuable for understanding cluster structure and its evolutionary state.

The theoretical isochrones of \cite{marigo2017new}, computed at the adopted cluster ages, provide the relation between stellar mass and luminosity required for this transformation. An analytical mass–luminosity relation (MLR) was obtained by fitting the isochrone mass and absolute magnitude data with a fourth-degree polynomial over the magnitude range sampled by the cluster members
\begin{equation}
M_{\rm c} = a_0 + a_1 M_{\rm G} + a_2 M_{\rm G}^2 + a_3 M_{\rm G}^3 + a_4 M_{\rm G}^4,
\label{Eq:ML_new}
\end{equation}
where $a_0$,~$a_1$,~$a_2$,~$a_3$,~and~$a_4$ are the coefficients obtained from the fitting procedure \citep{2017SerAJ.194...59A}.  The absolute magnitude intervals considered in the fitting span approximately 0.4360 to 12.8241 mag for NGC 1662, -0.7946 to 10.1147 mag for NGC 1778, -2.5915 to 12.2696 mag for NGC 2422, and -2.1023 to 9.7468 mag for NGC 2437.

Using the derived MLR, the luminosity functions were converted into mass functions by estimating the mass of each probable member \citep{eker2024, 2019Ap&SS.364..152Y, 2016Ap&SS.361..126A}. The total cluster mass $(M_{\rm C})$ was obtained by summing the individual stellar masses, and the mean stellar mass $\langle M_{\rm C} \rangle$ was calculated by dividing the total mass by the number of members. The polynomial coefficients together with the total and mean cluster masses are listed in Table~\ref{MLR_results}.

\begin{table*}[htbp]
\centering
\caption{MF and MLR results for the target OCs. Listed are the total derived cluster mass ($M_{\text{C}}$), the mean stellar mass ($\langle M_C \rangle$), the derived MF slope ($\alpha$), and the coefficients ($a_0...a_4$) for the polynomial MLR fit.}
\label{MLR_results}
\scalebox{0.8}{
\begin{tabular}{lccccccccc} 
\hline\hline
Cluster & $M_{\text{C}}\;(M_\odot)$ & $\langle M_C \rangle \;(M_\odot)$ & $\alpha$ & $a _0$ & $a _1$ & $a _2$ & $a _3$ & $a _4$ \\ 
\hline
NGC 1662 & 207  &  0.855  &  3.23 $\pm$ 0.01  &  2.4525  &  -0.3484  &  0.0090  & 0.0021 &  -0.0001 \\ 
NGC 1778 & 507  &  1.00  &  2.39 $\pm$ 0.10  &  2.9230  &  -0.6396  &  0.0675  & -0.0024 &  -0.0001 \\
NGC 2422 & 1108  &  0.904  &  3.62 $\pm$ 0.02  & 3.1630 &  -0.7243  &  0.0800  &  -0.0034  &  0.0002 \\
NGC 2437 & 3044  &  1.169 &  2.49 $\pm$ 0.10  &  2.6960  &  -0.4079  &  -0.0150  &  0.0096  &  -0.0001 \\

\hline 
\end{tabular}}
\end{table*}

\begin{figure*}
\centering
\includegraphics[width=0.35\linewidth]{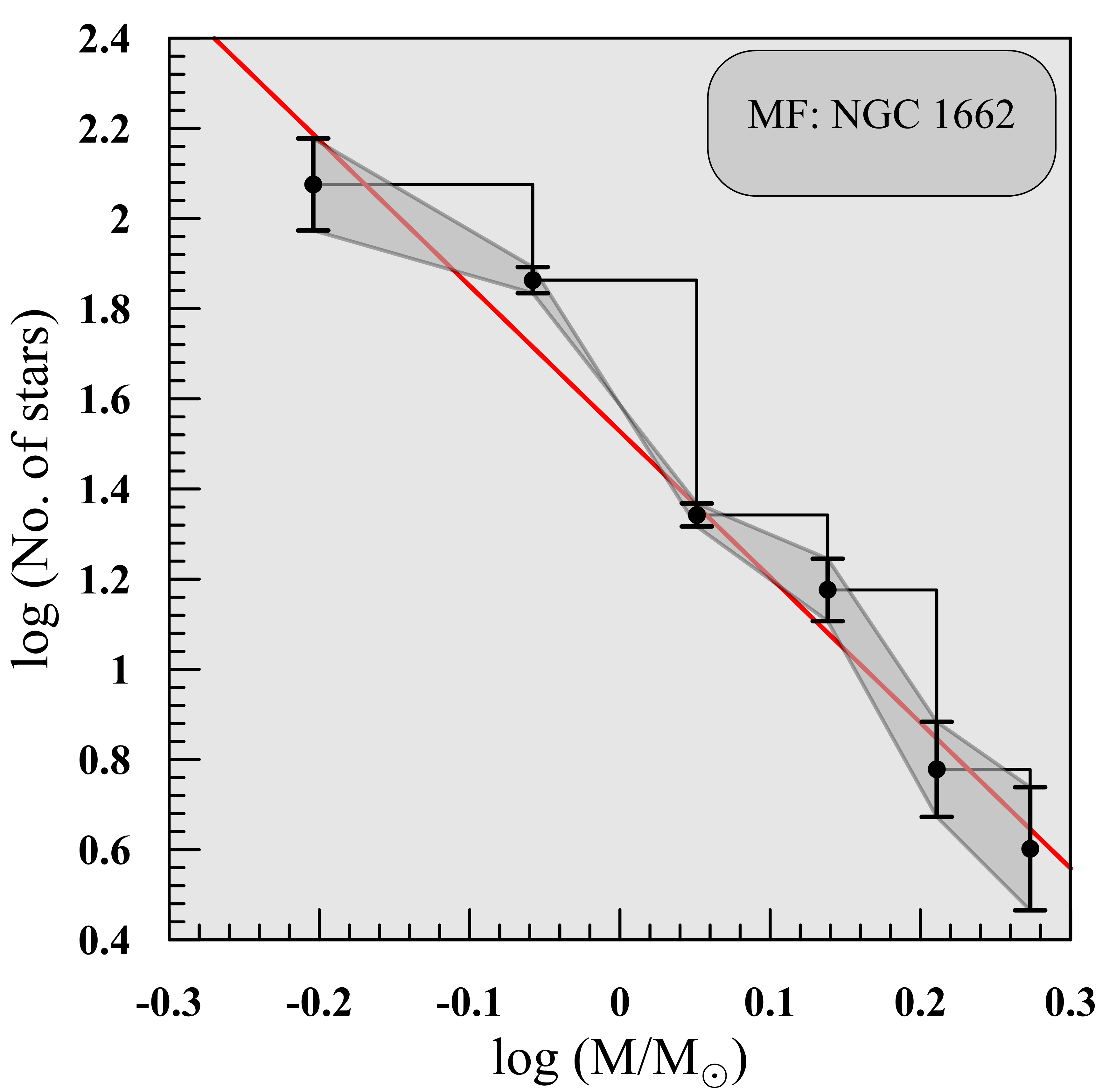}    \includegraphics[width=0.35\linewidth]{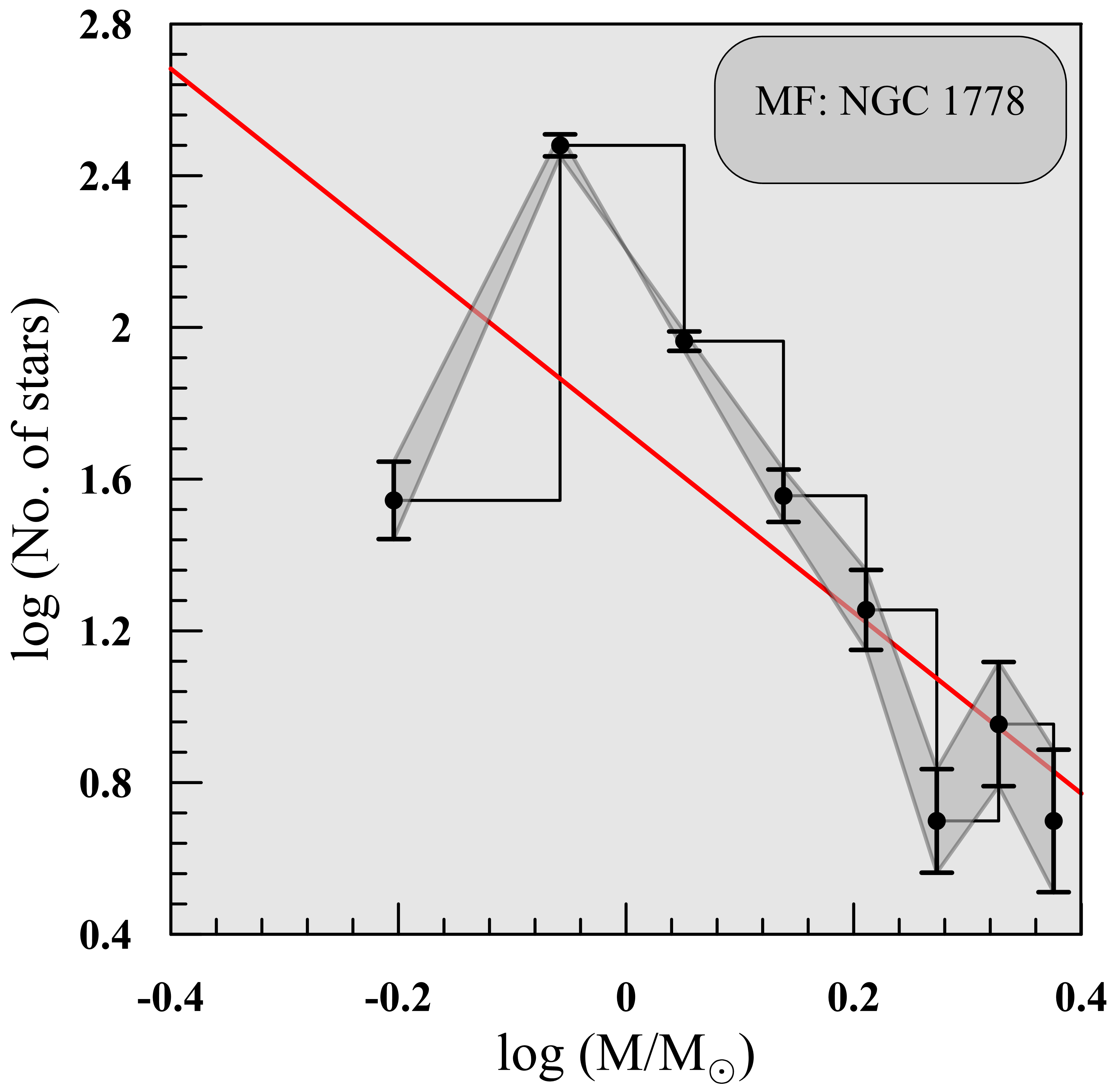}\\
\includegraphics[width=0.35\linewidth]{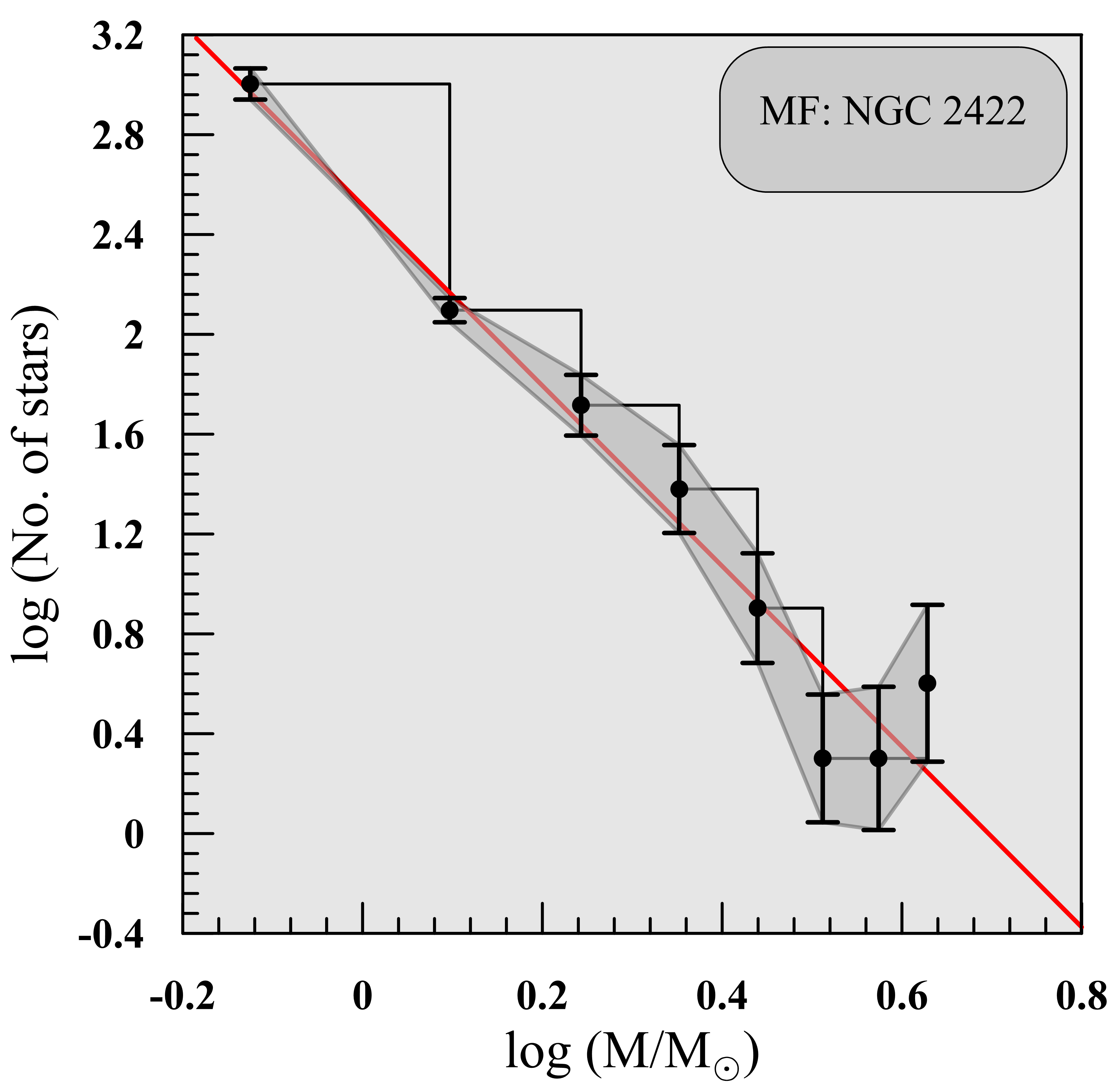}
\includegraphics[width=0.35\linewidth]{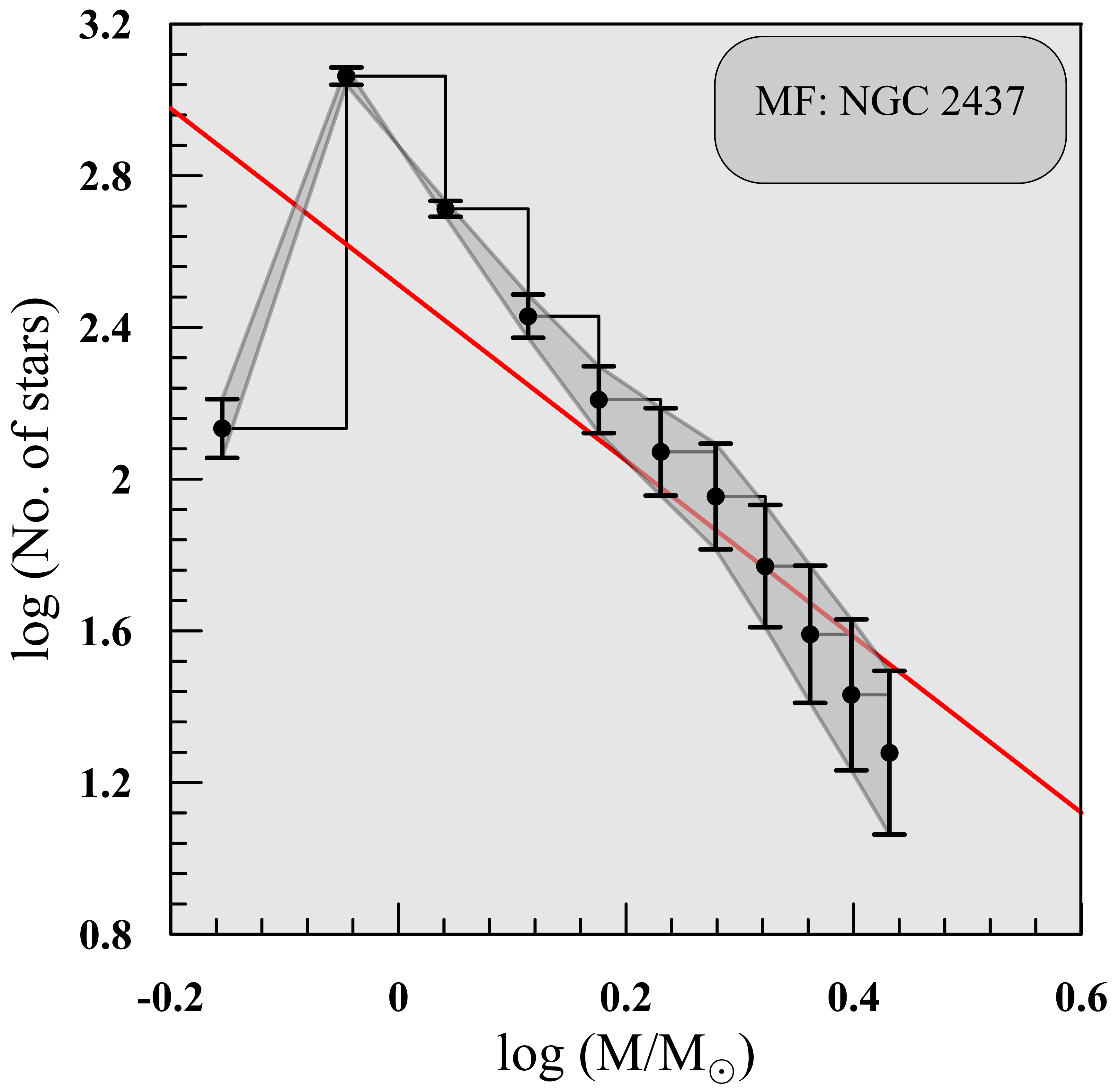}
\caption{Derivation of stellar mass functions for the OCs NGC 1662, NGC 1778, NGC 2422, and NGC 2437. Each was modelled using a power-law fit, showing a slope that was consistent with the anticipated initial mass distribution and around the classical \citep{salpeter1955} value.}
\label{MFs}
\end{figure*}

The stellar mass function is commonly expressed as a power law \citep{salpeter1955}
\begin{equation}
\frac{dN}{dM} \propto M^{-\alpha},
\label{eq:MF_new}
\end{equation}
where $\alpha$ denotes the slope of the mass function. The observed MFs of the four clusters are shown in Figure~\ref{MFs}. Linear fits were applied to the logarithmic mass distributions, and the resulting $\alpha$ values are provided in Table~\ref{MLR_results}. 

To reduce possible biases related to incompleteness at the bright end or small-number statistics in sparsely populated bins, the first magnitude bin was excluded from the linear fitting when necessary. The derived slopes range from $\alpha \simeq 2.4$ to $3.6$ across the sample. While NGC 1778 and NGC 2437 yield values close to the classical Salpeter slope of $\alpha = 2.35$ \citep{salpeter1955}, NGC 1662 and NGC 2422 show steeper slopes, indicating a relative deficit of low-mass members with respect to the canonical initial mass function. Such departures are commonly attributed to dynamical evolution (e.g., preferential loss of low-mass stars through two-body relaxation) or to residual incompleteness at the faint end of the sampled magnitude range.

\section{Evolution Times and the Kinematical Structures}

\subsection{Mean Radial Velocity Estimates}\label{sec:mean_rv}

Radial velocities constitute a key observable for constraining both the internal dynamical structure and the Galactic motion of OCs. In this study, we determined the mean radial velocities, $\langle V_{\rm r} \rangle$, for NGC 1662, NGC 1778, NGC 2422, and NGC 2437 using radial velocity measurements from Gaia DR3. The stellar samples were restricted to high-probability members selected through our astrometric and photometric membership analysis. In addition, strict quality constraints were imposed on the astrometric solutions, including a threshold on \texttt{RUWE}, to ensure that only reliable measurements were included in the analysis.

Radial velocity data are available for only a subset of the confirmed members in each cluster. Therefore, we considered those stars exclusively with published {\it Gaia} DR3 radial velocities \citep{GaiaDR3} and additional measurements from the Survey of Surveys (SoS) compilation \citep{Tsantaki2022} that also satisfy our membership and quality-selection criteria. The mean cluster radial velocities were computed as weighted averages, where individual stellar velocities were weighted by the inverse square of their formal uncertainties, following standard procedures commonly adopted in the literature \citep{Ahmed2026, Tasdemir2026}.

The derived mean radial velocities, together with their corresponding standard errors and the total number of stars used in each calculation, are as follows: NGC 1662 has a mean radial velocity of $\langle V_{\rm r} \rangle = -13.27 \pm 0.13$ km s$^{-1}$, calculated from 78 member stars. For NGC 1778, we obtain $\langle V_{\rm r} \rangle = -9.37 \pm 7.49$ km s$^{-1}$ based on 16 member stars. In the case of NGC 2422, the mean radial velocity is $\langle V_{\rm r} \rangle = 37.43 \pm 0.65$ km s$^{-1}$, derived from 231 stars. Finally, NGC 2437 has a mean radial velocity of $\langle V_{\rm r} \rangle = 48.46 \pm 0.17$ km s$^{-1}$, calculated from 175 stars. The quoted uncertainties correspond to the standard error of the weighted mean.

The \textit{Gaia} source identifiers, equatorial coordinates, $G$-band magnitudes, radial velocities, \texttt{RUWE} values, and membership probabilities of the stars included in the computations are listed in Table~\ref{tab:RV_info}. For brevity, only a representative portion of the dataset is presented in the printed version of the table, while the complete list for all four clusters is provided in electronic form\footnote{The journal will assign a URL}.

\begin{table*}[ht]
\centering
\footnotesize
\caption{Radial velocity measurements for the member stars in the four studied clusters. The full version of this table is provided in electronic format (Note: The journal will assign a URL).}
\setlength\tabcolsep{2pt}
\renewcommand{\arraystretch}{1.1}
\label{tab:RV_info}
\begin{tabular}{l c c c c c c c c c}
\hline \hline
ID & $\alpha$ (hh:mm:ss) & $\delta$ (dd:mm:ss) & $G$ (mag) & $P$ & $V_r$ (km s$^{-1}$)& $\sigma_{V_r}$ & & RUWE & Survey \\
\hline
\multicolumn{10}{c}{NGC 1662} \\
\hline
1 & 04:46:38.00 & +10:39:53.9 & 14.81 & 0.96 & -1.07 & 6.78 & & 1.02 & Gaia DR3 \\
2 & 04:46:42.44 & +11:08:50.5 & 11.81 & 1.00 & -12.67 & 1.90 & & 1.14 & LAMOST \\
3 & 04:46:47.24 & +10:37:55.2 & 12.71 & 1.00 & -13.11 & 2.08 & & 1.00 & LAMOST \\
\dots & \dots & \dots & \dots & \dots & \dots & \dots & & \dots & \dots \\
78 & 04:51:20.55 & +11:04:04.6 & 14.65 & 0.97 & -8.35 & 3.76 & & 1.02 & Gaia DR3 \\
\midrule

\multicolumn{10}{c}{NGC 1778} \\
\hline
1 & 05:07:00.89 & +37:04:34.3 & 14.33 & 0.97 & -14.41 & 2.09 & & 1.03 & Gaia DR3 \\
2 & 05:07:02.46 & +36:55:26.5 & 14.74 & 0.92 & -15.06 & 2.58 & & 1.03 & Gaia DR3 \\
3 & 05:07:48.17 & +37:13:50.3 & 11.85 & 0.99 & -16.85 & 18.42 & & 1.35 & Gaia DR3 \\
\dots & \dots & \dots & \dots & \dots & \dots & \dots & & \dots & \dots \\
16 & 05:09:58.65 & +36:39:45.9 & 14.60 & 1.00 & -39.35 & 2.55 & & 1.08 & LAMOST \\
\midrule

\multicolumn{10}{c}{NGC 2422} \\
\hline
1 & 07:32:11.90 & -14:08:20.8 & 14.84 & 0.80 & -43.28 & 8.04 & & 0.95 & Gaia DR3 \\
2 & 07:32:52.49 & -13:41:58.4 & 10.57 & 0.80 & 28.17 & 3.16 & & 0.91 & Gaia DR3 \\
3 & 07:32:56.34 & -14:50:24.9 & 10.59 & 0.97 & 24.80 & 4.06 & & 1.03 & Gaia DR3 \\
\dots & \dots & \dots & \dots & \dots & \dots & \dots & & \dots & \dots \\
231 & 07:41:03.37 & -14:34:07.2 & 10.14 & 0.73 & 30.82 & 6.03 & & 0.93 & Gaia DR3 \\
\midrule

\multicolumn{10}{c}{NGC 2437} \\
\hline
1 & 07:39:36.19 & -14:40:19.3 & 14.40 & 0.73 & 71.66 & 28.70 & & 1.09 & Gaia DR3 \\
2 & 07:39:57.78 & -14:51:27.5 & 11.24 & 0.87 & 50.42 & 6.39 & & 0.95 & Gaia DR3 \\
3 & 07:40:22.56 & -14:39:06.3 & 11.77 & 1.00 & 33.70 & 6.71 & & 0.94 & Gaia DR3 \\
\dots & \dots & \dots & \dots & \dots & \dots & \dots & & \dots & \dots \\
175 & 07:43:56.02 & -15:07:14.3 & 9.35 & 0.62 & 49.16 & 0.37 & & 0.93 & Gaia DR3 \\
\hline
\end{tabular}
\end{table*}

The resulting $\langle V_{\rm r} \rangle$ values for NGC~1662, NGC~1778, NGC~2422, and NGC~2437 are consistent, within uncertainties, with previously published determinations. Minor differences with respect to earlier studies can be attributed to variations in membership selection, sample size, and data release versions. The overall agreement with recent \textit{Gaia}-based analyses confirms the stability of the derived cluster systemic velocities and supports their use in the subsequent kinematic and orbital calculations presented in this work.

\subsection{Convergent Point and Space Motion Analysis}\label{sec_CP}
In order to investigate the kinematic coherence and gravitational binding of the stellar populations associated with NGC 1662, NGC 1778, NGC 2422, and NGC 2437, we computed their velocity ellipsoid parameters and space motion characteristics. The procedure follows the standard formalism adopted in recent kinematic studies of OCs.

The heliocentric space velocity components ($V_x$, $V_y$, $V_z$) were derived from the equatorial coordinates ($\alpha$, $\delta$), heliocentric distance ($d$), proper motion components ($\mu_{\alpha}\cos\delta$, $\mu_\delta$), and the mean radial velocity ($V_{\rm R}$). The transformation from astrometric observables to Cartesian velocity components was performed using
\begin{equation}
\label{eq:Vs_new}
\resizebox{0.9\linewidth}{!}{$
\begin{pmatrix}
V_x \\
V_y \\
V_z
\end{pmatrix}
=
\begin{pmatrix}
-4.74\, d\, \mu_{\alpha}\cos\delta \sin\alpha 
-4.74\, d\, \mu_{\delta}\sin\delta \cos\alpha 
+ V_{\rm R}\cos\delta \cos\alpha \\
+4.74\, d\, \mu_{\alpha}\cos\delta \cos\alpha 
-4.74\, d\, \mu_{\delta}\sin\delta \sin\alpha 
+ V_{\rm R}\cos\delta \sin\alpha \\
+4.74\, d\, \mu_{\delta}\cos\delta 
+ V_{\rm R}\sin\delta
\end{pmatrix}
$}
\end{equation}

\begin{figure*}[!t]
\centering
\includegraphics[width=0.75\linewidth]{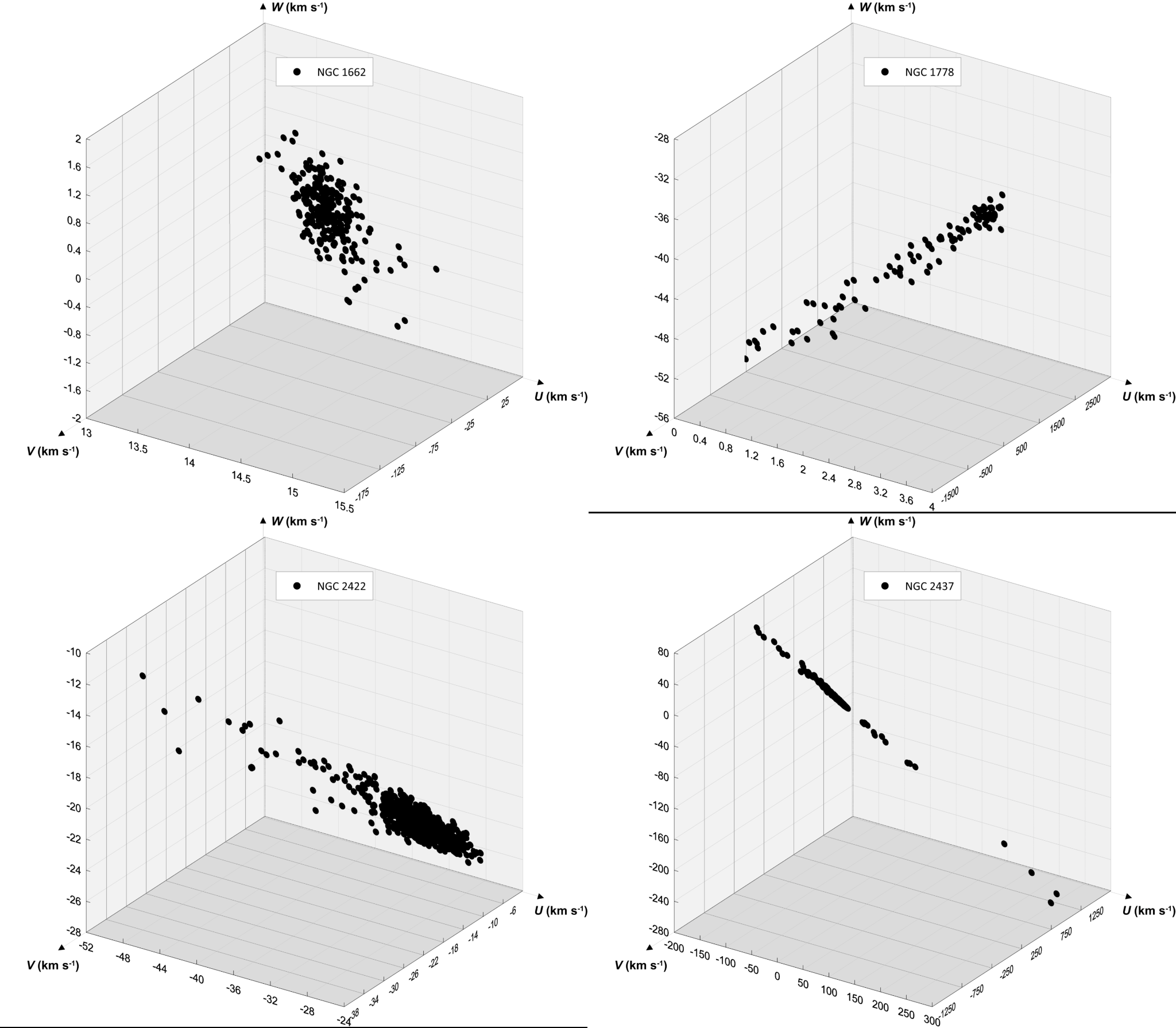}
\caption{Distribution of the heliocentric space velocity components ($U,~V,~W$) for the most probable members of NGC 1662, NGC 1778, NGC 2422, and NGC 2437.}
\label{figure:uvw}
\end{figure*}

The mean radial velocities adopted in the present kinematic analysis are given in Section~\ref{sec:mean_rv}. These values were used as input parameters in the computation of the heliocentric space velocity components for NGC 1662, NGC 1778, NGC 2422, and NGC 2437. Figure~\ref{figure:uvw} presents the distribution of the computed space velocity components for the probable members of each cluster.

The heliocentric velocity components were subsequently transformed into Galactic space velocity components ($U$, $V$, $W$) through the standard equatorial-to-Galactic rotation matrix
\begin{equation}
\label{eq:uvw_new}
\resizebox{0.88\linewidth}{!}{$
\begin{pmatrix}
U \\
V \\
W
\end{pmatrix}
=
\begin{pmatrix}
-0.05188074 & -0.87222264 & -0.48634972 \\
+0.48469224 & -0.44779209 & +0.75136921 \\
-0.87314490 & -0.19674834 & +0.44599133
\end{pmatrix}
\begin{pmatrix}
V_x \\
V_y \\
V_z
\end{pmatrix}
$}
\end{equation}
To place the clusters in the Local Standard of Rest (LSR) reference frame, the solar motion was subtracted from the heliocentric velocity components. The corrected mean velocities $(U,~V,~W)_{\rm LSR}$ and the total space velocity 
\[
S_{\rm LSR} = \sqrt{U^2 + V^2 + W^2}
\]
were calculated for each system. The resulting values are summarized in Table~\ref{tab:VEP}. The derived $S_{\rm LSR}$ values indicate that all four clusters are dynamically consistent with the Galactic thin-disc population.

\begin{table*}[!t]
\centering
\caption{Derived kinematic and relaxation times parameters for NGC 1662, NGC 1778, NGC 2422, and NGC 2437.}
\label{tab:VEP}
\begin{tabular}{lcccc}
\hline \hline
Parameter & NGC 1662 & NGC 1778 & NGC 2422 & NGC 2437 \\
\hline
$\overline{V}_X$ (km s$^{-1}$) & -0.14 $\pm$ 0.01 & -2.77 $\pm$ 0.60 & -0.11 $\pm$ 0.01 & 10.39 $\pm$ 3.22 \\
$\overline{V}_Y$ (km s$^{-1}$) & -12.24 $\pm$ 3.50 & 10.57 $\pm$ 3.25 & 40.32 $\pm$ 6.35 & 57.76 $\pm$ 7.60 \\
$\overline{V}_Z$ (km s$^{-1}$) & -9.30 $\pm$ 3.05 & -28.43 $\pm$ 5.33 & -7.05 $\pm$ 2.66 & -9.20 $\pm$ 3.03 \\
$A_{\rm o}$ ($^\circ$) & -90.64 $\pm$ 0.11 & 104.70 $\pm$ 0.10 & 90.15 $\pm$ 0.11 & 79.80$\pm$ 0.11 \\
$D_{\rm o}$ ($^\circ$) & -37.20 $\pm$ 0.16 & -68.99 $\pm$ 0.12 & -9.92 $\pm$ 0.32 & -8.91 $\pm$ 0.34 \\
$\overline{U}$ (km s$^{-1}$) & 15.21 $\pm$ 3.90 & 4.76 $\pm$ 0.46 & -31.73 $\pm$ 5.63 & -46.45 $\pm$ 6.18 \\
$\overline{V}$ (km s$^{-1}$) & -1.57 $\pm$ 0.07 & -27.44 $\pm$ 5.24 & -23.41 $\pm$ 4.84 & -27.74 $\pm$ 5.27 \\
$\overline{W}$ (km s$^{-1}$) & -1.62 $\pm$ 0.06 & -12.34 $\pm$ 3.51 & -10.98 $\pm$ 3.31 & -24.54 $\pm$ 4.95 \\
$S_{\rm LSR}$ (km s$^{-1}$) & 15.37 $\pm$ 3.92 & 30.46 $\pm$ 5.52 & 40.93 $\pm$ 6.40 & 59.41 $\pm$ 7.71 \\
\hline
\multicolumn{5}{c}{Evolving times} \\
\hline
$T_{relax}$ (Myr) & 11 & 69 & 32 & 133\\
$\tau_{ev}$ (Myr) & 11 $\times~10^2$ & 69 $\times~10^2$ & 32 $\times~10^2$ & 133 $\times~10^2$\\
$\tau$ & 74.00 & 2.30 & 1.97 & 2.73\\
\hline
\multicolumn{5}{c}{Dynamical parameter} \\
\hline
$Z_{\rm max}$ (kpc)         & $0.191\pm0.033$  & $0.089\pm0.005$  & $0.065\pm0.008$  & $0.238\pm0.082$ \\
$R_{\rm a}$ (kpc)           & $10.202\pm0.188$ & $9.997\pm0.066$  & $8.601\pm0.040$  & $9.241\pm0.190$ \\
$R_{\rm p}$ (kpc)           & $8.248\pm0.088$  & $9.143\pm0.241$  & $7.684\pm0.071$  & $8.283\pm0.563$ \\
$e$                         & $0.106\pm0.002$  & $0.045\pm0.010$  & $0.056\pm0.002$  & $0.055\pm0.021$ \\
$R_{\rm m}$ (kpc)           & $9.225\pm0.154$  & $9.570\pm0.111$  & $8.143\pm0.044$  & $8.762\pm0.307$ \\
$T_{\rm p}$ (Myr)           & $242\pm4.00$        & $251\pm3.00$        & $210\pm1.00$        & $228\pm9.00$ \\
$R_{\rm birth}$ (kpc)       & $10.003\pm0.592$ & $9.510\pm0.772$  & $7.686\pm0.146$  & $9.033\pm1.472$ \\
\hline
\end{tabular}
\end{table*}

\begin{figure*}[!h]
\centering
\includegraphics[width=0.60\linewidth]{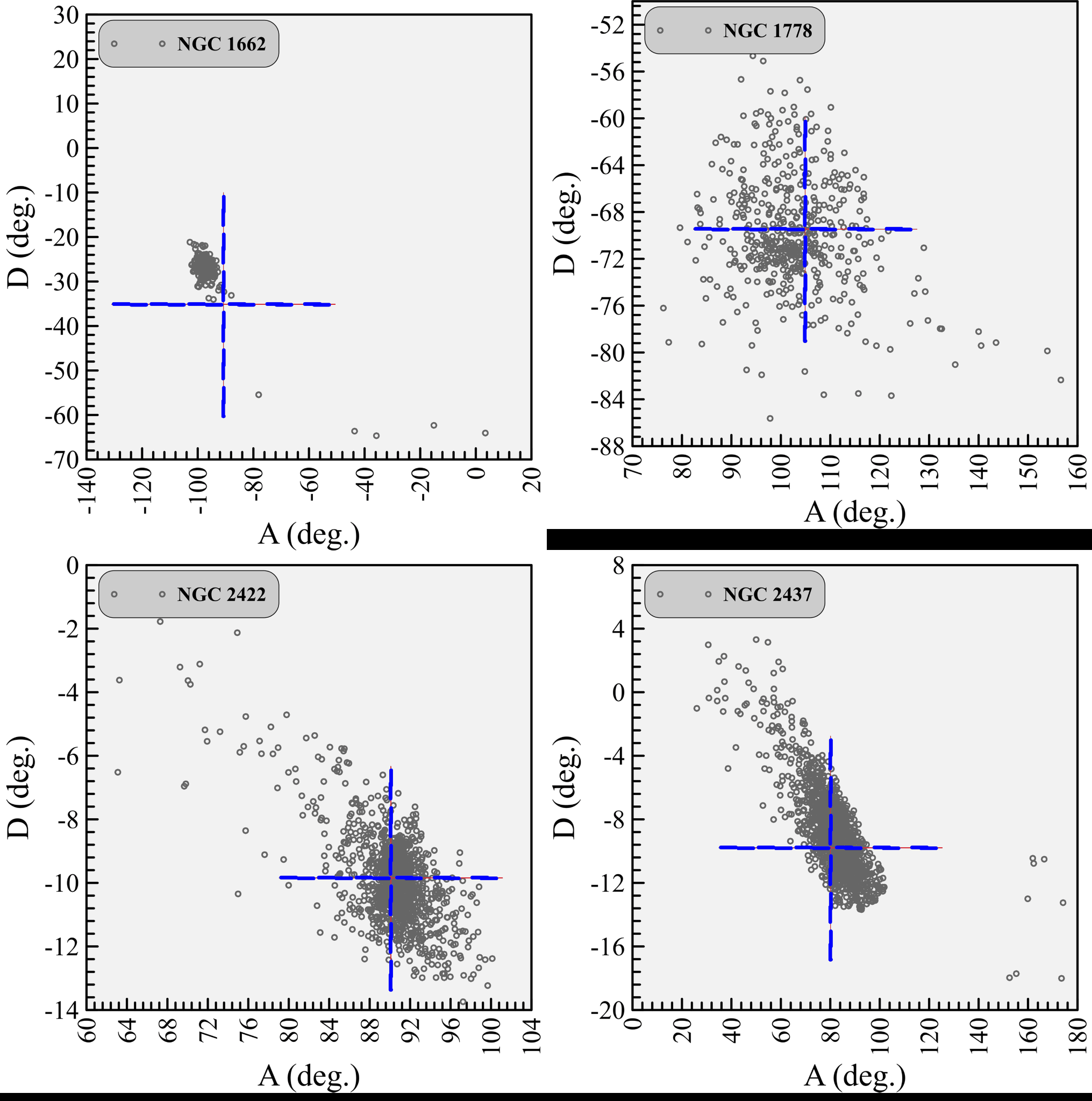}
\caption{Apex diagrams (AD-diagrams) for NGC 1662, NGC 1778, NGC 2422, and NGC 2437 OCs, showing the distribution of stellar apex coordinates $(A_0,~D_0)$ for cluster members. Crosses indicate the mean apex positions.}
\label{fig:CP}
\end{figure*}

The convergent point, or apex ($A_{\rm o}$, $D_{\rm o}$), defines the direction of the bulk motion of a stellar aggregate projected onto the celestial sphere. It was determined from the mean Cartesian velocity components using
\begin{equation}
\label{eq:ad_new}
A_{\rm o} = \tan^{-1}\left(\frac{\overline{V_y}}{\overline{V_x}}\right), \qquad
D_{\rm o} = \tan^{-1}\left(\frac{\overline{V_z}}
{\sqrt{\overline{V_x}^2 + \overline{V_y}^2}}\right)
\end{equation}
The positions of the apex points for NGC 1662, NGC 1778, NGC 2422, and NGC 2437 are indicated in Figure~\ref{fig:CP}. The corresponding numerical values are listed in Table~\ref{tab:VEP}.

\subsection{Dynamical Relaxation Time}

OCs, unlike their more compact globular counterparts, exhibit relatively loose stellar configurations. In such systems, repeated stellar encounters lead to kinetic energy exchange among members, gradually driving the cluster toward dynamical equilibrium \citep{Inagaki85, Baumgardt03}. This internal dynamical evolution often manifests as mass segregation, where more massive stars tend to concentrate toward the central regions, a behaviour reported for numerous OCs in recent studies \citep{Dib18, Bisht20, Bisht2026, Bisht2026b, Joshi20}.

The characteristic timescale over which a cluster approaches energy equipartition is described by the dynamical relaxation time, $T_{\rm relax}$. Following \citet{Spitzer71}, this quantity can be expressed as
\begin{equation}
\label{eq:Trelax}
T_{\rm relax} = \frac{8.9\times10^5\,~\sqrt{N}~ R_h^{3/2}}
{\sqrt{\overline{M_C}}~\log(0.4~N)},
\end{equation}
where $N$ is the total number of cluster members, $\overline{M_C}$ is the mean stellar mass, and $R_h$ is the half-mass radius in parsecs. The half-mass radius was estimated using the empirical relation proposed by \citet{Sableviciute_2006}
\begin{equation}
R_h = 0.547 \, r_c \left( \frac{r_t}{r_c} \right)^{0.486},
\end{equation}
where $r_c$ denotes the core radius and $r_t$ is the tidal radius. The latter was calculated following \citet{Jeffries2001}, adopting the relation $r_t = 1.46\, M_C^{1/3}$. The derived tidal radii are 8.64 $\pm$ 2.94 pc for NGC 1662, 11.64 $\pm$ 3.41 pc for NGC 1778, 15.11 $\pm$ 3.88 pc for NGC 2422, and 21.16 $\pm$ 4.60 pc for NGC 2437. The corresponding half-mass radii are 1.25, 3.97, 1.91, and 4.49 pc, respectively. Using these structural parameters, the dynamical relaxation times were estimated as 11 Myr for NGC 1662, 69 Myr for NGC 1778, 32 Myr for NGC 2422, and 133 Myr for NGC 2437.

The relaxation time also provides an estimate of the evaporation timescale, $\tau_{\rm ev}$, which represents the time required for the cluster to lose a substantial fraction of its members through internal stellar encounters \citep{Adams01}. This timescale is typically of the order of $10^2 \, T_{\rm relax}$. During this process, low-mass stars preferentially escape from the system, often through the Lagrange points, with relatively low velocities \citep{Kupper08}.

The dynamical state of each cluster was assessed through the dynamical evolution parameter, defined as $\tau = \text{age} / T_{\rm relax}$. Systems with $\tau \gg 1$ are considered dynamically relaxed, whereas values close to or below unity indicate that significant dynamical evolution is still ongoing. Our analysis shows that NGC 1662 has a $\tau$ value exceeding unity, suggesting that it has reached an advanced stage of dynamical relaxation. In contrast, the younger clusters NGC 1778, NGC 2422, and NGC 2437 exhibit $\tau$ values closer to unity, indicating that they are still progressing toward dynamical equilibrium. The derived dynamical parameters are summarized in Table~\ref{tab:VEP}.

\begin{figure*}[htb!]
\includegraphics[width=0.99\linewidth]{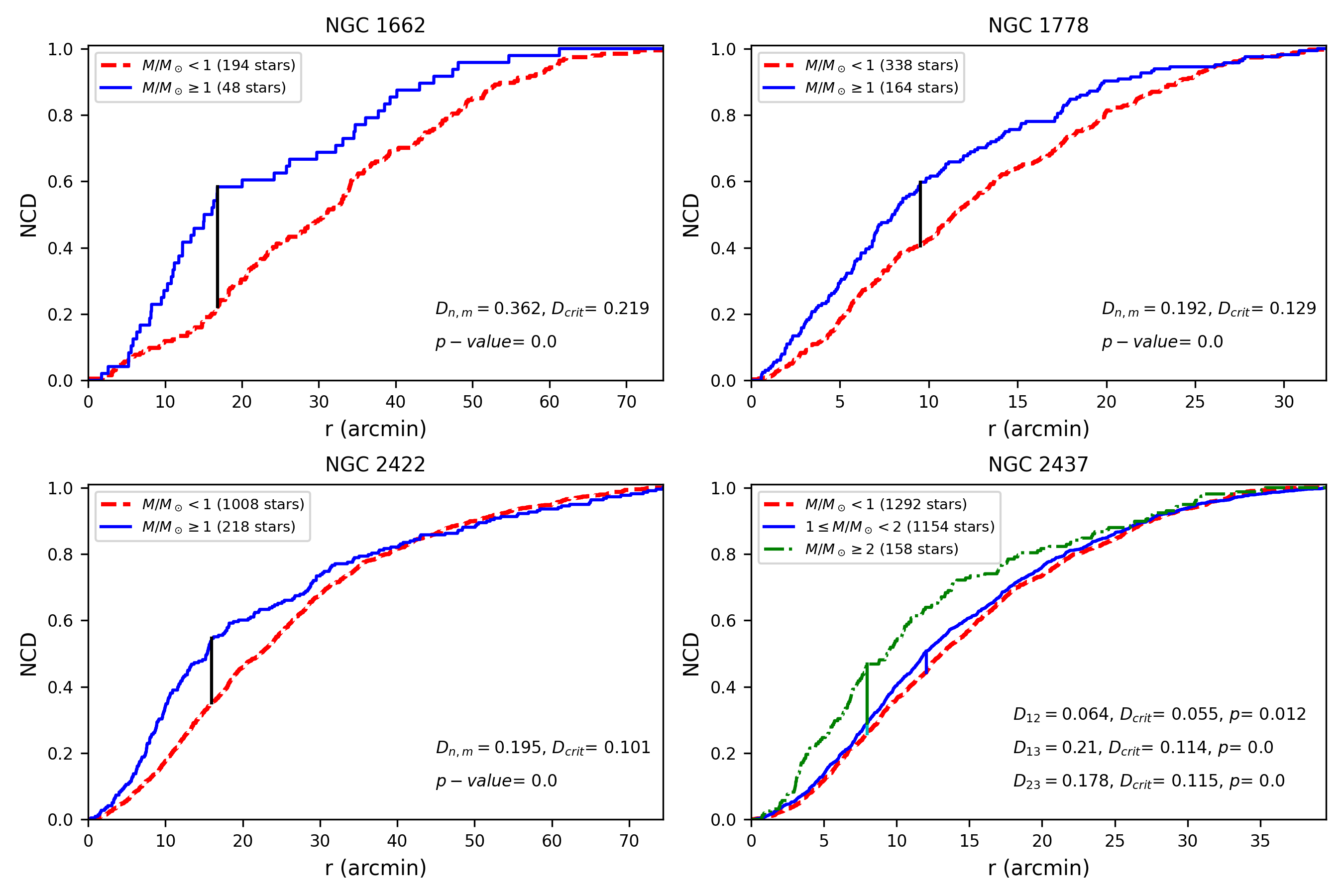}
\caption{NCDs of the estimated masses of the selected members of each cluster.}
\label{KS_tests}
\end{figure*}

\subsection{Mass Segregation}
Mass segregation refers to the tendency of more massive stars to be preferentially located toward the central regions of a cluster, while lower-mass members are distributed at larger radii. This effect can arise from early cluster conditions or from long-term dynamical evolution driven by two-body relaxation.

To test for mass segregation in NGC 1662, NGC 1778, NGC 2422, and NGC 2437, we examined the radial distributions of member stars in different mass intervals. For NGC 1662, NGC 1778, and NGC 2422, the sample was divided into two groups using a threshold of $M = 1\,M_{\odot}$. In the case of NGC 2437, three mass bins were defined with limits at $M = 1$ and $2\,M_{\odot}$, separating low-, intermediate-, and high-mass stars.
For each mass group, we constructed normalized cumulative radial distributions as a function of projected distance from the cluster center. Differences between the distributions were evaluated using the Kolmogorov–Smirnov (KS) test. The KS statistic measures the maximum absolute separation between two cumulative distributions,
\begin{equation}
D_{n,m} = \sup_x \left| F(x) - G(x) \right|,
\end{equation}
where $F(x)$ and $G(x)$ correspond to the cumulative distributions of two samples of sizes $n$ and $m$. The null hypothesis assumes that both samples are drawn from the same parent distribution. It is rejected when $D_{n,m}$ exceeds the critical value at a chosen significance level $\alpha$.
Figure~\ref{KS_tests} presents the cumulative radial distributions for the adopted mass ranges together with the corresponding KS statistics and $p$-values. In all four clusters, the higher-mass stars exhibit a more centrally concentrated distribution compared to lower-mass members at the $\alpha = 0.05$ significance level. Given that the cluster ages exceed their estimated relaxation times, the observed segregation is consistent with dynamical evolution.

\section{Dynamical Orbit Analyses}

To characterize the orbital dynamics of the four target clusters, we performed backward orbital integrations of their trajectories using the \textsc{galpy} suite \citep{Bovy2015}. The gravitational field was represented by the axisymmetric \textsc{MWPotential2014} model, which is physically motivated by a three-component architecture: a spherical bulge with a power-law density profile and exponential cut-off, a Miyamoto--Nagai stellar disc \citep{Miyamoto_Nagai1975}, and a spherical Navarro--Frenk--White (NFW) dark matter halo \citep{Navarro1996}. The composite potential is defined as:
\begin{equation}
\Phi_{\rm MW}(R,z) = \Phi_{\rm bulge}(r) + \Phi_{\rm disc}(R,z) + \Phi_{\rm halo}(r)
\end{equation}
where $r=\sqrt{R^{2}+z^{2}}$ denotes the Galactocentric radius. This model reproduces the principal dynamical constraints of the Milky Way, including the local circular velocity and the rotation curve, and is therefore widely adopted for studies of Galactic disc populations \citep{Bovy2015, Mackereth_Bovy2018}.

\begin{figure*}[!ht]
\centering
\includegraphics[width=0.95\linewidth]{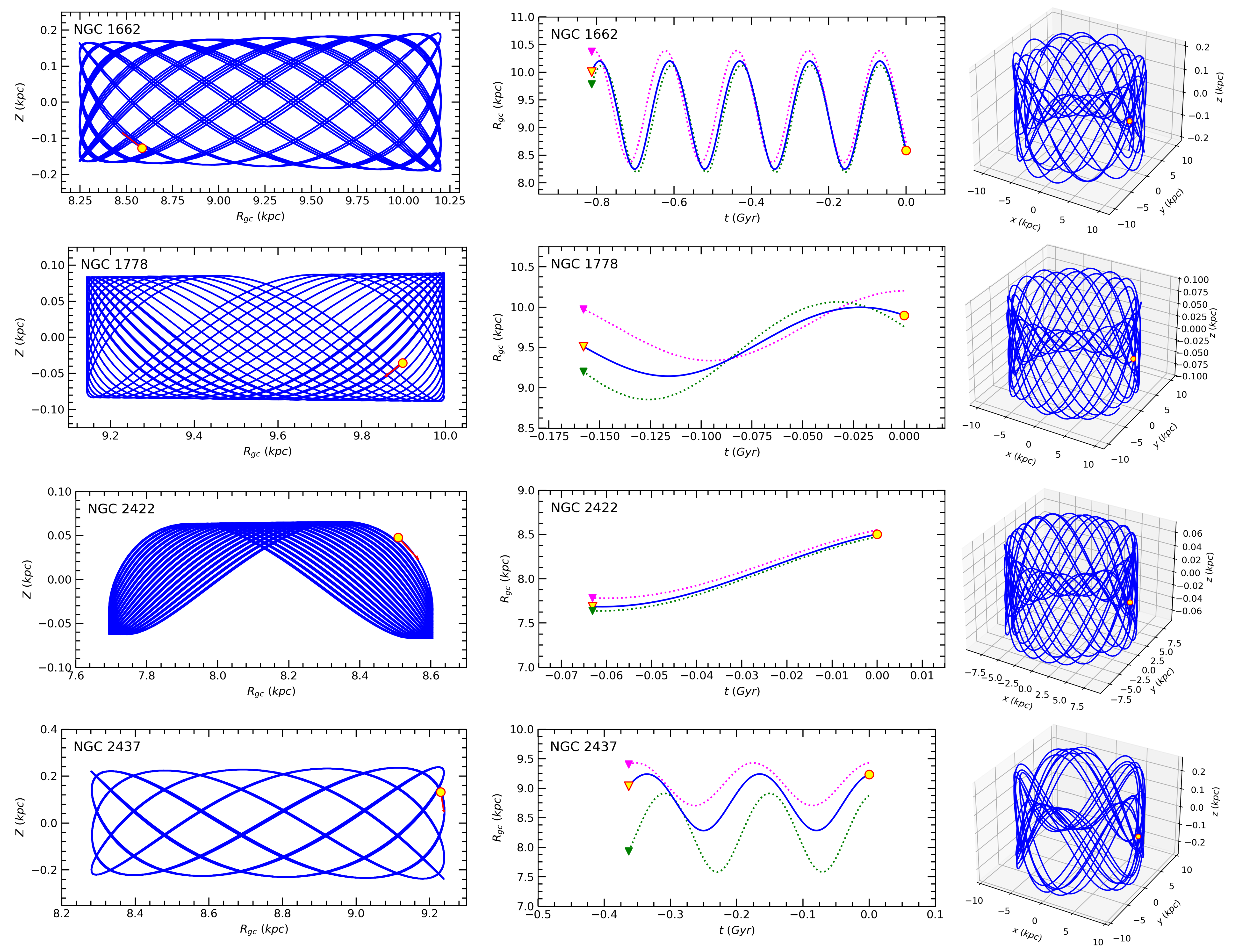}
\caption{Galactic orbits of NGC\,1662, NGC\,1778, NGC\,2422, and NGC\,2437 under the \textsc{MWPotential2014} model. \textit{Left panels:} Meridional plane projections ($R_{\rm gc}$ vs.\ $Z$) illustrating the orbital confinement and stability over a 3-Gyr period. \textit{Middle panels:} Time evolution of the Galactocentric distance ($R_{\rm gc}$) evaluated up to the clusters' respective fundamental ages; additional curves represent uncertainty propagation. \textit{Right panels:} Projections of the trajectories onto the three-dimensional Galactic $X$-$Y$-$Z$ plane. In all panels, yellow filled circles mark the present-day positions of clusters. }
\label{fig:orbital}
\end{figure*}

Phase-space initial conditions were derived from the current observational parameters of each cluster: equatorial coordinates $(\alpha,\delta)$, mean proper motions ($\langle\mu_{\alpha}\cos\delta\rangle$, $\langle\mu_{\delta}\rangle$), distance, and mean radial velocity $V_{\rm r}$. To explore the orbital phase space and determine the secular boundaries of the trajectories, the integrations were performed over a 3~Gyr time interval with a fixed time-step of 1.5~Myr. This time span enables the determination of time-averaged orbital descriptors, including the apogalactic ($R_{\rm a}$) and perigalactic ($R_{\rm p}$) radii, orbital eccentricity ($e$), maximum vertical excursion from the Galactic plane ($Z_{\rm max}$), mean Galactocentric radius ($R_{\rm m}$), and the radial period ($T_{\rm p}$). In addition, the primordial birth radii ($R_{\rm birth}$) were estimated by evaluating the integrated orbital trajectories at a look-back time corresponding to the estimated cluster ages \citep{YontanCanbay2023, Alzahrani2025a, Cinar2025, Cinar2026, Cinar2026b, Tasdemir2025}. The resulting orbital parameters are summarized in Table~\ref{tab:VEP}, while the reconstructed trajectories are shown in Fig.~\ref{fig:orbital}.

The derived orbital eccentricities ($e \simeq 0.045$--$0.106$) indicate nearly circular trajectories, consistent with the kinematic properties commonly associated with thin-disc stellar populations \citep{Wu2009, CantatGaudin2020}. NGC\,1662 shows the largest radial excursion ($e=0.106\pm0.002$), whereas the remaining clusters exhibit lower eccentricities ($e \approx 0.045$--$0.056$). The vertical amplitudes of the orbits remain modest ($Z_{\rm max} < 0.3$ kpc), suggesting that all four clusters are largely confined to the Galactic thin disc.

A comparison between the present-day Galactocentric distances and the inferred birth radii indicates moderate radial displacements along the Galactic disc. NGC\,2422 and NGC\,2437 may have radial displacement outward from their birth radii ($R_{\rm birth}=7.686$ and $9.033$~kpc) to their current locations at $R_{\rm GC}=8.508$ and $9.329$~kpc, respectively. A similar outward shift is inferred for NGC\,1778, which evolved from $R_{\rm birth}=9.510$~kpc to $R_{\rm GC}=9.934$~kpc. In contrast, NGC\,1662 suggests a possible inward displacement from a birth radius of $R_{\rm birth}=10.003$~kpc to its present position at $R_{\rm GC}=8.581$~kpc. Overall, the inferred radial variations ($\Delta R \approx 0.3$--$1.4$~kpc) are relatively small and are consistent with epicyclic oscillations expected for OCs evolving within a stable Galactic potential.

\section{Summary and Conclusions}

In this study, we presented a comprehensive photometric, astrometric, and dynamical investigation of four Galactic OCs utilizing the data from the \textit{Gaia} DR3 catalog. Our analysis yields fundamental parameters such as distances, ages, reddening values, mass functions, and orbital characteristics that provide a refined view compared to previous literature estimates. By applying robust membership selection criteria within the extraction radii of 40 to 80 arcmin, we identified highly probable cluster members for each system.  

From the astrometric solutions of the selected members, we derived mean parallaxes and proper motions and used them to estimate cluster distances and spatial locations. The resulting astrometric distances are 410 $\pm$ 12 pc (NGC~1662), 1761 $\pm$ 335 pc (NGC~1778), 477 $\pm$ 18 pc (NGC~2422), and 1681 $\pm$ 172 pc (NGC~2437), placing the clusters at Galactocentric distances of $R_{\rm GC}\simeq 8.508$--$9.934$ kpc with modest vertical offsets from the Galactic plane. These values are broadly consistent, within uncertainties, with recent \textit{Gaia}-based literature determinations.

The basic parameters derived in the present study were compared with the representative values listed in the Unified Cluster Catalog (UCC; \citet{Perren_2023}) in order to place our results within the context of recent catalog-level open-cluster determinations. The UCC parameters provide an external benchmark for assessing the consistency of our Gaia DR3-based analysis. The comparison shows that the astrometric parameters obtained in this work are in very good agreement with the UCC values. The mean differences, defined as this study minus UCC, are $\Delta\varpi=-0.0006$ mas, $\Delta\mu_{\alpha}\cos\delta=+0.0037$ mas yr$^{-1}$, and $\Delta\mu_{\delta}=+0.0034$ mas yr$^{-1}$. The corresponding median absolute differences are 0.0042 mas, 0.0092 mas yr$^{-1}$, and 0.0063 mas yr$^{-1}$, respectively. The photometric distances also show good agreement, with a median absolute difference of 0.025 kpc and a median relative absolute difference of about 2.9\%. The metallicities are consistent at the level of a median absolute difference of approximately 0.05 dex, while the systemic radial velocities show a median absolute difference of about 2.5 km s$^{-1}$. These small differences support the reliability of the present membership selection and parameter determination. At the same time, the present study extends beyond the catalog-level UCC parameters by providing a homogeneous member-based analysis of the structural, kinematical, internal dynamical, and Galactic orbital properties of NGC~1662, NGC~1778, NGC~2422, and NGC~2437.

The structural properties of the clusters were quantified through radial density profiles constructed around refined cluster centers. Gaussian fits to one-dimensional stellar-count histograms in right ascension and declination were used to determine the cluster centers, and the observed radial density profiles were modelled with a King-like surface-density prescription. The resulting core and cluster radii, central and background densities, and concentration parameters provide a consistent structural characterization of the four systems within the same \textit{Gaia} DR3 framework.

Fundamental astrophysical parameters were obtained via Bayesian isochrone fitting using \textsc{PARSEC} evolutionary models and an \texttt{emcee}-based MCMC sampling scheme. The derived ages ($\log(t/\text{yr})$) for the studied open clusters were determined as $8.91_{-0.33}^{+0.24}$ ($\approx 813$ Myr) for NGC~1662, $8.20_{-0.11}^{+0.18}$ ($\approx 158$ Myr) for NGC~1778, $7.80_{-0.11}^{+0.06}$ ($\approx 63$ Myr) for NGC~2422, and $8.56_{-0.34}^{+0.27}$ ($\approx 363$ Myr) for NGC~2437. The photometric distances ($d_{(m-M)}$) obtained from the MCMC isochrone fitting were determined to be $415_{-15}^{+12}$ pc for NGC~1662, $1725_{-21}^{+46}$ pc for NGC~1778, $471_{-96}^{+56}$ pc for NGC~2422, and $1544_{-507}^{+273}$ pc for NGC~2437. The derived $G$-band photometric extinction ($A_G$) values for the clusters were estimated as $0.58_{-0.35}^{+0.09}$ mag for NGC~1662, $0.96_{-0.09}^{+0.39}$ mag for NGC~1778, $0.14_{-0.10}^{+0.12}$ mag for NGC~2422, and $0.21_{-0.26}^{+0.33}$ mag for NGC~2437. Photometric distances inferred from the fitted moduli are consistent within uncertainties with the astrometric distances, supporting an internally coherent set of parameters for each cluster.

Using the cleaned member samples, we constructed luminosity functions in the \textit{Gaia} $G$ absolute band and transformed them into present-day mass functions through an isochrone-based mass--luminosity relation. The inferred mass-function slopes are broadly consistent, within uncertainties, with a Salpeter-like form over the magnitude ranges sampled, suggesting that the stellar mass distributions of the clusters follow the general behaviour expected for Galactic OCs.

We also derived kinematic properties by combining \textit{Gaia} DR3 astrometry with radial velocities available for subsets of member stars, supplemented by additional measurements from the Survey of Surveys compilation. Mean systemic radial velocities were obtained for the clusters and used to compute heliocentric and LSR-corrected Galactic space-velocity components $(U,V,W)$. The resulting total space velocities, $S_{\rm LSR}$, are consistent with disc-like kinematics, and the derived convergent-point (apex) directions provide an additional description of the bulk motions on the sky. Internal dynamical timescales were assessed through estimates of relaxation and evaporation times, and a mass-segregation analysis based on cumulative radial distributions indicates that higher-mass members are more centrally concentrated at conventional significance levels.

Backward orbital integrations over a 3-Gyr baseline within an axisymmetric Milky Way potential were used to characterize the long-term orbital morphology of the clusters. The inferred low eccentricities ($e \le 0.106$) and the limited vertical excursions ($Z_{\rm max} < 0.3$~kpc) are consistent with dynamically cold, thin-disc kinematics, suggesting that all four clusters predominantly reside within the Galactic disc.

By evaluating the integrated orbits at a look-back time corresponding to the cluster ages, we estimated their primordial birth radii ($R_{\rm birth}$). A comparison between the present-day Galactocentric distances ($R_{\rm GC}$) and the inferred birth radii indicates moderate radial displacements within the Galactic disc. NGC\,2422 and NGC\,2437 show possible outward radial displacements from $R_{\rm birth}=7.686\pm0.146$ and $9.033\pm1.472$~kpc to their current positions at $R_{\rm GC}=8.508$ and $9.329$~kpc, respectively. A similar outward shift is inferred for NGC\,1778, which evolved from $R_{\rm birth}=9.510\pm0.772$~kpc to $R_{\rm GC}=9.934$~kpc. 

In contrast, NGC\,1662 suggests a possible inward radial displacement, evolving from $R_{\rm birth}=10.003\pm0.592$~kpc to its present Galactocentric distance of $R_{\rm GC}=8.581$~kpc. These relatively small radial variations are consistent with epicyclic oscillations expected for OCs evolving within a stable Galactic potential.

\section*{Acknowledgments}
We would like to thank the anonymous referee for their helpful recommendations and feedback, which have contributed meaningfully to the improvement of our paper. This study presents results derived from the European Space Agency (ESA) space mission Gaia. The data from $Gaia$ are processed by the $Gaia$ Data Processing and Analysis Consortium (DPAC). Financial support for DPAC is provided by national institutions, primarily those participating in the $Gaia$ Multi-Lateral Agreement (MLA). For additional information, the official $Gaia$ mission website can be accessed at \url{https://www.cosmos.esa.int/gaia}, and the $Gaia$ archive is available at \url{https://archives.esac.esa.int/gaia}. The authors extend their appreciation to the Deanship of Scientific Research at Northern Border University, Arar, KSA for funding this research work through the project number “NBU-FFR-2026-1519-06”.


\appendix
\section{The RDPs of NGC 1778, NGC 2422 and NGC 2437}\label{rdps-append}

\begin{figure*}[ht!]
    \centering
   \includegraphics[width=0.5\linewidth]{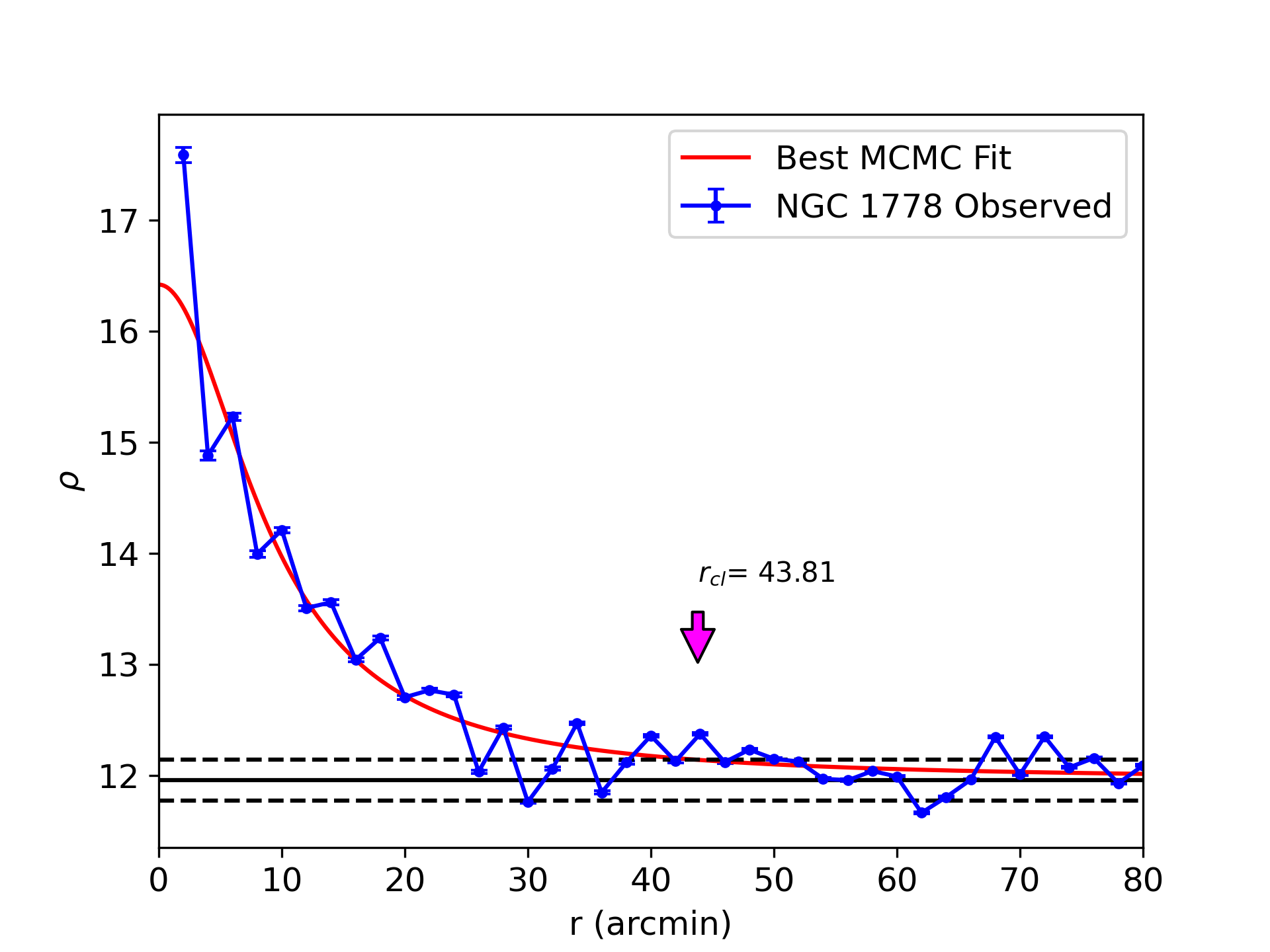}
    \includegraphics[width=0.45\linewidth]{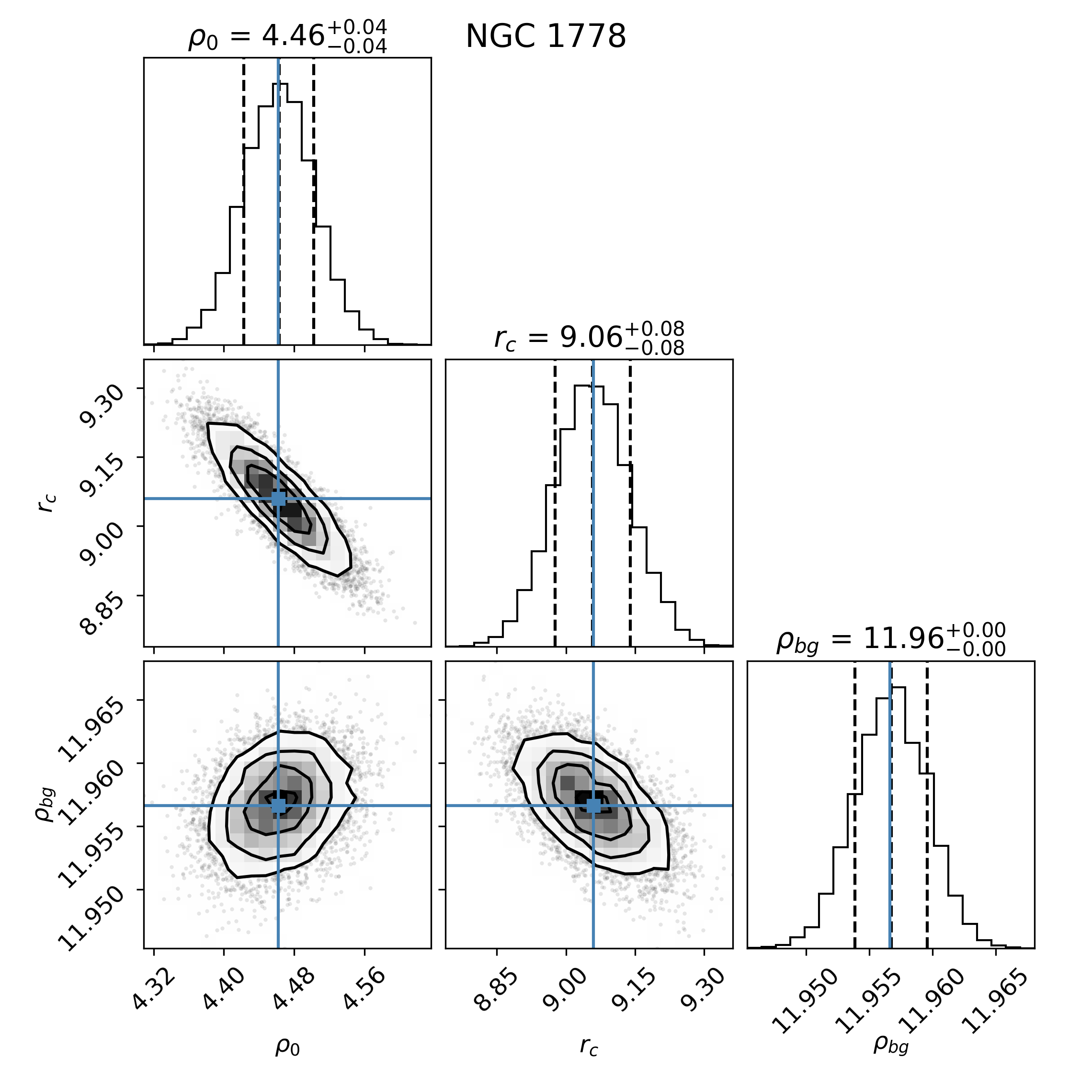}
    \includegraphics[width=0.5\linewidth]{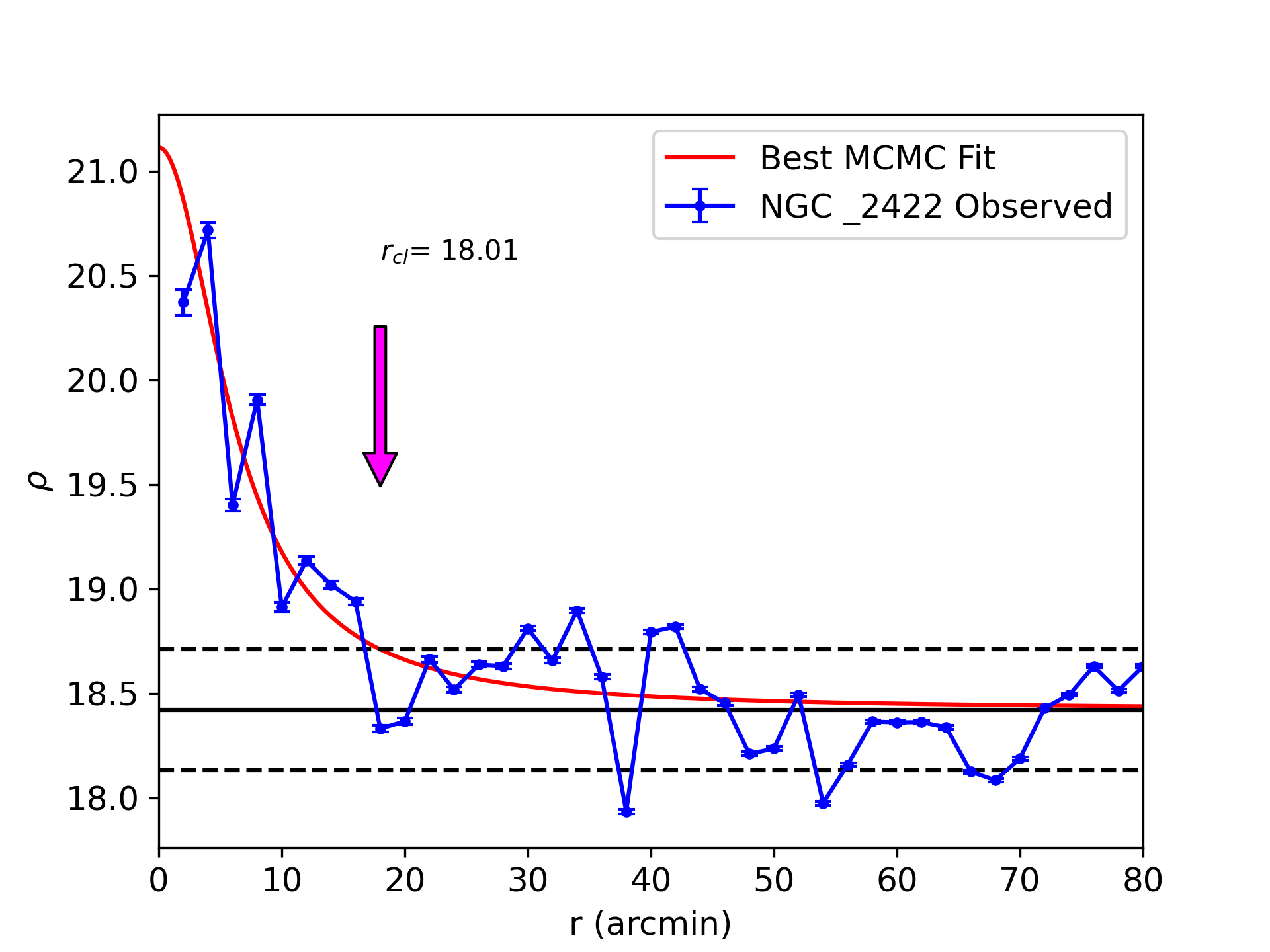}
    \includegraphics[width=0.45\linewidth]{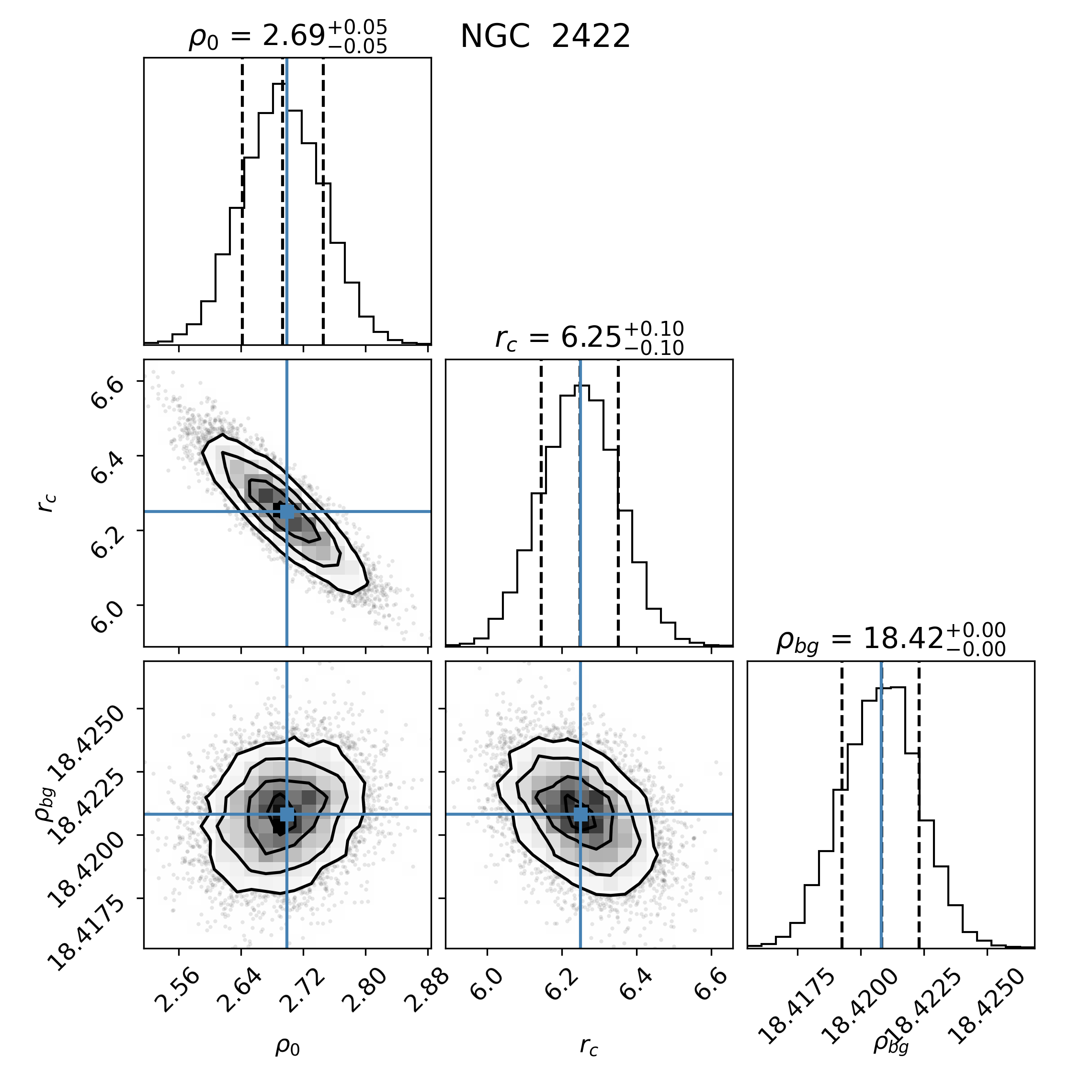}
   \includegraphics[width=0.5\linewidth]{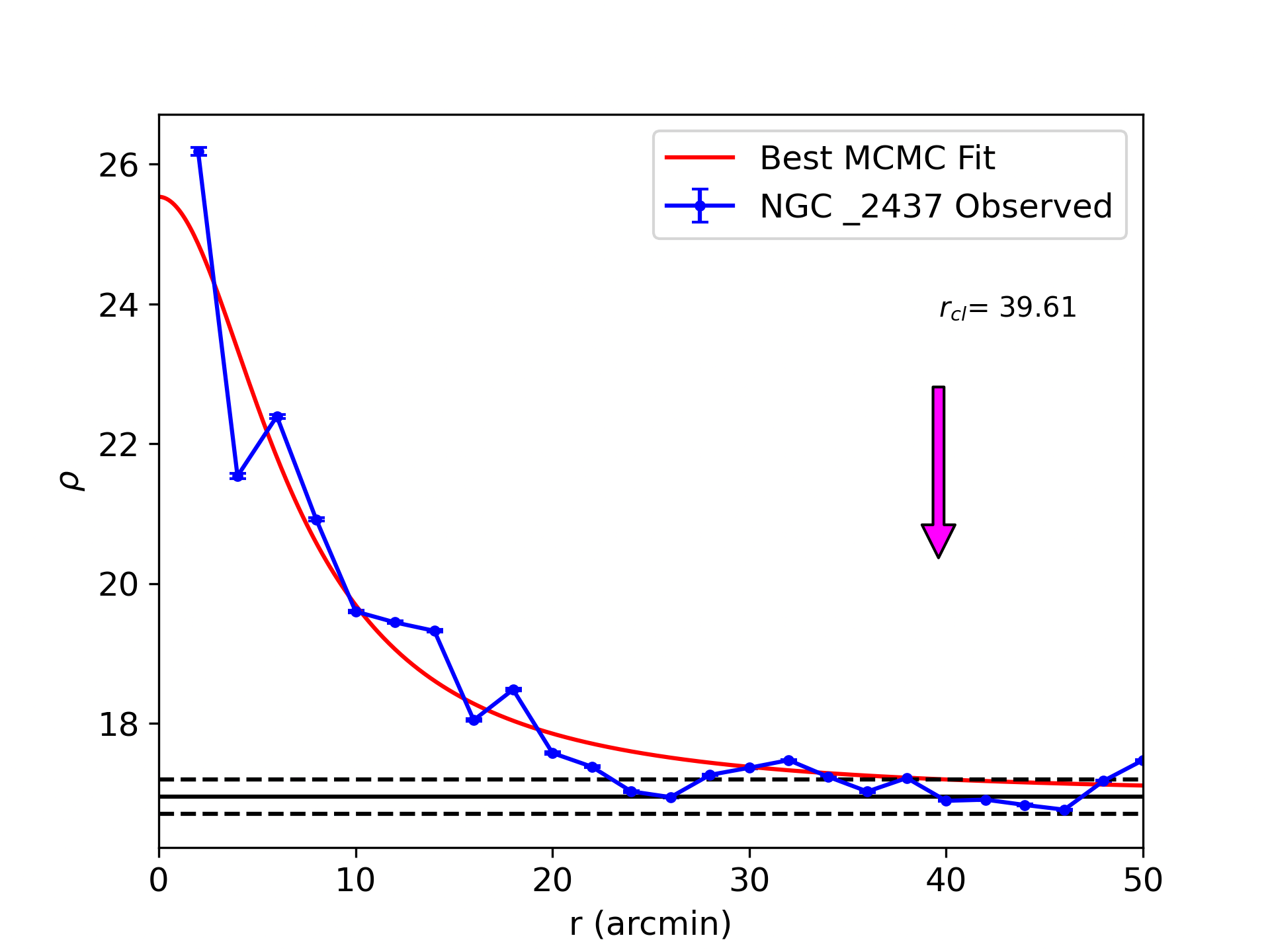}
    \includegraphics[width=0.45\linewidth]{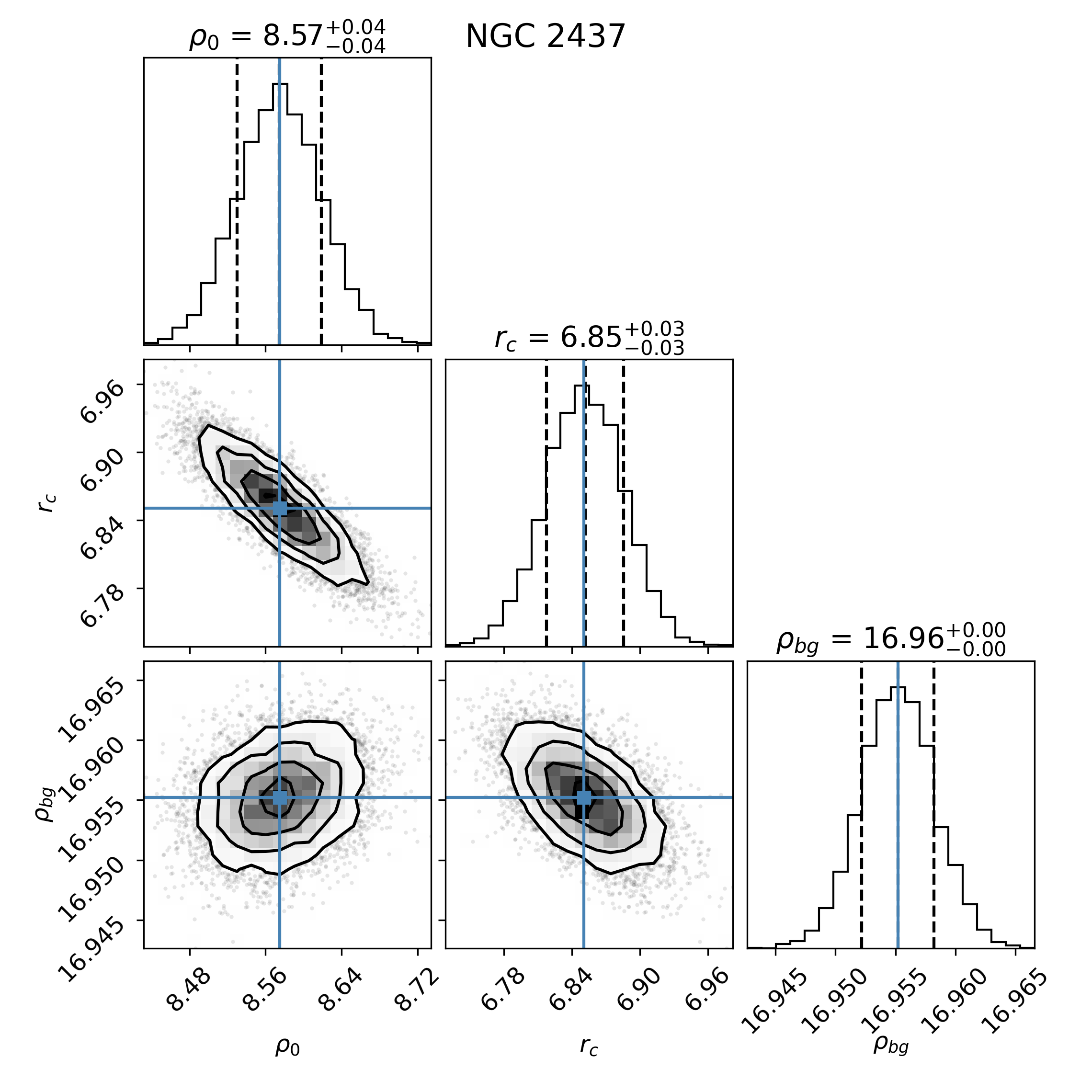}
    \caption{The same as Figure \ref{fig:rdps} for  NGC 1778, NGC 2422, and NGC 2437 OCs.}
    \label{rdps2}
\end{figure*}

In this appendix, we present the radial density profiles (RDPs) of the remaining three clusters analyzed in this study. The profiles were constructed following the same procedure described in Section~\ref{sec:RDP}, including the use of concentric annuli and Poisson-based uncertainties. For each cluster, the observed stellar surface densities are shown as a function of radius, together with the estimated background levels and the corresponding best-fitting King models (See Figure~\ref{rdps2}). These profiles are provided for completeness and allow a direct comparison of the structural properties across the full cluster sample.

\bibliographystyle{jasr-model5-names}
\biboptions{authoryear}
\bibliography{Reference}

\begin{thebibliography}{99}
\expandafter\ifx\csname natexlab\endcsname\relax\def\natexlab#1{#1}\fi
\ifx\xfnm\relax \def\xfnm[#1]{\unskip,\space#1}\fi

\bibitem[{{Adams} \& {Myers}(2001)}]{Adams01}
\bibinfo{author}{{Adams}, F.~C.},  \& \bibinfo{author}{{Myers}, P.~C.} (\bibinfo{year}{2001}).
\newblock \bibinfo{title}{{Modes of Multiple Star Formation}}.
\newblock {\it \bibinfo{journal}{\apj}\/},  {\it \bibinfo{volume}{553}\/}\bibinfo{issue}{(2)}, \bibinfo{pages}{744--753}. \DOIprefix\doi{10.1086/320941}. \href{http://arxiv.org/abs/astro-ph/0102039}{\tt arXiv:astro-ph/0102039}.

\bibitem[{{Ahmed} et~al.(2026){Ahmed}, {Elsanhoury}, {{\c{C}}{\i}nar}, {Haroon}, {Alenazi} \& {Alkholy}}]{Ahmed2026}
\bibinfo{author}{{Ahmed}, A.}, \bibinfo{author}{{Elsanhoury}, W.~H.}, \bibinfo{author}{{{\c{C}}{\i}nar}, D.~C.} et~al. (\bibinfo{year}{2026}).
\newblock \bibinfo{title}{{Structure and Evolution of Multi-Cluster within Galactic Disc: Gaia DR3 Insights into Eight Open Clusters}}.
\newblock {\it \bibinfo{journal}{arXiv e-prints}\/},  (p. \bibinfo{pages}{arXiv:2602.02669}). \DOIprefix\doi{10.48550/arXiv.2602.02669}. \href{http://arxiv.org/abs/2602.02669}{\tt arXiv:2602.02669}.

\bibitem[{{Ak} et~al.(2016){Ak}, {Bostanc{\i}}, {Yontan}, {Bilir}, {G{\"u}ver}, {Ak}, {{\"U}rg{\"u}p} \& {Paunzen}}]{2016Ap&SS.361..126A}
\bibinfo{author}{{Ak}, T.}, \bibinfo{author}{{Bostanc{\i}}, Z.~F.}, \bibinfo{author}{{Yontan}, T.} et~al. (\bibinfo{year}{2016}).
\newblock \bibinfo{title}{{CCD UBV photometry of the open cluster NGC 6819}}.
\newblock {\it \bibinfo{journal}{Astrophysics and Space Science}\/},  {\it \bibinfo{volume}{361}\/}, \bibinfo{pages}{126}. \DOIprefix\doi{10.1007/s10509-016-2707-2}. \href{http://arxiv.org/abs/1603.00474}{\tt arXiv:1603.00474}.

\bibitem[{{Alzhrani} et~al.(2025{\natexlab{a}}){Alzhrani}, {Haroon}, {Elsanhoury} \& {{\c{C}}{\i}nar}}]{Alzahrani2025b}
\bibinfo{author}{{Alzhrani}, A.~Y.}, \bibinfo{author}{{Haroon}, A.~A.}, \bibinfo{author}{{Elsanhoury}, W.~H.} et~al. (\bibinfo{year}{2025}{\natexlab{a}}).
\newblock \bibinfo{title}{{Enhancing SED-based astrometric, photometric, and kinematic studies of SAI 72 and SAI 75 using Gaia DR3}}.
\newblock {\it \bibinfo{journal}{Journal of Astrophysics and Astronomy}\/},  {\it \bibinfo{volume}{46}\/}\bibinfo{issue}{(2)}, \bibinfo{pages}{50}. \DOIprefix\doi{10.1007/s12036-025-10076-6}. \href{http://arxiv.org/abs/2504.15341}{\tt arXiv:2504.15341}.

\bibitem[{{Alzhrani} et~al.(2025{\natexlab{b}}){Alzhrani}, {Haroon}, {Elsanhoury} \& {{\c{C}}{\i}nar}}]{Alzahrani2025a}
\bibinfo{author}{{Alzhrani}, A.~Y.}, \bibinfo{author}{{Haroon}, A.~A.}, \bibinfo{author}{{Elsanhoury}, W.~H.} et~al. (\bibinfo{year}{2025}{\natexlab{b}}).
\newblock \bibinfo{title}{{In-depth analysis of photometric and kinematic characteristics of SAI 16, SAI 81 and SAI 86 open clusters utilizing Gaia DR3}}.
\newblock {\it \bibinfo{journal}{Journal of Astrophysics and Astronomy}\/},  {\it \bibinfo{volume}{46}\/}\bibinfo{issue}{(2)}, \bibinfo{pages}{58}. \DOIprefix\doi{10.1007/s12036-025-10083-7}. \href{http://arxiv.org/abs/2504.15344}{\tt arXiv:2504.15344}.

\bibitem[{{Amin} \& {Elsanhoury}(2017)}]{2017SerAJ.194...59A}
\bibinfo{author}{{Amin}, M.~Y.},  \& \bibinfo{author}{{Elsanhoury}, W.~H.} (\bibinfo{year}{2017}).
\newblock \bibinfo{title}{{Astrometric and Photometric Study of the Open Cluster NGC 2323}}.
\newblock {\it \bibinfo{journal}{Serbian Astronomical Journal}\/},  {\it \bibinfo{volume}{194}\/}, \bibinfo{pages}{59--69}. \DOIprefix\doi{10.2298/SAJ160802003A}.

\bibitem[{{Barbon} \& {Hassan}(1973)}]{1973A&AS...10....1B}
\bibinfo{author}{{Barbon}, R.},  \& \bibinfo{author}{{Hassan}, S.~M.} (\bibinfo{year}{1973}).
\newblock \bibinfo{title}{{A study of the open cluster NGC 1778}}.
\newblock {\it \bibinfo{journal}{Astronomy and Astrophysics Supplement}\/},  {\it \bibinfo{volume}{10}\/}, \bibinfo{pages}{1}.

\bibitem[{{Baumgardt} \& {Makino}(2003)}]{Baumgardt03}
\bibinfo{author}{{Baumgardt}, H.},  \& \bibinfo{author}{{Makino}, J.} (\bibinfo{year}{2003}).
\newblock \bibinfo{title}{{Dynamical evolution of star clusters in tidal fields}}.
\newblock {\it \bibinfo{journal}{\mnras}\/},  {\it \bibinfo{volume}{340}\/}\bibinfo{issue}{(1)}, \bibinfo{pages}{227--246}. \DOIprefix\doi{10.1046/j.1365-8711.2003.06286.x}. \href{http://arxiv.org/abs/astro-ph/0211471}{\tt arXiv:astro-ph/0211471}.

\bibitem[{{Bilir} et~al.(2026){Bilir}, {Ta{\textcommabelow s}demir}, {Erayd{\i}n}, {{\c{C}}{\i}nar}, {Alfonso} \& {Canbay}}]{Bilir2026}
\bibinfo{author}{{Bilir}, S.}, \bibinfo{author}{{Ta{\textcommabelow s}demir}, S.}, \bibinfo{author}{{Erayd{\i}n}, E.} et~al. (\bibinfo{year}{2026}).
\newblock \bibinfo{title}{{Gaia DR3 Analysis of Four Open Clusters Toward the Galactic Anticenter}}.
\newblock {\it \bibinfo{journal}{Research in Astronomy and Astrophysics}\/},  {\it \bibinfo{volume}{26}\/}\bibinfo{issue}{(8)}, \bibinfo{pages}{085002}. \DOIprefix\doi{10.1088/1674-4527/ae587d}. \href{http://arxiv.org/abs/2602.22304}{\tt arXiv:2602.22304}.

\bibitem[{{Bisht} et~al.(2020){Bisht}, {Elsanhoury}, {Zhu}, {Sariya}, {Yadav}, {Rangwal}, {Durgapal} \& {Jiang}}]{Bisht20}
\bibinfo{author}{{Bisht}, D.}, \bibinfo{author}{{Elsanhoury}, W.~H.}, \bibinfo{author}{{Zhu}, Q.} et~al. (\bibinfo{year}{2020}).
\newblock \bibinfo{title}{{An Investigation of Poorly Studied Open Cluster NGC 4337 Using Multicolor Photometric and Gaia DR2 Astrometric Data}}.
\newblock {\it \bibinfo{journal}{\aj}\/},  {\it \bibinfo{volume}{160}\/}\bibinfo{issue}{(3)}, \bibinfo{pages}{119}. \DOIprefix\doi{10.3847/1538-3881/ab9ffd}. \href{http://arxiv.org/abs/2006.13618}{\tt arXiv:2006.13618}.

\bibitem[{{Bisht} et~al.(2026{\natexlab{a}}){Bisht}, {Jiang}, {Belwal}, {{\c{C}}inar}, {Dattatrey}, {Rangwal}, {Raj}, {Biswas}, {Bisht} \& {Durgapal}}]{Bisht2026}
\bibinfo{author}{{Bisht}, D.}, \bibinfo{author}{{Jiang}, I.-G.}, \bibinfo{author}{{Belwal}, K.} et~al. (\bibinfo{year}{2026}{\natexlab{a}}).
\newblock \bibinfo{title}{{Multiwavelength Study of Blue Straggler Stars in Tombaugh 2: Evidence for Binary Mass Transfer and Constraints on Cluster Dynamical State}}.
\newblock {\it \bibinfo{journal}{\apj}\/},  {\it \bibinfo{volume}{1003}\/}\bibinfo{issue}{(2)}, \bibinfo{pages}{182}. \DOIprefix\doi{10.3847/1538-4357/ae6330}. \href{http://arxiv.org/abs/2604.12494}{\tt arXiv:2604.12494}.

\bibitem[{{Bisht} et~al.(2026{\natexlab{b}}){Bisht}, {Jiang}, {Belwal}, {{\c{C}}{\i}nar}, {Durgapal}, {Biswas}, {Raj}, {Rangwal}, {Bisht} \& {Manu}}]{Bisht2026b}
\bibinfo{author}{{Bisht}, D.}, \bibinfo{author}{{Jiang}, I.-G.}, \bibinfo{author}{{Belwal}, K.} et~al. (\bibinfo{year}{2026}{\natexlab{b}}).
\newblock \bibinfo{title}{{Blue straggler stars in Berkeley 18: a multiwavelength study of their physical properties and dynamical evolution}}.
\newblock {\it \bibinfo{journal}{\mnras}\/},  {\it \bibinfo{volume}{550}\/}\bibinfo{issue}{(1)}, \bibinfo{pages}{stag1068}. \DOIprefix\doi{10.1093/mnras/stag1068}. \href{http://arxiv.org/abs/2606.04624}{\tt arXiv:2606.04624}.

\bibitem[{Bisht et~al.(2026)Bisht, Jiang, Elsanhoury, Belwal, Çinar, Raj, Biswas, Dattatrey, Rangwal, Sariya, Bisht \& Durgapal}]{Bisht2025}
\bibinfo{author}{Bisht, D.}, \bibinfo{author}{Jiang, I.-G.}, \bibinfo{author}{Elsanhoury, W.~H.} et~al. (\bibinfo{year}{2026}).
\newblock \bibinfo{title}{Dynamical and photometric analysis of ngc 146 and king 14: Evidence for a comoving, unbound cluster pair}.
\newblock {\it \bibinfo{journal}{The Astronomical Journal}\/},  {\it \bibinfo{volume}{171}\/}\bibinfo{issue}{(2)}, \bibinfo{pages}{72}. \URLprefix \url{https://doi.org/10.3847/1538-3881/ae285b}. \DOIprefix\doi{10.3847/1538-3881/ae285b}.

\bibitem[{{Bisht} et~al.(2021){Bisht}, {Zhu}, {Yadav}, {Rangwal}, {Durgapal}, {Sariya} \& {Jiang}}]{Bisht2021}
\bibinfo{author}{{Bisht}, D.}, \bibinfo{author}{{Zhu}, Q.}, \bibinfo{author}{{Yadav}, R.~K.~S.} et~al. (\bibinfo{year}{2021}).
\newblock \bibinfo{title}{{A Deep Study of an Intermediate-age Open Cluster SAI 35 (Juchert 20) Using Ground-based Imaging and Gaia EDR3 Astrometry}}.
\newblock {\it \bibinfo{journal}{\aj}\/},  {\it \bibinfo{volume}{161}\/}\bibinfo{issue}{(4)}, \bibinfo{pages}{182}. \DOIprefix\doi{10.3847/1538-3881/abe383}. \href{http://arxiv.org/abs/2102.03735}{\tt arXiv:2102.03735}.

\bibitem[{{Bland-Hawthorn, J.} et~al.(2019){Bland-Hawthorn, J.}, {Sharma, S.} \& {Tepper-Garcia, T.}}]{bland2019galah}
\bibinfo{author}{{Bland-Hawthorn, J.}}, \bibinfo{author}{{Sharma, S.}},  \& \bibinfo{author}{{Tepper-Garcia, T.}} (\bibinfo{year}{2019}).
\newblock {\it \bibinfo{journal}{Monthly Notices of the Royal Astronomical Society}\/},  {\it \bibinfo{volume}{486}\/}, \bibinfo{pages}{1167}.

\bibitem[{{Bostanc{\i}} et~al.(2015){Bostanc{\i}}, {Ak}, {Yontan}, {Bilir}, {G{\"u}ver}, {Ak}, {{\c{C}}ak{\i}rl{\i}}, {{\"O}zdarcan}, {Paunzen}, {De Cat}, {Fu}, {Zhang}, {Hou}, {Li}, {Wang}, {Zhang}, {Shi} \& {Wu}}]{2015MNRAS.453.1095B}
\bibinfo{author}{{Bostanc{\i}}, Z.~F.}, \bibinfo{author}{{Ak}, T.}, \bibinfo{author}{{Yontan}, T.} et~al. (\bibinfo{year}{2015}).
\newblock \bibinfo{title}{{A comprehensive study of the open cluster NGC 6866}}.
\newblock {\it \bibinfo{journal}{Monthly Notices of the Royal Astronomical Society}\/},  {\it \bibinfo{volume}{453}\/}\bibinfo{issue}{(1)}, \bibinfo{pages}{1095--1107}. \DOIprefix\doi{10.1093/mnras/stv1665}. \href{http://arxiv.org/abs/1507.05968}{\tt arXiv:1507.05968}.

\bibitem[{{Bostanc{\i}} et~al.(2018){Bostanc{\i}}, {Yontan}, {Bilir}, {Ak}, {G{\"u}ver}, {Ak}, {Paunzen}, {Ba{\c{s}}aran}, {Vurgun}, {Akti}, {{\c{C}}elebi} \& {{\"U}rg{\"u}p}}]{Bostanci2018}
\bibinfo{author}{{Bostanc{\i}}, Z.~F.}, \bibinfo{author}{{Yontan}, T.}, \bibinfo{author}{{Bilir}, S.} et~al. (\bibinfo{year}{2018}).
\newblock \bibinfo{title}{{CCD UBV photometric study of five open clusters{\textemdash}Dolidze 36, NGC 6728, NGC 6800, NGC 7209, and Platais 1}}.
\newblock {\it \bibinfo{journal}{Astrophysics and Space Science}\/},  {\it \bibinfo{volume}{363}\/}\bibinfo{issue}{(7)}, \bibinfo{pages}{143}. \DOIprefix\doi{10.1007/s10509-018-3364-4}. \href{http://arxiv.org/abs/1806.00947}{\tt arXiv:1806.00947}.

\bibitem[{{Bovy}(2015)}]{Bovy2015}
\bibinfo{author}{{Bovy}, J.} (\bibinfo{year}{2015}).
\newblock \bibinfo{title}{{galpy: A python Library for Galactic Dynamics}}.
\newblock {\it \bibinfo{journal}{The Astrophysical Journal Supplement Series}\/},  {\it \bibinfo{volume}{216}\/}\bibinfo{issue}{(2)}, \bibinfo{pages}{29}. \DOIprefix\doi{10.1088/0067-0049/216/2/29}. \href{http://arxiv.org/abs/1412.3451}{\tt arXiv:1412.3451}.

\bibitem[{Bovy \& Tremaine(2012)}]{Bovy2012}
\bibinfo{author}{Bovy, J.},  \& \bibinfo{author}{Tremaine, S.} (\bibinfo{year}{2012}).
\newblock \bibinfo{title}{On the local dark matter density}.
\newblock {\it \bibinfo{journal}{The Astrophysical Journal}\/},  {\it \bibinfo{volume}{756}\/}\bibinfo{issue}{(1)}, \bibinfo{pages}{89}. \DOIprefix\doi{10.1088/0004-637X/756/1/89}.

\bibitem[{{Bressan} et~al.(2012{\natexlab{a}}){Bressan}, {Marigo}, {Girardi}, {Salasnich}, {Dal Cero}, {Rubele} \& {Nanni}}]{Bressan12}
\bibinfo{author}{{Bressan}, A.}, \bibinfo{author}{{Marigo}, P.}, \bibinfo{author}{{Girardi}, L.} et~al. (\bibinfo{year}{2012}{\natexlab{a}}).
\newblock \bibinfo{title}{{PARSEC: stellar tracks and isochrones with the PAdova and TRieste Stellar Evolution Code}}.
\newblock {\it \bibinfo{journal}{\mnras}\/},  {\it \bibinfo{volume}{427}\/}\bibinfo{issue}{(1)}, \bibinfo{pages}{127--145}. \DOIprefix\doi{10.1111/j.1365-2966.2012.21948.x}. \href{http://arxiv.org/abs/1208.4498}{\tt arXiv:1208.4498}.

\bibitem[{{Bressan} et~al.(2012{\natexlab{b}}){Bressan}, {Marigo}, {Girardi}, {Salasnich}, {Dal Cero}, {Rubele} \& {Nanni}}]{Bressan2012}
\bibinfo{author}{{Bressan}, A.}, \bibinfo{author}{{Marigo}, P.}, \bibinfo{author}{{Girardi}, L.} et~al. (\bibinfo{year}{2012}{\natexlab{b}}).
\newblock \bibinfo{title}{{PARSEC: stellar tracks and isochrones with the PAdova and TRieste Stellar Evolution Code}}.
\newblock {\it \bibinfo{journal}{Monthly Notices of the Royal Astronomical Society}\/},  {\it \bibinfo{volume}{427}\/}\bibinfo{issue}{(1)}, \bibinfo{pages}{127--145}. \DOIprefix\doi{10.1111/j.1365-2966.2012.21948.x}. \href{http://arxiv.org/abs/1208.4498}{\tt arXiv:1208.4498}.

\bibitem[{{Canbay} et~al.(2023){Canbay}, {Bilir}, {{\"O}zd{\"o}nmez} \& {Ak}}]{2023AJ....165..163C}
\bibinfo{author}{{Canbay}, R.}, \bibinfo{author}{{Bilir}, S.}, \bibinfo{author}{{{\"O}zd{\"o}nmez}, A.} et~al. (\bibinfo{year}{2023}).
\newblock \bibinfo{title}{{Galactic Model Parameters and Spatial Density of Cataclysmic Variables in the Gaia Era: New Constraints on Population Models}}.
\newblock {\it \bibinfo{journal}{The Astronomical Journal}\/},  {\it \bibinfo{volume}{165}\/}\bibinfo{issue}{(4)}, \bibinfo{pages}{163}. \DOIprefix\doi{10.3847/1538-3881/acbead}. \href{http://arxiv.org/abs/2302.11568}{\tt arXiv:2302.11568}.

\bibitem[{Cantat-Gaudin et~al.(2020)Cantat-Gaudin, Anders, Castro-Ginard, Jordi, Romero-G{\'o}mez, Soubiran, Casamiquela, Tarricq, Moitinho, Vallenari et~al.}]{CantatGaudin2020}
\bibinfo{author}{Cantat-Gaudin, T.}, \bibinfo{author}{Anders, F.}, \bibinfo{author}{Castro-Ginard, A.} et~al. (\bibinfo{year}{2020}).
\newblock \bibinfo{title}{{Painting a portrait of the Galactic disc with its stellar clusters}}.
\newblock {\it \bibinfo{journal}{Astronomy \& Astrophysics}\/},  {\it \bibinfo{volume}{640}\/}, \bibinfo{pages}{A1}.

\bibitem[{Cantat-Gaudin et~al.(2018)Cantat-Gaudin, Jordi, Vallenari, Bragaglia, Balaguer-N{\'u}{\~n}ez, Soubiran, Bossini, Moitinho, Castro-Ginard, Krone-Martins et~al.}]{Cantat-Gaudin2018}
\bibinfo{author}{Cantat-Gaudin, T.}, \bibinfo{author}{Jordi, C.}, \bibinfo{author}{Vallenari, A.} et~al. (\bibinfo{year}{2018}).
\newblock \bibinfo{title}{{A Gaia DR2 view of the open cluster population in the Milky Way}}.
\newblock {\it \bibinfo{journal}{Astronomy \& Astrophysics}\/},  {\it \bibinfo{volume}{618}\/}, \bibinfo{pages}{A93}.

\bibitem[{{Cantat-Gaudin, T.} \& {Anders, F.}(2020)}]{Cant20a}
\bibinfo{author}{{Cantat-Gaudin, T.}},  \& \bibinfo{author}{{Anders, F.}} (\bibinfo{year}{2020}).
\newblock \bibinfo{title}{{Clusters and mirages: Cataloguing stellar aggregates in the Milky Way}}.
\newblock {\it \bibinfo{journal}{Astronomy \& Astrophysics}\/},  {\it \bibinfo{volume}{633}\/}, \bibinfo{pages}{A99}.

\bibitem[{{Cardelli, J. A.} et~al.(1989){Cardelli, J. A.}, {Clayton, G. C.} \& {Mathis, J. S.}}]{cardelli1989relationship}
\bibinfo{author}{{Cardelli, J. A.}}, \bibinfo{author}{{Clayton, G. C.}},  \& \bibinfo{author}{{Mathis, J. S.}} (\bibinfo{year}{1989}).
\newblock {\it \bibinfo{journal}{Astrophysical Journal, Part 1}\/},  {\it \bibinfo{volume}{345}\/}, \bibinfo{pages}{245}.

\bibitem[{{Casagrande, L.} \& {VandenBerg, D. A.}(2018)}]{casagrande2018use}
\bibinfo{author}{{Casagrande, L.}},  \& \bibinfo{author}{{VandenBerg, D. A.}} (\bibinfo{year}{2018}).
\newblock {\it \bibinfo{journal}{Monthly Notices of the Royal Astronomical Society: Letters}\/},  {\it \bibinfo{volume}{479}\/}, \bibinfo{pages}{L102}.

\bibitem[{{{\c{C}}akmak} et~al.(2024){{\c{C}}akmak}, {Yontan}, {Bilir}, {Banks}, {Michel}, {Soydugan}, {Ko{\c{c}}} \& {Er{\c{c}}ay}}]{2024AN....34540054C}
\bibinfo{author}{{{\c{C}}akmak}, H.}, \bibinfo{author}{{Yontan}, T.}, \bibinfo{author}{{Bilir}, S.} et~al. (\bibinfo{year}{2024}).
\newblock \bibinfo{title}{{Photometric and Kinematic Studies of Open Clusters Ruprecht 1 and Ruprecht 171}}.
\newblock {\it \bibinfo{journal}{Astronomische Nachrichten}\/},  {\it \bibinfo{volume}{345}\/}\bibinfo{issue}{(8)}, \bibinfo{pages}{e20240054}. \DOIprefix\doi{10.1002/asna.20240054}. \href{http://arxiv.org/abs/2409.02298}{\tt arXiv:2409.02298}.

\bibitem[{{{\c{C}}{\i}nar} et~al.(2025){{\c{C}}{\i}nar}, {Bilir}, {{\c{S}}ahin} \& {Plevne}}]{Cinar2025}
\bibinfo{author}{{{\c{C}}{\i}nar}, D.~C.}, \bibinfo{author}{{Bilir}, S.}, \bibinfo{author}{{{\c{S}}ahin}, T.} et~al. (\bibinfo{year}{2025}).
\newblock \bibinfo{title}{{Tracing the Galactic Origins of Selected Four G-type Stars in the Solar Neighborhood}}.
\newblock {\it \bibinfo{journal}{The Astronomical Journal}\/},  {\it \bibinfo{volume}{170}\/}\bibinfo{issue}{(1)}, \bibinfo{pages}{13}. \DOIprefix\doi{10.3847/1538-3881/add343}. \href{http://arxiv.org/abs/2504.17844}{\tt arXiv:2504.17844}.

\bibitem[{{{\c{C}}{\i}nar} et~al.(2026){{\c{C}}{\i}nar}, {Elsanhoury} \& {Haroon}}]{Cinar2026}
\bibinfo{author}{{{\c{C}}{\i}nar}, D.~C.}, \bibinfo{author}{{Elsanhoury}, W.~H.},  \& \bibinfo{author}{{Haroon}, A.~A.} (\bibinfo{year}{2026}).
\newblock \bibinfo{title}{{Exploring the near galactic centre: A comprehensive study of bulge OCs HSC 25, HSC 37, HSC 2878 utilising Gaia DR3 data}}.
\newblock {\it \bibinfo{journal}{\na}\/},  {\it \bibinfo{volume}{125}\/}, \bibinfo{pages}{102532}. \DOIprefix\doi{10.1016/j.newast.2026.102532}.

\bibitem[{{Collinder}(1931{\natexlab{a}})}]{1931AnLun...2....1C}
\bibinfo{author}{{Collinder}, P.} (\bibinfo{year}{1931}{\natexlab{a}}).
\newblock \bibinfo{title}{{On Structural Properties of Open Galactic Clusters and their Spatial Distribution. Catalog of Open Galactic Clusters.}}
\newblock {\it \bibinfo{journal}{Annals of the Observatory of Lund}\/},  {\it \bibinfo{volume}{2}\/}, \bibinfo{pages}{B1--B46}.

\bibitem[{{Collinder}(1931{\natexlab{b}})}]{1931AnLun...2...64C}
\bibinfo{author}{{Collinder}, P.} (\bibinfo{year}{1931}{\natexlab{b}}).
\newblock \bibinfo{title}{{On Structural Properties of Open Galactic Clusters and their Spatial Distribution. Plates.}}
\newblock {\it \bibinfo{journal}{Annals of the Observatory of Lund}\/},  {\it \bibinfo{volume}{2}\/}, \bibinfo{pages}{B64--B79}.

\bibitem[{{Conrad} et~al.(2017){Conrad}, {Scholz}, {Kharchenko}, {Piskunov}, {R{\"o}ser}, {Schilbach}, {de Jong}, {Schnurr}, {Steinmetz}, {Grebel}, {Zwitter}, {Bienaym{\'e}}, {2422Bland-Hawthorn}, {Gibson}, {Gilmore}, {Kordopatis}, {Kunder}, {Navarro}, {Parker}, {Reid}, {Seabroke}, {Siviero}, {Watson} \& {Wyse}}]{2017A&A...600A.106C}
\bibinfo{author}{{Conrad}, C.}, \bibinfo{author}{{Scholz}, R.-D.}, \bibinfo{author}{{Kharchenko}, N.~V.} et~al. (\bibinfo{year}{2017}).
\newblock \bibinfo{title}{{A RAVE investigation on Galactic open clusters . II. Open cluster pairs, groups and complexes}}.
\newblock {\it \bibinfo{journal}{\aap}\/},  {\it \bibinfo{volume}{600}\/}, \bibinfo{pages}{A106}. \DOIprefix\doi{10.1051/0004-6361/201630012}.

\bibitem[{{Cuffey}(1941)}]{1941ApJ....94...55C}
\bibinfo{author}{{Cuffey}, J.} (\bibinfo{year}{1941}).
\newblock \bibinfo{title}{{The Galactic Clusters Messier 46, Messier 50, and NGC 2324.}}
\newblock {\it \bibinfo{journal}{\apj}\/},  {\it \bibinfo{volume}{94}\/}, \bibinfo{pages}{55}. \DOIprefix\doi{10.1086/144312}.

\bibitem[{{Daflon} \& {Cunha}(2004)}]{2004ApJ...617.1115D}
\bibinfo{author}{{Daflon}, S.},  \& \bibinfo{author}{{Cunha}, K.} (\bibinfo{year}{2004}).
\newblock \bibinfo{title}{{Galactic Metallicity Gradients Derived from a Sample of OB Stars}}.
\newblock {\it \bibinfo{journal}{\apj}\/},  {\it \bibinfo{volume}{617}\/}\bibinfo{issue}{(2)}, \bibinfo{pages}{1115--1126}. \DOIprefix\doi{10.1086/425607}. \href{http://arxiv.org/abs/astro-ph/0409084}{\tt arXiv:astro-ph/0409084}.

\bibitem[{{Dambis}(1999)}]{1999AstL...25....7D}
\bibinfo{author}{{Dambis}, A.~K.} (\bibinfo{year}{1999}).
\newblock \bibinfo{title}{{Space-age distribution of young open clusters and observational selection}}.
\newblock {\it \bibinfo{journal}{Astronomy Letters}\/},  {\it \bibinfo{volume}{25}\/}\bibinfo{issue}{(1)}, \bibinfo{pages}{7--13}.

\bibitem[{{Davidge}(2013)}]{2013PASP..125..115D}
\bibinfo{author}{{Davidge}, T.~J.} (\bibinfo{year}{2013}).
\newblock \bibinfo{title}{{The Open Cluster NGC 2437 (Messier 46)}}.
\newblock {\it \bibinfo{journal}{\pasp}\/},  {\it \bibinfo{volume}{125}\/}\bibinfo{issue}{(924)}, \bibinfo{pages}{115}. \DOIprefix\doi{10.1086/669823}. \href{http://arxiv.org/abs/1301.6806}{\tt arXiv:1301.6806}.

\bibitem[{{Dias} et~al.(2000){Dias}, {Boczko}, {Camargo}, {Teixeira} \& {Benevides-Soares}}]{2000A&A...357..149D}
\bibinfo{author}{{Dias}, W.~S.}, \bibinfo{author}{{Boczko}, R.}, \bibinfo{author}{{Camargo}, J.~I.~B.} et~al. (\bibinfo{year}{2000}).
\newblock \bibinfo{title}{{Proper motion and membership determination in the young open cluster NGC 1662}}.
\newblock {\it \bibinfo{journal}{\aap}\/},  {\it \bibinfo{volume}{357}\/}, \bibinfo{pages}{149--152}. \DOIprefix\doi{10.48550/arXiv.astro-ph/0003440}. \href{http://arxiv.org/abs/astro-ph/0003440}{\tt arXiv:astro-ph/0003440}.

\bibitem[{Dias et~al.(2021)Dias, Monteiro, Moitinho, L{\'e}pine, Carraro, Paunzen, Alessi \& Villela}]{Dias2021}
\bibinfo{author}{Dias, W.~S.}, \bibinfo{author}{Monteiro, H.}, \bibinfo{author}{Moitinho, A.} et~al. (\bibinfo{year}{2021}).
\newblock \bibinfo{title}{{Updated parameters of 1743 open clusters based on Gaia DR2}}.
\newblock {\it \bibinfo{journal}{Monthly Notices of the Royal Astronomical Society}\/},  {\it \bibinfo{volume}{504}\/}\bibinfo{issue}{(1)}, \bibinfo{pages}{356--371}.

\bibitem[{{Dias} et~al.(2021){Dias}, {Monteiro}, {Moitinho}, {L{\'e}pine}, {Carraro}, {Paunzen}, {Alessi} \& {Villela}}]{2021MNRAS.504..356D}
\bibinfo{author}{{Dias}, W.~S.}, \bibinfo{author}{{Monteiro}, H.}, \bibinfo{author}{{Moitinho}, A.} et~al. (\bibinfo{year}{2021}).
\newblock \bibinfo{title}{{Updated parameters of 1743 open clusters based on Gaia DR2}}.
\newblock {\it \bibinfo{journal}{\mnras}\/},  {\it \bibinfo{volume}{504}\/}\bibinfo{issue}{(1)}, \bibinfo{pages}{356--371}. \DOIprefix\doi{10.1093/mnras/stab770}. \href{http://arxiv.org/abs/2103.12829}{\tt arXiv:2103.12829}.

\bibitem[{{Dias, W.} et~al.(2014){Dias, W.}, {Monteiro, H.} \& {Caetano, T.}}]{dias2014proper}
\bibinfo{author}{{Dias, W.}}, \bibinfo{author}{{Monteiro, H.}},  \& \bibinfo{author}{{Caetano, T.}} (\bibinfo{year}{2014}).
\newblock {\it \bibinfo{journal}{Astronomy \& Astrophysics}\/},  {\it \bibinfo{volume}{564}\/}, \bibinfo{pages}{A79}.

\bibitem[{{Dib} \& {Basu}(2018)}]{Dib18}
\bibinfo{author}{{Dib}, S.},  \& \bibinfo{author}{{Basu}, S.} (\bibinfo{year}{2018}).
\newblock \bibinfo{title}{{The emergence of the galactic stellar mass function from a non-universal IMF in clusters}}.
\newblock {\it \bibinfo{journal}{\aap}\/},  {\it \bibinfo{volume}{614}\/}, \bibinfo{pages}{A43}. \DOIprefix\doi{10.1051/0004-6361/201732490}. \href{http://arxiv.org/abs/1711.07487}{\tt arXiv:1711.07487}.

\bibitem[{{Eker} et~al.(2024){Eker}, {Soydugan} \& {Bilir}}]{eker2024}
\bibinfo{author}{{Eker}, Z.}, \bibinfo{author}{{Soydugan}, F.},  \& \bibinfo{author}{{Bilir}, S.} (\bibinfo{year}{2024}).
\newblock \bibinfo{title}{{Fundamentals of Stars: A Critical Look at Mass-Luminosity Relations and Beyond}}.
\newblock {\it \bibinfo{journal}{Physics and Astronomy Reports}\/},  {\it \bibinfo{volume}{2}\/}\bibinfo{issue}{(1)}, \bibinfo{pages}{41--57}. \DOIprefix\doi{10.26650/PAR.2024.00001}. \href{http://arxiv.org/abs/2402.07947}{\tt arXiv:2402.07947}.

\bibitem[{{Elsanhoury}(2025)}]{Elsanhoury2025}
\bibinfo{author}{{Elsanhoury}, W.~H.} (\bibinfo{year}{2025}).
\newblock \bibinfo{title}{{Enhanced Kinematics and Distribution Characteristics of Upper Scorpius and Ophiuchus Associations Based on Gaia DR3}}.
\newblock {\it \bibinfo{journal}{Astronomische Nachrichten}\/},  {\it \bibinfo{volume}{346}\/}\bibinfo{issue}{(2)}, \bibinfo{pages}{e20240082}. \DOIprefix\doi{10.1002/asna.20240082}. \href{http://arxiv.org/abs/2501.10732}{\tt arXiv:2501.10732}.

\bibitem[{{Fernie}(1962)}]{1962AJ.....67..224F}
\bibinfo{author}{{Fernie}, J.~D.} (\bibinfo{year}{1962}).
\newblock \bibinfo{title}{{Interstellar absorption in the galactic neighborhood of the sun}}.
\newblock {\it \bibinfo{journal}{\aj}\/},  {\it \bibinfo{volume}{67}\/}, \bibinfo{pages}{224}. \DOIprefix\doi{10.1086/108698}.

\bibitem[{{Foreman-Mackey} et~al.(2013){Foreman-Mackey}, {Hogg}, {Lang} \& {Goodman}}]{Foreman-Mackey2013}
\bibinfo{author}{{Foreman-Mackey}, D.}, \bibinfo{author}{{Hogg}, D.~W.}, \bibinfo{author}{{Lang}, D.} et~al. (\bibinfo{year}{2013}).
\newblock \bibinfo{title}{{emcee: The MCMC Hammer}}.
\newblock {\it \bibinfo{journal}{Publications of the Astronomical Society of the Pacific}\/},  {\it \bibinfo{volume}{125}\/}\bibinfo{issue}{(925)}, \bibinfo{pages}{306}. \DOIprefix\doi{10.1086/670067}. \href{http://arxiv.org/abs/1202.3665}{\tt arXiv:1202.3665}.

\bibitem[{{Friel}(1995)}]{Friel95}
\bibinfo{author}{{Friel}, E.~D.} (\bibinfo{year}{1995}).
\newblock \bibinfo{title}{{The Old Open Clusters Of The Milky Way}}.
\newblock {\it \bibinfo{journal}{\araa}\/},  {\it \bibinfo{volume}{33}\/}, \bibinfo{pages}{381--414}. \DOIprefix\doi{10.1146/annurev.aa.33.090195.002121}.

\bibitem[{{Gaia Collaboration} et~al.(2018){Gaia Collaboration}, {Brown}, {Vallenari}, {Prusti}, {de Bruijne}, {Babusiaux}, {Bailer-Jones}, {Biermann}, {Evans}, {Eyer}, {Jansen}, {Jordi}, {Klioner}, {Lammers}, {Lindegren}, {Luri}, {Mignard}, {Panem}, {Pourbaix}, {Randich}, {Sartoretti}, {Siddiqui}, {Soubiran}, {van Leeuwen}, {Walton}, {Arenou}, {Bastian}, {Cropper}, {Drimmel}, {Katz}, {Lattanzi}, {Bakker}, {Cacciari}, {Casta{\~n}eda}, {Chaoul}, {Cheek}, {De Angeli}, {Fabricius}, {Guerra}, {Holl}, {Masana}, {Messineo}, {Mowlavi}, {Nienartowicz}, {Panuzzo}, {Portell}, {Riello}, {Seabroke}, {Tanga}, {Th{\'e}venin}, {Gracia-Abril}, {Comoretto}, {Garcia-Reinaldos}, {Teyssier}, {Altmann}, {Andrae}, {Audard}, {Bellas-Velidis}, {Benson}, {Berthier}, {Blomme}, {Burgess}, {Busso}, {Carry}, {Cellino}, {Clementini}, {Clotet}, {Creevey}, {Davidson}, {De Ridder}, {Delchambre}, {Dell'Oro}, {Ducourant}, {Fern{\'a}ndez-Hern{\'a}ndez}, {Fouesneau}, {Fr{\'e}mat}, {Galluccio}, {Garc{\'\i}a-Torres}, {Gonz{\'a}lez-N{\'u}{\~n}ez},
  {Gonz{\'a}lez-Vidal}, {Gosset}, {Guy}, {Halbwachs}, {Hambly}, {Harrison}, {Hern{\'a}ndez}, {Hestroffer}, {Hodgkin}, {Hutton}, {Jasniewicz}, {Jean-Antoine-Piccolo}, {Jordan}, {Korn}, {Krone-Martins}, {Lanzafame}, {Lebzelter}, {L{\"o}ffler}, {Manteiga}, {Marrese} et~al.}]{evans2018}
\bibinfo{author}{{Gaia Collaboration}}, \bibinfo{author}{{Brown}, A.~G.~A.}, \bibinfo{author}{{Vallenari}, A.} et~al. (\bibinfo{year}{2018}).
\newblock \bibinfo{title}{{Gaia Data Release 2. Summary of the contents and survey properties}}.
\newblock {\it \bibinfo{journal}{Astronomy \& Astrophysics}\/},  {\it \bibinfo{volume}{616}\/}, \bibinfo{pages}{A1}. \DOIprefix\doi{10.1051/0004-6361/201833051}. \href{http://arxiv.org/abs/1804.09365}{\tt arXiv:1804.09365}.

\bibitem[{{Gaia Collaboration} et~al.(2023{\natexlab{a}}){Gaia Collaboration}, {Drimmel}, {Romero-G{\'o}mez}, {Chemin}, {Ramos}, {Poggio}, {Ripepi}, {Andrae}, {Blomme}, {Cantat-Gaudin}, {Castro-Ginard}, {Clementini}, {Figueras}, {Fouesneau}, {Fr{\'e}mat}, {Jardine}, {Khanna}, {Lobel}, {Marshall}, {Muraveva}, {Brown}, {Vallenari}, {Prusti}, {de Bruijne}, {Arenou}, {Babusiaux}, {Biermann}, {Creevey}, {Ducourant}, {Evans}, {Eyer}, {Guerra}, {Hutton}, {Jordi}, {Klioner}, {Lammers}, {Lindegren}, {Luri}, {Mignard}, {Panem}, {Pourbaix}, {Randich}, {Sartoretti}, {Soubiran}, {Tanga}, {Walton}, {Bailer-Jones}, {Bastian}, {Jansen}, {Katz}, {Lattanzi}, {van Leeuwen}, {Bakker}, {Cacciari}, {Casta{\~n}eda}, {De Angeli}, {Fabricius}, {Galluccio}, {Guerrier}, {Heiter}, {Masana}, {Messineo}, {Mowlavi}, {Nicolas}, {Nienartowicz}, {Pailler}, {Panuzzo}, {Riclet}, {Roux}, {Seabroke}, {Sordo}, {Th{\'e}venin}, {Gracia-Abril}, {Portell}, {Teyssier}, {Altmann}, {Audard}, {Bellas-Velidis}, {Benson}, {Berthier}, {Burgess}, {Busonero},
  {Busso}, {C{\'a}novas}, {Carry}, {Cellino}, {Cheek}, {Damerdji}, {Davidson}, {de Teodoro}, {Nu{\~n}ez Campos}, {Delchambre}, {Dell'Oro}, {Esquej}, {Fern{\'a}ndez-Hern{\'a}ndez}, {Fraile}, {Garabato}, {Garc{\'\i}a-Lario}, {Gosset} et~al.}]{2023A&A...674A..37G}
\bibinfo{author}{{Gaia Collaboration}}, \bibinfo{author}{{Drimmel}, R.}, \bibinfo{author}{{Romero-G{\'o}mez}, M.} et~al. (\bibinfo{year}{2023}{\natexlab{a}}).
\newblock \bibinfo{title}{{Gaia Data Release 3. Mapping the asymmetric disc of the Milky Way}}.
\newblock {\it \bibinfo{journal}{\aap}\/},  {\it \bibinfo{volume}{674}\/}, \bibinfo{pages}{A37}. \DOIprefix\doi{10.1051/0004-6361/202243797}. \href{http://arxiv.org/abs/2206.06207}{\tt arXiv:2206.06207}.

\bibitem[{{Gaia Collaboration} et~al.(2023{\natexlab{b}}){Gaia Collaboration}, {Vallenari}, {Brown}, {Prusti}, {de Bruijne}, {Arenou}, {Babusiaux}, {Biermann}, {Creevey}, {Ducourant}, {Evans}, {Eyer}, {Guerra}, {Hutton}, {Jordi}, {Klioner}, {Lammers}, {Lindegren}, {Luri}, {Mignard}, {Panem}, {Pourbaix}, {Randich}, {Sartoretti}, {Soubiran}, {Tanga}, {Walton}, {Bailer-Jones}, {Bastian}, {Drimmel}, {Jansen}, {Katz}, {Lattanzi}, {van Leeuwen}, {Bakker}, {Cacciari}, {Casta{\~n}eda}, {De Angeli}, {Fabricius}, {Fouesneau}, {Fr{\'e}mat}, {Galluccio}, {Guerrier}, {Heiter}, {Masana}, {Messineo}, {Mowlavi}, {Nicolas}, {Nienartowicz}, {Pailler}, {Panuzzo}, {Riclet}, {Roux}, {Seabroke}, {Sordo}, {Th{\'e}venin}, {Gracia-Abril}, {Portell}, {Teyssier}, {Altmann}, {Andrae}, {Audard}, {Bellas-Velidis}, {Benson}, {Berthier}, {Blomme}, {Burgess}, {Busonero}, {Busso}, {C{\'a}novas}, {Carry}, {Cellino}, {Cheek}, {Clementini}, {Damerdji}, {Davidson}, {de Teodoro}, {Nu{\~n}ez Campos}, {Delchambre}, {Dell'Oro}, {Esquej},
  {Fern{\'a}ndez-Hern{\'a}ndez}, {Fraile}, {Garabato}, {Garc{\'\i}a-Lario}, {Gosset}, {Haigron}, {Halbwachs}, {Hambly}, {Harrison}, {Hern{\'a}ndez}, {Hestroffer}, {Hodgkin}, {Holl}, {Jan{\ss}en}, {Jevardat de Fombelle}, {Jordan}, {Krone-Martins}, {Lanzafame} et~al.}]{GaiaDR3}
\bibinfo{author}{{Gaia Collaboration}}, \bibinfo{author}{{Vallenari}, A.}, \bibinfo{author}{{Brown}, A.~G.~A.} et~al. (\bibinfo{year}{2023}{\natexlab{b}}).
\newblock \bibinfo{title}{{Gaia Data Release 3. Summary of the content and survey properties}}.
\newblock {\it \bibinfo{journal}{Astronomy \& Astrophysics}\/},  {\it \bibinfo{volume}{674}\/}, \bibinfo{pages}{A1}. \DOIprefix\doi{10.1051/0004-6361/202243940}. \href{http://arxiv.org/abs/2208.00211}{\tt arXiv:2208.00211}.

\bibitem[{{Gelman} \& {Rubin}(1992)}]{gelman1992}
\bibinfo{author}{{Gelman}, A.},  \& \bibinfo{author}{{Rubin}, D.~B.} (\bibinfo{year}{1992}).
\newblock \bibinfo{title}{{Inference from Iterative Simulation Using Multiple Sequences}}.
\newblock {\it \bibinfo{journal}{Statistical Science}\/},  {\it \bibinfo{volume}{7}\/}, \bibinfo{pages}{457--472}. \DOIprefix\doi{10.1214/ss/1177011136}.

\bibitem[{{Gossage} et~al.(2018){Gossage}, {Conroy}, {Dotter}, {Choi}, {Rosenfield}, {Cargile} \& {Dolphin}}]{2018ApJ...863...67G}
\bibinfo{author}{{Gossage}, S.}, \bibinfo{author}{{Conroy}, C.}, \bibinfo{author}{{Dotter}, A.} et~al. (\bibinfo{year}{2018}).
\newblock \bibinfo{title}{{Age Determinations of the Hyades, Praesepe, and Pleiades via MESA Models with Rotation}}.
\newblock {\it \bibinfo{journal}{\apj}\/},  {\it \bibinfo{volume}{863}\/}\bibinfo{issue}{(1)}, \bibinfo{pages}{67}. \DOIprefix\doi{10.3847/1538-4357/aad0a0}. \href{http://arxiv.org/abs/1804.06441}{\tt arXiv:1804.06441}.

\bibitem[{{Haroon} et~al.(2025){Haroon}, {Elsanhoury}, {Elkholy}, {Saad} \& {{\c{C}}{\i}nar}}]{Haroon2025}
\bibinfo{author}{{Haroon}, A.~A.}, \bibinfo{author}{{Elsanhoury}, W.~H.}, \bibinfo{author}{{Elkholy}, E.~A.} et~al. (\bibinfo{year}{2025}).
\newblock \bibinfo{title}{{Study of open star clusters using the gaia DR3: I-poorly studied king 2 and king 5}}.
\newblock {\it \bibinfo{journal}{Physica Scripta}\/},  {\it \bibinfo{volume}{100}\/}\bibinfo{issue}{(5)}, \bibinfo{pages}{055006}. \DOIprefix\doi{10.1088/1402-4896/adbf71}.

\bibitem[{{Hassan}(1972)}]{1972MmSAI..43..117H}
\bibinfo{author}{{Hassan}, S.~M.} (\bibinfo{year}{1972}).
\newblock \bibinfo{title}{{A new photometric investigation of NGC 1662.}}
\newblock {\it \bibinfo{journal}{\memsai}\/},  {\it \bibinfo{volume}{43}\/}, \bibinfo{pages}{117--122}.

\bibitem[{{Hoag} \& {Applequist}(1965)}]{1965ApJS...12..215H}
\bibinfo{author}{{Hoag}, A.~A.},  \& \bibinfo{author}{{Applequist}, N.~L.} (\bibinfo{year}{1965}).
\newblock \bibinfo{title}{{Distance Moduli of Open Clusters.}}
\newblock {\it \bibinfo{journal}{\apjs}\/},  {\it \bibinfo{volume}{12}\/}, \bibinfo{pages}{215}. \DOIprefix\doi{10.1086/190125}.

\bibitem[{{Hoag} et~al.(1961){Hoag}, {Johnson}, {Iriarte}, {Mitchell}, {Hallam} \& {Sharpless}}]{1961PUSNO..17..343H}
\bibinfo{author}{{Hoag}, A.~A.}, \bibinfo{author}{{Johnson}, H.~L.}, \bibinfo{author}{{Iriarte}, B.} et~al. (\bibinfo{year}{1961}).
\newblock \bibinfo{title}{{Photometry of stars in galactic cluster fields}}.
\newblock {\it \bibinfo{journal}{Publications of the U.S. Naval Observatory Second Series}\/},  {\it \bibinfo{volume}{17}\/}, \bibinfo{pages}{344--542}.

\bibitem[{{Hunt} \& {Reffert}(2024)}]{Hunt2024}
\bibinfo{author}{{Hunt}, E.~L.},  \& \bibinfo{author}{{Reffert}, S.} (\bibinfo{year}{2024}).
\newblock \bibinfo{title}{{Improving the open cluster census. III. Using cluster masses, radii, and dynamics to create a cleaned open cluster catalogue}}.
\newblock {\it \bibinfo{journal}{Astronomy \& Astrophysics}\/},  {\it \bibinfo{volume}{686}\/}, \bibinfo{pages}{A42}. \DOIprefix\doi{10.1051/0004-6361/202348662}. \href{http://arxiv.org/abs/2403.05143}{\tt arXiv:2403.05143}.

\bibitem[{{Inagaki} \& {Saslaw}(1985)}]{Inagaki85}
\bibinfo{author}{{Inagaki}, S.},  \& \bibinfo{author}{{Saslaw}, W.~C.} (\bibinfo{year}{1985}).
\newblock \bibinfo{title}{{Equipartition in multicomponent gravitational systems}}.
\newblock {\it \bibinfo{journal}{\apj}\/},  {\it \bibinfo{volume}{292}\/}, \bibinfo{pages}{339--347}. \DOIprefix\doi{10.1086/163164}.

\bibitem[{{Janes}(1979)}]{1979ApJS...39..135J}
\bibinfo{author}{{Janes}, K.~A.} (\bibinfo{year}{1979}).
\newblock \bibinfo{title}{{Evidence for an abundance gradient in the galactic disk.}}
\newblock {\it \bibinfo{journal}{\apjs}\/},  {\it \bibinfo{volume}{39}\/}, \bibinfo{pages}{135--156}. \DOIprefix\doi{10.1086/190568}.

\bibitem[{{Jeffries} et~al.(2001){Jeffries}, {Thurston} \& {Hambly}}]{Jeffries2001}
\bibinfo{author}{{Jeffries}, R.~D.}, \bibinfo{author}{{Thurston}, M.~R.},  \& \bibinfo{author}{{Hambly}, N.~C.} (\bibinfo{year}{2001}).
\newblock \bibinfo{title}{{Photometry and membership for low mass stars in the young open cluster NGC 2516}}.
\newblock {\it \bibinfo{journal}{\aap}\/},  {\it \bibinfo{volume}{375}\/}, \bibinfo{pages}{863--889}. \DOIprefix\doi{10.1051/0004-6361:20010918}. \href{http://arxiv.org/abs/astro-ph/0107097}{\tt arXiv:astro-ph/0107097}.

\bibitem[{{Joshi} et~al.(2020){Joshi}, {Maurya}, {John}, {Panchal}, {Joshi} \& {Kumar}}]{Joshi20}
\bibinfo{author}{{Joshi}, Y.~C.}, \bibinfo{author}{{Maurya}, J.}, \bibinfo{author}{{John}, A.~A.} et~al. (\bibinfo{year}{2020}).
\newblock \bibinfo{title}{{Photometric, kinematic, and variability study in the young open cluster NGC 1960}}.
\newblock {\it \bibinfo{journal}{\mnras}\/},  {\it \bibinfo{volume}{492}\/}\bibinfo{issue}{(3)}, \bibinfo{pages}{3602--3621}. \DOIprefix\doi{10.1093/mnras/staa029}. \href{http://arxiv.org/abs/2001.04068}{\tt arXiv:2001.04068}.

\bibitem[{{Karag{\"o}z} et~al.(2025){Karag{\"o}z}, {{\c{C}}{\i}nar}, {Ta{\c{s}}demir} \& {Canbay}}]{Karagoz2025}
\bibinfo{author}{{Karag{\"o}z}, H.}, \bibinfo{author}{{{\c{C}}{\i}nar}, D.~C.}, \bibinfo{author}{{Ta{\c{s}}demir}, S.} et~al. (\bibinfo{year}{2025}).
\newblock \bibinfo{title}{{Close Binary Open Cluster Systems: A Gaia DR3 Perspective on Alessi 36 and Collinder 135}}.
\newblock {\it \bibinfo{journal}{arXiv e-prints}\/},  (p. \bibinfo{pages}{arXiv:2509.14308}). \DOIprefix\doi{10.48550/arXiv.2509.14308}. \href{http://arxiv.org/abs/2509.14308}{\tt arXiv:2509.14308}.

\bibitem[{{Kharchenko, N.} et~al.(2013){Kharchenko, N.}, {Piskunov, A.} \& {Schilbach, E.}}]{kharchenko2013global}
\bibinfo{author}{{Kharchenko, N.}}, \bibinfo{author}{{Piskunov, A.}},  \& \bibinfo{author}{{Schilbach, E.}} (\bibinfo{year}{2013}).
\newblock {\it \bibinfo{journal}{Astronomy \& Astrophysics}\/},  {\it \bibinfo{volume}{558}\/}, \bibinfo{pages}{A53}.

\bibitem[{King(1962)}]{King1962}
\bibinfo{author}{King, I.} (\bibinfo{year}{1962}).
\newblock \bibinfo{title}{{The structure of star clusters. I. an empirical density law}}.
\newblock {\it \bibinfo{journal}{The Astronomical Journal}\/},  {\it \bibinfo{volume}{67}\/}, \bibinfo{pages}{471}.

\bibitem[{King(1966)}]{King1966}
\bibinfo{author}{King, I.~R.} (\bibinfo{year}{1966}).
\newblock \bibinfo{title}{{The structure of star clusters. III. Some simple dynamical models}}.
\newblock {\it \bibinfo{journal}{The Astronomical Journal}\/},  {\it \bibinfo{volume}{71}\/}, \bibinfo{pages}{64}.

\bibitem[{Krone-Martins \& Moitinho(2014)}]{Krone-Martins_2014}
\bibinfo{author}{Krone-Martins, A.},  \& \bibinfo{author}{Moitinho, A.} (\bibinfo{year}{2014}).
\newblock \bibinfo{title}{{UPMASK}: unsupervised photometric membership assignment in stellar clusters}.
\newblock {\it \bibinfo{journal}{A\&A}\/},  {\it \bibinfo{volume}{561}\/}, \bibinfo{pages}{A57}. \DOIprefix\doi{10.1051/0004-6361/201321143}.

\bibitem[{{K{\"u}pper} et~al.(2008){K{\"u}pper}, {MacLeod} \& {Heggie}}]{Kupper08}
\bibinfo{author}{{K{\"u}pper}, A. H.~W.}, \bibinfo{author}{{MacLeod}, A.},  \& \bibinfo{author}{{Heggie}, D.~C.} (\bibinfo{year}{2008}).
\newblock \bibinfo{title}{{On the structure of tidal tails}}.
\newblock {\it \bibinfo{journal}{\mnras}\/},  {\it \bibinfo{volume}{387}\/}\bibinfo{issue}{(3)}, \bibinfo{pages}{1248--1252}. \DOIprefix\doi{10.1111/j.1365-2966.2008.13323.x}. \href{http://arxiv.org/abs/0804.2476}{\tt arXiv:0804.2476}.

\bibitem[{{Mackereth} \& {Bovy}(2018)}]{Mackereth_Bovy2018}
\bibinfo{author}{{Mackereth}, J.~T.},  \& \bibinfo{author}{{Bovy}, J.} (\bibinfo{year}{2018}).
\newblock \bibinfo{title}{{Fast Estimation of Orbital Parameters in Milky Way-like Potentials}}.
\newblock {\it \bibinfo{journal}{\pasp}\/},  {\it \bibinfo{volume}{130}\/}\bibinfo{issue}{(993)}, \bibinfo{pages}{114501}. \DOIprefix\doi{10.1088/1538-3873/aadcdd}. \href{http://arxiv.org/abs/1802.02592}{\tt arXiv:1802.02592}.

\bibitem[{{Marigo, P.} et~al.(2017){Marigo, P.}, {Girardi, L.} \& {Bressan, A.}}]{marigo2017new}
\bibinfo{author}{{Marigo, P.}}, \bibinfo{author}{{Girardi, L.}},  \& \bibinfo{author}{{Bressan, A.}} (\bibinfo{year}{2017}).
\newblock {\it \bibinfo{journal}{The Astrophysical Journal}\/},  {\it \bibinfo{volume}{835}\/}, \bibinfo{pages}{77}.

\bibitem[{{Miyamoto} \& {Nagai}(1975)}]{Miyamoto_Nagai1975}
\bibinfo{author}{{Miyamoto}, M.},  \& \bibinfo{author}{{Nagai}, R.} (\bibinfo{year}{1975}).
\newblock \bibinfo{title}{{Three-Dimensional Models for the Distribution of Mass in Galaxies}}.
\newblock {\it \bibinfo{journal}{\pasj}\/},  {\it \bibinfo{volume}{27}\/}\bibinfo{issue}{(4)}, \bibinfo{pages}{533--543}. \DOIprefix\doi{10.1093/pasj/27.4.533}.

\bibitem[{{Navarro}(1996)}]{Navarro1996}
\bibinfo{author}{{Navarro}, J.} (\bibinfo{year}{1996}).
\newblock \bibinfo{title}{{The Structure of Cold Dark Matter Halos}}.
\newblock In \bibinfo{editor}{R.~{Bender}}, \& \bibinfo{editor}{R.~L. {Davies}} (Eds.), {\it \bibinfo{booktitle}{New Light on Galaxy Evolution}\/} (p. \bibinfo{pages}{255}).
\newblock volume \bibinfo{volume}{171} of {\it \bibinfo{series}{IAU Symposium}\/}.
\newblock \DOIprefix\doi{10.48550/arXiv.astro-ph/9511016}. \href{http://arxiv.org/abs/astro-ph/9511016}{\tt arXiv:astro-ph/9511016}.

\bibitem[{{Pandey}(1998)}]{1998BASI...26..539P}
\bibinfo{author}{{Pandey}, A.~K.} (\bibinfo{year}{1998}).
\newblock \bibinfo{title}{{A report on the studies of star clusters with the UPSO 104-cm Sampurnanand telescope during last 25 years}}.
\newblock {\it \bibinfo{journal}{Bulletin of the Astronomical Society of India}\/},  {\it \bibinfo{volume}{26}\/}, \bibinfo{pages}{539}.

\bibitem[{Perren et~al.(2023)Perren, Pera, Navone \& Vázquez}]{Perren_2023}
\bibinfo{author}{Perren, G.~I.}, \bibinfo{author}{Pera, M.~S.}, \bibinfo{author}{Navone, H.~D.} et~al. (\bibinfo{year}{2023}).
\newblock \bibinfo{title}{The unified cluster catalogue: towards a comprehensive and homogeneous data base of stellar clusters}.
\newblock {\it \bibinfo{journal}{Monthly Notices of the Royal Astronomical Society}\/},  {\it \bibinfo{volume}{526}\/}\bibinfo{issue}{(3)}, \bibinfo{pages}{4107--4119}. \URLprefix \url{https://doi.org/10.1093/mnras/stad2826}. \DOIprefix\doi{10.1093/mnras/stad2826}. \href{http://arxiv.org/abs/https://academic.oup.com/mnras/article-pdf/526/3/4107/52189100/stad2826.pdf}{\tt arXiv:https://academic.oup.com/mnras/article-pdf/526/3/4107/52189100/stad2826.pdf}.

\bibitem[{{Poggio, E.}(2021)}]{Pog21}
\bibinfo{author}{{Poggio, E.}} (\bibinfo{year}{2021}).
\newblock {\it \bibinfo{journal}{Astronomy \& Astrophysics}\/},  {\it \bibinfo{volume}{651}\/}, \bibinfo{pages}{A104}.

\bibitem[{{Robichon} et~al.(1999){Robichon}, {Arenou}, {Mermilliod} \& {Turon}}]{1999A&A...345..471R}
\bibinfo{author}{{Robichon}, N.}, \bibinfo{author}{{Arenou}, F.}, \bibinfo{author}{{Mermilliod}, J.-C.} et~al. (\bibinfo{year}{1999}).
\newblock \bibinfo{title}{{Open clusters with Hipparcos. I. Mean astrometric parameters}}.
\newblock {\it \bibinfo{journal}{\aap}\/},  {\it \bibinfo{volume}{345}\/}, \bibinfo{pages}{471--484}. \DOIprefix\doi{10.48550/arXiv.astro-ph/9903131}. \href{http://arxiv.org/abs/astro-ph/9903131}{\tt arXiv:astro-ph/9903131}.

\bibitem[{{Rojo Arellano} et~al.(1997){Rojo Arellano}, {Pena} \& {Gonzalez}}]{1997A&AS..123...25R}
\bibinfo{author}{{Rojo Arellano}, E.}, \bibinfo{author}{{Pena}, J.~H.},  \& \bibinfo{author}{{Gonzalez}, D.} (\bibinfo{year}{1997}).
\newblock \bibinfo{title}{{Photoelectric uvby-beta photometry of the open cluster NGC 2422}}.
\newblock {\it \bibinfo{journal}{\aaps}\/},  {\it \bibinfo{volume}{123}\/}, \bibinfo{pages}{25--30}. \DOIprefix\doi{10.1051/aas:1997303}.

\bibitem[{{Salaris} et~al.(2004){Salaris}, {Weiss} \& {Percival}}]{2004A&A...414..163S}
\bibinfo{author}{{Salaris}, M.}, \bibinfo{author}{{Weiss}, A.},  \& \bibinfo{author}{{Percival}, S.~M.} (\bibinfo{year}{2004}).
\newblock \bibinfo{title}{{The age of the oldest Open Clusters}}.
\newblock {\it \bibinfo{journal}{\aap}\/},  {\it \bibinfo{volume}{414}\/}, \bibinfo{pages}{163--174}. \DOIprefix\doi{10.1051/0004-6361:20031578}. \href{http://arxiv.org/abs/astro-ph/0310363}{\tt arXiv:astro-ph/0310363}.

\bibitem[{{Salpeter, E. E.}(1955)}]{salpeter1955}
\bibinfo{author}{{Salpeter, E. E.}} (\bibinfo{year}{1955}).
\newblock {\it \bibinfo{journal}{Astrophysical Journal}\/},  {\it \bibinfo{volume}{121}\/}, \bibinfo{pages}{161}.

\bibitem[{{Sampedro, L.} et~al.(2017){Sampedro, L.}, {Dias, W.} \& {Alfaro, E. J.}}]{sampedro2017multimembership}
\bibinfo{author}{{Sampedro, L.}}, \bibinfo{author}{{Dias, W.}},  \& \bibinfo{author}{{Alfaro, E. J.}} (\bibinfo{year}{2017}).
\newblock {\it \bibinfo{journal}{Monthly Notices of the Royal Astronomical Society}\/},  {\it \bibinfo{volume}{470}\/}, \bibinfo{pages}{3937}.

\bibitem[{{Spitzer} \& {Hart}(1971)}]{Spitzer71}
\bibinfo{author}{{Spitzer}, J., Lyman},  \& \bibinfo{author}{{Hart}, M.~H.} (\bibinfo{year}{1971}).
\newblock \bibinfo{title}{{Random Gravitational Encounters and the Evolution of Spherical Systems. I. Method}}.
\newblock {\it \bibinfo{journal}{\apj}\/},  {\it \bibinfo{volume}{164}\/}, \bibinfo{pages}{399}. \DOIprefix\doi{10.1086/150855}.

\bibitem[{{Ta{\c{s}}demir} \& {{\c{C}}{\i}nar}(2025)}]{Tasdemir2025a}
\bibinfo{author}{{Ta{\c{s}}demir}, S.},  \& \bibinfo{author}{{{\c{C}}{\i}nar}, D.~C.} (\bibinfo{year}{2025}).
\newblock \bibinfo{title}{{A detailed analysis of close binary OCs}}.
\newblock {\it \bibinfo{journal}{Contributions of the Astronomical Observatory Skalnaté Pleso}\/},  {\it \bibinfo{volume}{55}\/}\bibinfo{issue}{(3)}, \bibinfo{pages}{506--511}. \URLprefix \url{https://doi.org/10.31577/caosp.2025.55.3.506}. \DOIprefix\doi{10.31577/caosp.2025.55.3.506}. \href{http://arxiv.org/abs/2501.17235}{\tt arXiv:2501.17235}.

\bibitem[{{Ta{\c{s}}demir} \& {Yontan}(2023)}]{Tasdemir2023}
\bibinfo{author}{{Ta{\c{s}}demir}, S.},  \& \bibinfo{author}{{Yontan}, T.} (\bibinfo{year}{2023}).
\newblock \bibinfo{title}{{Analysis of the Young Open Cluster Trumpler 2 Using Gaia DR3 Data}}.
\newblock {\it \bibinfo{journal}{Physics and Astronomy Reports}\/},  {\it \bibinfo{volume}{1}\/}\bibinfo{issue}{(1)}, \bibinfo{pages}{1--10}. \DOIprefix\doi{10.26650/PAR.2023.00001}. \href{http://arxiv.org/abs/2304.14270}{\tt arXiv:2304.14270}.

\bibitem[{{Tarricq} et~al.(2021){Tarricq}, {Soubiran}, {Casamiquela}, {Cantat-Gaudin}, {Chemin}, {Anders}, {Antoja}, {Romero-G{\'o}mez}, {Figueras}, {Jordi}, {Bragaglia}, {Balaguer-N{\'u}{\~n}ez}, {Carrera}, {Castro-Ginard}, {Moitinho}, {Ramos} \& {Bossini}}]{2021A&A...647A..19T}
\bibinfo{author}{{Tarricq}, Y.}, \bibinfo{author}{{Soubiran}, C.}, \bibinfo{author}{{Casamiquela}, L.} et~al. (\bibinfo{year}{2021}).
\newblock \bibinfo{title}{{3D kinematics and age distribution of the open cluster population}}.
\newblock {\it \bibinfo{journal}{\aap}\/},  {\it \bibinfo{volume}{647}\/}, \bibinfo{pages}{A19}. \DOIprefix\doi{10.1051/0004-6361/202039388}. \href{http://arxiv.org/abs/2012.04017}{\tt arXiv:2012.04017}.

\bibitem[{{Tasdemir} et~al.(2025){Tasdemir}, {{\c{C}}{\i}nar}, {Canbay}, {Tastan}, {Elsanhoury} \& {Haroon}}]{Tasdemir2025}
\bibinfo{author}{{Tasdemir}, S.}, \bibinfo{author}{{{\c{C}}{\i}nar}, D.~C.}, \bibinfo{author}{{Canbay}, R.} et~al. (\bibinfo{year}{2025}).
\newblock \bibinfo{title}{{Comprehensive Analysis of Middle-Aged Open Cluster NGC 6793 in Vulpecula via Gaia DR3 Data}}.
\newblock {\it \bibinfo{journal}{Physics and Astronomy Reports}\/},  {\it \bibinfo{volume}{3}\/}, \bibinfo{pages}{1--12}. \DOIprefix\doi{10.48550/arXiv.2503.19015}. \href{http://arxiv.org/abs/2503.19015}{\tt arXiv:2503.19015}.

\bibitem[{{Tasdemir} et~al.(2026){Tasdemir}, {Elsanhoury}, {{\c{C}}{\i}nar}, {Haroon} \& {Bilir}}]{Tasdemir2026}
\bibinfo{author}{{Tasdemir}, S.}, \bibinfo{author}{{Elsanhoury}, W.}, \bibinfo{author}{{{\c{C}}{\i}nar}, D.~C.} et~al. (\bibinfo{year}{2026}).
\newblock \bibinfo{title}{{Comprehensive Gaia DR3-Based Astrometric, Photometric, and Kinematic Studies of the Binary Open Cluster h and X Persei}}.
\newblock {\it \bibinfo{journal}{arXiv e-prints}\/},  (p. \bibinfo{pages}{arXiv:2601.16252}). \DOIprefix\doi{10.48550/arXiv.2601.16252}. \href{http://arxiv.org/abs/2601.16252}{\tt arXiv:2601.16252}.

\bibitem[{{Trumpler}(1930)}]{1930LicOB..14..154T}
\bibinfo{author}{{Trumpler}, R.~J.} (\bibinfo{year}{1930}).
\newblock \bibinfo{title}{{Preliminary results on the distances, dimensions and space distribution of open star clusters}}.
\newblock {\it \bibinfo{journal}{Lick Observatory Bulletin}\/},  {\it \bibinfo{volume}{420}\/}, \bibinfo{pages}{154--188}. \DOIprefix\doi{10.5479/ADS/bib/1930LicOB.14.154T}.

\bibitem[{{Tsantaki} et~al.(2022){Tsantaki}, {Pancino}, {Marrese}, {Marinoni}, {Rainer}, {Sanna}, {Turchi}, {Randich}, {Gallart}, {Battaglia} \& {Masseron}}]{Tsantaki2022}
\bibinfo{author}{{Tsantaki}, M.}, \bibinfo{author}{{Pancino}, E.}, \bibinfo{author}{{Marrese}, P.} et~al. (\bibinfo{year}{2022}).
\newblock \bibinfo{title}{{Survey of Surveys. I. The largest compilation of radial velocities for the Galaxy}}.
\newblock {\it \bibinfo{journal}{\aap}\/},  {\it \bibinfo{volume}{659}\/}, \bibinfo{pages}{A95}. \DOIprefix\doi{10.1051/0004-6361/202141702}. \href{http://arxiv.org/abs/2110.09316}{\tt arXiv:2110.09316}.

\bibitem[{{Twarog} et~al.(1997){Twarog}, {Ashman} \& {Anthony-Twarog}}]{1997AJ....114.2556T}
\bibinfo{author}{{Twarog}, B.~A.}, \bibinfo{author}{{Ashman}, K.~M.},  \& \bibinfo{author}{{Anthony-Twarog}, B.~J.} (\bibinfo{year}{1997}).
\newblock \bibinfo{title}{{Some Revised Observational Constraints on the Formation and Evolution of the Galactic Disk}}.
\newblock {\it \bibinfo{journal}{\aj}\/},  {\it \bibinfo{volume}{114}\/}, \bibinfo{pages}{2556}. \DOIprefix\doi{10.1086/118667}. \href{http://arxiv.org/abs/astro-ph/9709122}{\tt arXiv:astro-ph/9709122}.

\bibitem[{{von Hippel}(2005)}]{2005astro.ph..9152V}
\bibinfo{author}{{von Hippel}, T.} (\bibinfo{year}{2005}).
\newblock \bibinfo{title}{{Galactic Open Clusters}}.
\newblock {\it \bibinfo{journal}{arXiv e-prints}\/},  (pp. \bibinfo{pages}{astro--ph/0509152}). \DOIprefix\doi{10.48550/arXiv.astro-ph/0509152}. \href{http://arxiv.org/abs/astro-ph/0509152}{\tt arXiv:astro-ph/0509152}.

\bibitem[{{{\v{S}}ablevi{\v{c}}i{\={u}}t{\.{e}}} et~al.(2006){{\v{S}}ablevi{\v{c}}i{\={u}}t{\.{e}}}, {Vansevi{\v{c}}ius}, {Kodaira}, {Narbutis}, {Stonkut{\.{e}}} \& {Brid{\v{z}}ius}}]{Sableviciute_2006}
\bibinfo{author}{{{\v{S}}ablevi{\v{c}}i{\={u}}t{\.{e}}}, I.}, \bibinfo{author}{{Vansevi{\v{c}}ius}, V.}, \bibinfo{author}{{Kodaira}, K.} et~al. (\bibinfo{year}{2006}).
\newblock \bibinfo{title}{{A Survey of Compact Star Clusters in the South-West Field of the M 31 Disk. Structural Parameters}}.
\newblock {\it \bibinfo{journal}{Baltic Astronomy}\/},  {\it \bibinfo{volume}{15}\/}, \bibinfo{pages}{547--560}. \DOIprefix\doi{10.48550/arXiv.astro-ph/0701774}. \href{http://arxiv.org/abs/astro-ph/0701774}{\tt arXiv:astro-ph/0701774}.

\bibitem[{{Wu} et~al.(2009){Wu}, {Zhou}, {Ma} \& {Du}}]{Wu2009}
\bibinfo{author}{{Wu}, Z.-Y.}, \bibinfo{author}{{Zhou}, X.}, \bibinfo{author}{{Ma}, J.} et~al. (\bibinfo{year}{2009}).
\newblock \bibinfo{title}{{The orbits of open clusters in the Galaxy}}.
\newblock {\it \bibinfo{journal}{\mnras}\/},  {\it \bibinfo{volume}{399}\/}\bibinfo{issue}{(4)}, \bibinfo{pages}{2146--2164}. \DOIprefix\doi{10.1111/j.1365-2966.2009.15416.x}. \href{http://arxiv.org/abs/0909.3737}{\tt arXiv:0909.3737}.

\bibitem[{{Xin} et~al.(2007){Xin}, {Deng} \& {Han}}]{2007ApJ...660..319X}
\bibinfo{author}{{Xin}, Y.}, \bibinfo{author}{{Deng}, L.},  \& \bibinfo{author}{{Han}, Z.~W.} (\bibinfo{year}{2007}).
\newblock \bibinfo{title}{{Blue Straggler Stars in Galactic Open Clusters and the Simple Stellar Population Model}}.
\newblock {\it \bibinfo{journal}{\apj}\/},  {\it \bibinfo{volume}{660}\/}\bibinfo{issue}{(1)}, \bibinfo{pages}{319--329}. \DOIprefix\doi{10.1086/512964}. \href{http://arxiv.org/abs/astro-ph/0609745}{\tt arXiv:astro-ph/0609745}.

\bibitem[{{Yontan}(2023)}]{2023AJ....165...79Y}
\bibinfo{author}{{Yontan}, T.} (\bibinfo{year}{2023}).
\newblock \bibinfo{title}{{An Investigation of Open Clusters Berkeley 68 and Stock 20 Using CCD UBV and Gaia DR3 Data}}.
\newblock {\it \bibinfo{journal}{The Astronomical Journal}\/},  {\it \bibinfo{volume}{165}\/}\bibinfo{issue}{(3)}, \bibinfo{pages}{79}. \DOIprefix\doi{10.3847/1538-3881/aca6f0}. \href{http://arxiv.org/abs/2211.09825}{\tt arXiv:2211.09825}.

\bibitem[{{Yontan} et~al.(2019){Yontan}, {Bilir}, {Bostanc{\i}}, {Ak}, {Ak}, {G{\"u}ver}, {Paunzen}, {{\"U}rg{\"u}p}, {{\c{C}}elebi}, {Akti} \& {G{\"o}kmen}}]{2019Ap&SS.364..152Y}
\bibinfo{author}{{Yontan}, T.}, \bibinfo{author}{{Bilir}, S.}, \bibinfo{author}{{Bostanc{\i}}, Z.~F.} et~al. (\bibinfo{year}{2019}).
\newblock \bibinfo{title}{{CCD UBV photometric and Gaia astrometric study of eight open clusters{\textemdash}ASCC 115, Collinder 421, NGC 6793, NGC 7031, NGC 7039, NGC 7086, Roslund 1 and Stock 21}}.
\newblock {\it \bibinfo{journal}{Astrophysics and Space Science}\/},  {\it \bibinfo{volume}{364}\/}\bibinfo{issue}{(9)}, \bibinfo{pages}{152}. \DOIprefix\doi{10.1007/s10509-019-3640-y}. \href{http://arxiv.org/abs/1909.00345}{\tt arXiv:1909.00345}.

\bibitem[{{Yontan} et~al.(2015){Yontan}, {Bilir}, {Bostanc{\i}}, {Ak}, {Karaali}, {G{\"u}ver}, {Ak}, {Duran} \& {Paunzen}}]{2015Ap&SS.355..267Y}
\bibinfo{author}{{Yontan}, T.}, \bibinfo{author}{{Bilir}, S.}, \bibinfo{author}{{Bostanc{\i}}, Z.~F.} et~al. (\bibinfo{year}{2015}).
\newblock \bibinfo{title}{{CCD UBVRI photometry of NGC 6811}}.
\newblock {\it \bibinfo{journal}{Astrophysics and Space Science}\/},  {\it \bibinfo{volume}{355}\/}\bibinfo{issue}{(2)}, \bibinfo{pages}{267--281}. \DOIprefix\doi{10.1007/s10509-014-2175-5}. \href{http://arxiv.org/abs/1410.3129}{\tt arXiv:1410.3129}.

\bibitem[{{Yontan} \& {Canbay}(2023)}]{YontanCanbay2023}
\bibinfo{author}{{Yontan}, T.},  \& \bibinfo{author}{{Canbay}, R.} (\bibinfo{year}{2023}).
\newblock \bibinfo{title}{{Comprehensive Analysis of the Open Cluster Collinder 74}}.
\newblock {\it \bibinfo{journal}{Physics and Astronomy Reports}\/},  {\it \bibinfo{volume}{1}\/}\bibinfo{issue}{(2)}, \bibinfo{pages}{65}. \DOIprefix\doi{10.26650/PAR.2023.00008}. \href{http://arxiv.org/abs/2310.13582}{\tt arXiv:2310.13582}.

\bibitem[{{Yontan} et~al.(2022){Yontan}, {{\c{C}}akmak}, {Bilir}, {Banks}, {Ra{\'u}l}, {Canbay}, {Ko{\c{c}}}, {Ta{\c{s}}demir}, {Er{\c{c}}ay}, {Tan{\i}k Ozt{\"u}rk} \& {Dursun}}]{Yontan2022}
\bibinfo{author}{{Yontan}, T.}, \bibinfo{author}{{{\c{C}}akmak}, T.}, \bibinfo{author}{{Bilir}, S.} et~al. (\bibinfo{year}{2022}).
\newblock \bibinfo{title}{{A Study of the NGC 1193 and NGC 1798 Open Clusters Using CCD UBV Photometric and Gaia EDR3 Data}}.
\newblock {\it \bibinfo{journal}{The Revista Mexicana de Astronomía y Astrofísica}\/},  {\it \bibinfo{volume}{58}\/}, \bibinfo{pages}{333--353}. \DOIprefix\doi{10.22201/ia.01851101p.2022.58.02.14}. \href{http://arxiv.org/abs/2207.06407}{\tt arXiv:2207.06407}.

\bibitem[{{Zhong, J.} et~al.(2019){Zhong, J.}, {Chen, L.} \& {Kouwenhoven, M.}}]{zhong2019substructure}
\bibinfo{author}{{Zhong, J.}}, \bibinfo{author}{{Chen, L.}},  \& \bibinfo{author}{{Kouwenhoven, M.}} (\bibinfo{year}{2019}).
\newblock {\it \bibinfo{journal}{Astronomy \& Astrophysics}\/},  {\it \bibinfo{volume}{624}\/}, \bibinfo{pages}{A34}.

\bibitem[{Çınar et~al.(2026)Çınar, Bisht, Bilir, Qin \& Saker}]{Cinar2026b}
\bibinfo{author}{Çınar, D.~C.}, \bibinfo{author}{Bisht, D.}, \bibinfo{author}{Bilir, S.} et~al. (\bibinfo{year}{2026}).
\newblock \bibinfo{title}{Hot degenerate components in blue stragglers: A multiwavelength spectral energy distribution analysis of nine open clusters with swift/uvot}.
\newblock {\it \bibinfo{journal}{The Astrophysical Journal}\/},  {\it \bibinfo{volume}{1006}\/}\bibinfo{issue}{(1)}, \bibinfo{pages}{109}. \URLprefix \url{https://doi.org/10.3847/1538-4357/ae7a33}. \DOIprefix\doi{10.3847/1538-4357/ae7a33}.

\end{thebibliography}

\end{document}